\documentclass[showpacs,amsmath,amsfonts,amssymb,aps,superscriptaddress]{revtex4}
\usepackage{tabularx}
\usepackage{comment}
\usepackage{tikz}
\usetikzlibrary{decorations.pathreplacing}
\usepackage{enumitem}
\usepackage{graphicx}
\usepackage[font=scriptsize]{caption} 
\usepackage{float}
\usepackage{dcolumn}
\usepackage{bm}
\usepackage{mathrsfs}
\usepackage{amsmath}

\newcommand{\be}{\begin{eqnarray}}
\newcommand{\ee}{\end{eqnarray}}
\newcommand{\om}{\alpha}
\usepackage{amsfonts} 
\usepackage{caption}
\usepackage[compatibility=false,font=small]{caption}

\begin{document}

\title{Quasinormal modes of Bonanno-Reuter black holes via the Spectral Method}
\author{Davide Batic}
\email{davide.batic@ku.ac.ae}
\affiliation{Mathematics Department, Khalifa University of Science and Technology, PO Box 127788, Abu Dhabi, United Arab Emirates}

\author{Denys Dutykh}
\email{denys.dutykh@ku.ac.ae}
\affiliation{Mathematics Department, Khalifa University of Science and Technology, PO Box 127788, Abu Dhabi, United Arab Emirates}

\author{Fabio Scardigli}
\email{fabio@phys.ntu.edu.tw}
\affiliation{Dipartimento di Matematica, Politecnico di Milano, Piazza Leonardo da Vinci 32, 20133 Milano, Italy}
\affiliation{Institute-Lorentz for Theoretical Physics, Leiden University, \\P.O. Box 9506, Leiden, The Netherlands}

\date{\today}

\begin{abstract}
In this work, we explore the quasinormal modes (QNMs) of the Bonanno-Reuter black hole, one of the first regular black hole metric suggested by the Asymptotically Safe Gravity (ASG) program. The running parameter $\alpha$ is set to a positive value, the related running Newton coupling vanishes at high energies, fully achieving an ultraviolet fixed point and eliminating non-physical UV divergences. This yields a singularity-free geometry. Hence, we focus on the resulting renormalisation-group-improved Schwarzschild metric, which naturally produces an (Anti)deSitter non-singular core. On the basis of this background, we compute the QNM spectrum for scalar, electromagnetic, and gravitational perturbations by employing the Spectral Method (SM). This method, recognised for its enhanced precision compared to high-order WKB methods, allows the identification of fundamental modes, extensive collections of overtones, and purely imaginary overdamped modes that were entirely missed in previous analyses. These characteristics, resolved here for the first time in the Bonanno-Reuter black hole, underscore the crucial importance of high-precision spectral methods in investigating delicate signatures of black hole models inspired by quantum gravity.
\end{abstract}

\pacs{XXX}
\maketitle

\section{Introduction}\label{Intro}

During the second half of the Seventies, it became increasingly clear that a standard quantisation of General Relativity (GR), namely a canonical or a path-integral covariant one, would have inevitably led to a perturbatively non-renormalisable quantum field theory \cite{Sagnotti1985}, which requires, in principle, the adjustment of infinitely many counterterms to make physical predictions. The theoretical physics community then reacted to this state of affairs by producing alternative schemes, all aimed at achieving what is, as then and now, seen as the holy grail of fundamental physics: a Quantum theory of Gravitation (QG). Among the (then) nascent theories, such as string theory, or supersymmetry theory which was immediately addressed towards a theory of supergravity, or also theories containing higher derivative terms (these, however, plagued by ghost degree of freedom), there was the Weinberg's suggestion \cite{Weinberg1, Weinberg2}, developed further by others (see, e.g., \cite{Wetterich93, Reuter98, Niedermaier2006}) about the possibility that a gravitational quantum theory could \textit{dynamically} sidestep the divergences found in perturbative gravity. Within this concept, referred to as Asymptotically Safe Gravity (ASG), the main idea is the existence of a non-Gaussian fixed point in the gravitational renormalisation group (RG) flow, which influences the high-energy dynamics of the theory and ensures the elimination of non-physical UV divergences. The RG-flow equation can then be integrated to derive a varying Newton's constant $G(k)$, dependent on the momentum/energy scale $k$.

In recent years, ASG has been revealed as one of the most promising avenues into the still largely uncharted territory of Quantum Gravity. Seen as an effective field theory, the ASG framework allows scholars to model and compute phenomena at energy scales which, if not coincident, are at least quite close to the \textit{magic} threshold of Planck energy ($10^{19}$ GeV). A precious fruit of the ASG approach has been the development of various black hole metrics free of central singularities, now built solely on the basis of the ASG general framework, rather than on ad hoc \textit{ansatzes}.

In black hole physics, the running coupling $G(k)$ can be integrated into the classical solution to create a refined lapse function. At this point, though, $G$ still depends on a chosen renormalisation scale $k$, and it is necessary to establish a physically meaningful connection between $k$ and the radial coordinate $r$ of spacetime. After this identification is carried out, the composite function $G(k(r))$ can be reintroduced into the classical metric, yielding a completely defined, quantum-enhanced lapse function.
However, a weak point is that this identification is phenomenological and not unique. Various prescriptions for the function $k(r)$, all consistent with fundamental physical principles, yield distinct RG-improved geometries. Such extensions, motivated by the asymptotic safety program, naturally modify classical black hole geometries and encode quantum-gravity effects.

Running couplings (gravitational, electromagnetic, and even the cosmological constant) are a generic feature of quantum field theory at the level of the effective action. The scale dependence is expected
to influence essential black hole properties, including the horizon structure, thermodynamics, and QNM spectra, as explored, for example, in \cite{Koch16, Ricon19}. Of course, once the ASG-corrected quantum metric is set back into Einstein equations, we should interpret the resulting right-hand side of the equations as an effective stress–energy tensor $T^{(eff)}_{\mu\nu}$, encoding short-distance quantum-gravity effects. As in other quantum-corrected black-hole spacetimes, $T^{(eff)}_{\mu\nu}$ is not expected to obey the standard classical energy conditions in the vicinity of the (Anti) de Sitter core and should not be regarded as arising from a conventional matter field with a well-defined equation of state (for a detailed analysis of the energy conditions with ASG metrics, see Ref.\cite{Hassannejad2025PRD}, Sec.VII).

In this paper, we investigate the Quasi Normal Modes (QNM) of the Bonanno-Reuter black hole \cite{Bonanno2000PRD}, one of the first examples of a singularity-free metric constructed on the basis of the ASG framework. In our setup, the parameter $\alpha$ is fixed to a positive value (as it was the parameter $\tilde{\omega}$ in the original paper of Bonanno and Reuter), so that the associated running Newton coupling $G$ realises the ultraviolet Gaussian fixed point of asymptotically safe gravity. We focus on the resulting renormalisation-group-improved Schwarzschild metric, which naturally yields a singularity-free (Anti) de Sitter core.

As in the companion paper \cite{BDS-EPJC2026}, we compute the QNM spectrum for scalar, electromagnetic, and gravitational perturbations on this background using the Spectral Method (SM). Unlike what was done in the Refs.\cite{Rincon2020PDU, Konoplya2022JCAP}, we consider both the non-extreme case, as well as the full extreme case, extending the analysis to the gravitational perturbations. It is well known that the spectral method offers greater precision than high-order WKB techniques and yields the identification of fundamental modes, extensive collections of overtones, and purely imaginary overdamped modes that were completely overlooked in earlier evaluations. 

Among the main physical motivations calling for the most accurate study of QNM, there is certainly the direct detection of gravitational waves from black hole mergers \cite{Abbott2016PRL, Abbott2019}, which has sparked interest in black hole perturbations \cite{Regge1957PR, Teukolsky1972} and associated QNMs.

The complex-frequency QNM oscillations are believed to dominate the \textit{ringdown} phase, where the merger object settles into equilibrium through damped spacetime oscillations and emission of gravitational waves \cite{Ferrari2008}. QNMs offer a distinct spectral signature of the black hole, since they depend only on the geometry of the background and on the type of perturbation considered (scalar, electromagnetic, gravitational, or fermionic) \cite{Cardoso2003, Rincon2020PDU, Berti2009CQG, Konoplya2011RMP}. As expected, for large astrophysical black holes, the observed QNMs closely follow those predicted for the Schwarzschild solution, with great accuracy. This makes the quantum-gravity modifications suggested by ASG (as well as by other QG models) practically unnoticeable \cite{Liu2012}.

On the contrary, for small or primordial black holes, quantum effects may play a major role, and QNMs provide a precise tool for differentiating between various regular black hole models. In comparison to alternative observational instruments, QNM spectroscopy is remarkably accurate \cite{Franchini2023}. Measurements of black hole shadows from the Event Horizon Telescope have uncertainties around $10\%$, while the fundamental QNM can independently constrain parameters with a precision larger than $90\%$. Future detectors, combined with the incorporation of overtone modes \cite{Giesler2019PRX, Spina24}, offer enhanced precision, positioning QNMs as a key tool for exploring alternatives to classical general relativity.

QNMs of regular black hole models have been studied in a large number of papers \cite{Konoplya2022PRL, Konoplya2022JCAP, KonoplyaPRD2023}, which focus in particular on Hayward metric \cite{Flachi2013, DuttaRoy2022} and on the Bonanno-Reuter black hole \cite{Rincon2020PDU, Liu2012}. Nevertheless, we believe that many important and fascinating aspects of the QNM spectrum have been neglected in previous studies, largely due to the systematic oversight of overtone modes. Common knowledge usually holds that the fundamental mode prevails in the gravitational-wave signal, numerical relativity simulations \cite{Giesler2019PRX} demonstrate that capturing the complete ringdown phase requires accounting for about ten overtones. This point highlights the physical significance of overtones and suggests that quasinormal ringing starts notably sooner in the post-merger signal than previously suspected.

The search for overtones has necessitated the choice of a computational method capable of accurately addressing both extensive sets of overtones and various other delicate spectral characteristics. We found that the SM is perfectly matched for this goal \cite{Batic2024CQG, Batic2024EPJC, Batic2024PRD, Batic2025EPJC, Batic2025CQG, BDS-EPJC2026}: SM accurately reproduces the benchmark Schwarzschild QNMs across a wide spectrum of spins and multipoles, reveals overdamped modes systematically overlooked by high-order WKB methods, clarifies broad overtone spectra showing spectral convergence, while also maintaining numerical stability under extreme parameter ranges. This blend of precision, thoroughness, and strength makes SM especially well-suited to the standard black hole scenario, where identifying isolated overdamped modes, unusual intermode gaps, and other intricate spectral details requires an outstanding resolution in the complex-frequency domain.

Comparison with earlier research further strengthens our choice for the Spectral Method, as we did in our previous publications on the topic. As a matter of fact, our analysis accurately reproduces the benchmark Schwarzschild QNM spectrum in the large-mass limit $M \gg 1$, a consistency test that was, for example, failed by Ref.~\cite {Lambiase2023EPJC}. Most importantly, high-order WKB methods (see, e.g., Ref.~\cite {Lambiase2023EPJC}) are unable to identify the purely imaginary, overdamped modes that, on the contrary, arise naturally in our spectral analysis. Similarly, Refs.\cite{Rincon2020PDU, Konoplya2022JCAP} depend on WKB-based methods, and they too overlook such modes, along with the intricate spectral structures uncovered in our study. Similar problems emerge in other recent studies \cite{Malik2024EPL, Stashko2024PRD} concerning regular or ASG-inspired black holes. These studies either do not confirm the large-mass Schwarzschild limit, confine their analysis to a few low-lying modes, or rely on semi-analytic approaches that ignore the overdamped sector completely. All WKB methods and other analytical literature usually share these recurrent defects. These techniques can identify the basic mode in straightforward situations, but they do not fully capture the entire spectral structure. It is clear that this kind of failure can be catastrophic, in particular in regular black hole metrics (such as Bonanno-Reuter), where quantum corrections are expected to significantly affect the high-overtone and purely imaginary modes. In contrast, the SM utilised in the present work not only accurately resolves the fundamental and higher-overtone modes but also detects the overdamped sector, even in parameter ranges where WKB approximations yield incomplete or misleading spectra.

\section{Asymptotically safe gravity}

Within the ASG program, renormalization-group--improved solutions in Newtonian gravity or general relativity are obtained by replacing the constant Newton coupling $G_N$ with a running coupling $G(k)$, depending on a RG energy/momentum scale $k$, and identifying a physically motivated relation $k=k(r)$ between the RG scale $k$ and the spacetime coordinate \cite{Bonanno2000PRD, Lambiase2022PRD}. In the formulation adopted here,
\be
\label{Gk}
  G(k)= \frac{G_N}{1+ \om G_N k^2/\hbar}, \qquad {\rm with} \qquad k(r) = \hbar\left(\frac{r+\gamma G_N M}{r^3} \right)^{1/2},
\ee
where $\om$ and $\gamma$ are dimensionless constants, $c = k_B = 1$, and $\hbar$ is retained explicitly. This scale-setting prescription is phenomenological. It is motivated by the analyses of \cite{Bonanno2000PRD, Bonanno2004} and by the requirement of reproducing the quantum-corrected Newtonian potential once $\alpha$ and $\gamma$ are fixed, but it is not unique. Alternatively, equally reasonable identifications of $k(r)$ would lead to different functions $G(r)$ and thus to different RG-improved black-hole geometries. Combining these expressions gives the scale-dependent Newton constant,
\be \label{Geffr}
  G(r) = \frac{G_N r^3}{r^3 + \om G_N \hbar \left(r + \gamma G_N M \right)},
\ee
which reduces to the standard Newtonian coupling in the classical limit $\hbar \to 0$. 
Clearly, the presence of $\hbar$ signals the quantum character of the correction that the ASG approach gives to the standard general relativity theory. In fully geometrized Planck units ($c = k_B = G_N = \hbar = 1$), this simplifies to
\be
\label{run_c}
G(r)=\frac{r^3}{r^3+\alpha(r+\gamma M)}\,.
\ee
Since the Planck length $\ell_p=\sqrt{G_N\hbar/c^3}$ and the Planck mass $m_p=\sqrt{\hbar c/4G_N}$, then in geometrized units we have $\ell_p=1$, and $m_p=1/2$, hence it follows that $r$ is measured in Planck lengths, and $M$ is the dimensionless geometric mass linked to the physical mass via $M_{\rm phys}=2M m_p$.

In the ASG literature, the parameter $\om$ is always taken to be positive. In fact, this ensures that $dG/dk<0$, $G(k) \geq 0$, and $G(k) \to 0$ as $k \to \infty$, preserving asymptotic safety and avoiding UV divergences. Negative $\om$ values, in contrast, lead to a divergence of $G(k)$ near the Planck scale, sign changes for $k > k_{\mathrm{Planck}}$, and $G(k) \to 0^{-}$ at high energies. As a result, for $\alpha<0$ the UV behaviour of $G(k)$ lies outside the asymptotically safe gravity paradigm in the strict sense, although a theory with $\alpha<0$ can still be classified as a Scale Dependent Gravity theory, as happens for example for the Planck-star metric considered in \cite{Scardigli2023PRD}.

\section{Prolegomena on Bonanno-Reuter black holes}

The renormalisation group (RG)‐improved Schwarzschild line element in Boyer–Lindquist coordinates becomes (see \cite{Bonanno2000PRD, Lambiase2022PRD, Scardigli2023PRD})  
\begin{equation}\label{LE}
  ds^2=-F(r)dt^2+\frac{dr^2}{F(r)}+r^2 d\vartheta^2+r^2\sin^2{\vartheta}d\varphi^2, \qquad\quad F(r)=1-\frac{2MG(r)}{r}=1-\frac{2 M r^2}{r^3 +  \om (r + \gamma  M)}.
\end{equation}  
Note that the usual Schwarzschild metric is recovered either for $\alpha=0$ or in the low energy scales, i.e. as $r\to\infty$ or $k\to 0$. For more information on how the energy scale $k$ arises from the identification of the infrared cutoff, we refer to \cite{Bonanno2000PRD}. As already observed by \cite{Scardigli2023PRD}, at high energy scales, i.e. $r\to 0$ or $k\to\infty$, the metric \eqref{LE} goes over to   
\begin{equation}
  F(r \to 0) \simeq 1 - \frac{2 r^2}{\om \gamma},
\end{equation}
which, depending on the sign of $\om \gamma$, yields a regular de Sitter ($\om\gamma > 0$) or anti–de Sitter ($\om\gamma < 0$) core. In other words, the RG-improved metric interpolates between Schwarzschild at large $r$ and a (A)dS-like core at small $r$. It is interesting to observe that whenever $\gamma=0$ the $g_{00}$ metric coefficient becomes
\begin{equation}
  F(r)=1-\frac{2M r}{r^2+\alpha}\,.
\end{equation}
In this case, although $F(0) = 1$ is finite, the Kretschmann invariant (for $\gamma=0$)
\begin{equation}
  R^{\mu\nu\rho\lambda}R_{\mu\nu\rho\lambda}=\frac{16 M^{2}\left(3r^{8}-2 \alpha r^{6}+13\alpha^{2}r^{4}+4\alpha^{3}r^{2}+2 \alpha^{4}\right)}{r^2\left(r^{2}+\alpha\right)^{6}}
\end{equation}
becomes singular at $r=0$ and therefore, such a black hole is still plagued by a central (conic) singularity. Concerning the parameters entering in \eqref{run_c}, we will assume $\alpha=118/(15\pi)$, as in \cite{Bonanno2000PRD}. To gain insights into the historical evolution of the numerical value of $\alpha$, we refer to \cite{Scardigli2023PRD, Hamber1995PLB, Bjerrum2003PRD, Khriplovich2002JETP, Bjerrum2003PRDa, Khriplovich2004JETP, Akhundov2008EJTP, Kiefer, Bjerrum2015PRL, Donoghue2015JPG, Batic2016EPJC, Batic2017EPL}. Moreover, following the approach in \cite{Bonanno2000PRD, Lambiase2023EPJC}, we adopt the value $\gamma = 9/2$.

\begin{figure}[h]
\centering
\includegraphics[scale=1.2]{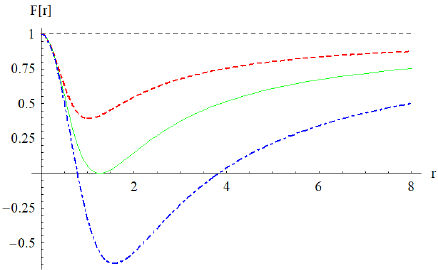}
\caption{Lapse function $F(r)$ for $\om > 0$: Horizons correspond to the zeros of $F(r)$. The curves are shown for a fixed $\om > 0$ and increasing values of the mass parameter $M$: $0 < M_{\text{red}} < M_{\text{green}} \equiv M_c < M_{\text{blue}}$. At the critical mass $M = M_c$ (green curve), the two horizons merge, corresponding to the extremal black hole configuration. For $M > M_c$ (blue curve), the lapse exhibits two distinct horizons $r_-$ and $r_+$, where $F(r_-) = F(r_+) = 0$.}
\label{fig1}
\end{figure}

As previously discussed in \cite{Rincon2020PDU, Konoplya2022JCAP}, the spacetime characterised by $\alpha > 0$ and $\gamma>0$ remains free of curvature singularities, indicating a regular geometry throughout. This is explicitly confirmed by the finiteness of the Kretschmann invariant, for $\alpha>0$ and $\gamma>0$
\begin{align}
\label{djjlwqqw}
&R_{\mu\nu\alpha\beta}R^{\mu\nu\alpha\beta}=\notag\\
&\frac{256 M^6 \left[ \left(r^3+\om(r +\gamma M) \right)^4
+\left(r^3+\om(r +\gamma M)\right)^2 \left(r^3-\om(r+2\gamma M)\right)^2+
\left(r^6-\om r^3(3r+7\gamma M)+(\om \gamma M)^2\right)^2 \right]}{\big[r^3+\om(r +\gamma M)\big]^6}\,.
\end{align}
Fig.~\ref{fig1} shows how the lapse function \eqref{LE} behaves for various values of $M>0$, remaining regular everywhere in the physical domain $r \geq 0$. A critical mass for the central object emerges, $M = M_c$. For $M > M_c$, the metric exhibits two distinct horizons ($r_-$, $r_+$), so the solution describes a non-extremal regular black hole. When $M = M_c$, we see a pair of coincident horizons, corresponding to an extremal configuration. In the subcritical regime, when $0<M<M_c$, no horizons at all is present, and the spacetime represents a self-gravitating droplet \footnote{We note however that considering values of the mass parameter $M<1$ (in Planck units) may lack physical justification within this framework, as such configurations would fall outside the regime where a semiclassical description remains valid, and the concept of a self-gravitating droplet becomes questionable.}. In the present work, we focus on the case $M\geq M_c$. A detailed mathematical proof of these general properties of $F(r)$ for $\alpha > 0$, together with explicit expressions for $r_\pm$ and $M_c$, can be found in \cite{Hassannejad2025PRD} (Appendix E). In the simpler special case $\gamma = 0$, the two horizons, i.e. the outer event horizon $r_+$ and the inner Cauchy horizon $r_-$, are given analytically by $r_\pm = M \pm \sqrt{M^2 - \alpha}$, satisfying $0 < r_- < r_+$ and $0 < \alpha \leq M^2$ as before.

The location $r=r_h$ of the event horizon is the largest positive solution of the equation $F(r_h)=0$, namely
\begin{equation}
  r_h^3-2Mr_h^2+\alpha r_h+M\gamma\alpha=0.    
\end{equation}
As shown in Table~\ref{table:event}, increasing the mass $M$ causes $r_h$ to asymptotically approach the Schwarzschild radius $r_h=2M$, indicating convergence toward the classical black hole limit.
\begin{table}[ht]
\caption{Representative numerical values of the event horizon radius $r_h$ for various values of the mass parameter $M$, $\alpha=118/(15\pi)$, and $\gamma=9/2$.}
\label{table:event}
\vspace*{1em}
\begin{tabular}{||c|c|c|c||}
\hline\hline
$M$               & $r_h$           & $M$       & $r_h$\\ [0.5ex]
\hline\hline
$M_c$              & $4.484183919$   & $6.503$   & $12.31997141$\\
$3.503$           & $4.518502309$   & $7.503$   & $14.42618139$\\
$4.0$             & $6.577475949$   & $8.0$     & $15.46092721$\\
$4.503$           & $7.868129752$   & $8.503$   & $16.50243526$\\
$5.0$             & $9.032138747$   & $9.503$   & $18.56023867$\\
$5.503$           & $10.15863403$   & $10^3$    & $1999.995931$\\[1ex]
\hline\hline 
\end{tabular}
\end{table}

\section{Quasinormal modes of Bonanno-Reuter black holes}

For completeness, we briefly recall the origin of the radial master equation used below. For a static and spherically symmetric line element of the form \eqref{LE}, it is convenient to introduce the tortoise coordinate $r_{*}$ defined by
\begin{equation}\label{tort}
  \frac{dr_*}{dr}=\frac{1}{F(r)} .
\end{equation}
Massless perturbing fields of spin $s=0,1,2$ can then be decomposed into spherical harmonics, or into the corresponding vector and tensor spherical harmonics in the electromagnetic and gravitational cases. In the scalar case, one starts from $\nabla_\mu\nabla^\mu\Psi=0$ and writes  $\Psi=e^{-i\omega t}Y_{\ell m}(\vartheta,\varphi)\psi_{\omega\ell}(r)/r$. For electromagnetic perturbations, the Maxwell equations are expanded in vector spherical harmonics, while for vector-type gravitational perturbations, one uses the standard odd-parity Regge-Wheeler decomposition \cite{Regge1957PR}. In each case, the angular dependence separates, and the remaining radial amplitudes can be combined into a single master 
function obeying a Schr\"odinger-type equation in the tortoise coordinate \cite{Batic2019EPJC}
\begin{equation}
  \frac{d^2\psi_{\omega\ell s}}{dr_*^2} + \left[\omega^2-U_s(r)\right]\psi_{\omega\ell s} = 0.
\end{equation}
The spin dependence is encoded in the effective potential
\begin{equation}
    U_\epsilon(r)=F(r)\left[\frac{\epsilon}{r}\frac{dF}{dr}+\frac{\ell(\ell+1)}{r^2}\right],
\end{equation}
which reduces to the usual Regge--Wheeler potential in the Schwarzschild limit. Equivalently, by using $d/dr_*=F(r)d/dr$ and defining $\epsilon=1-s^2$, the radial equation can be written directly in terms of the coordinate $r$ as
\begin{equation}\label{ODE01}
    F(r)\frac{d}{dr}\left(F(r)\frac{d\psi_{\omega\ell\epsilon}}{dr}\right)
    +\left[\omega^2-U_\epsilon(r)\right]\psi_{\omega\ell\epsilon}(r)=0,\qquad
    U_\epsilon(r)=F(r)\left[
    \frac{\epsilon}{r}\frac{dF}{dr}
    +\frac{\ell(\ell+1)}{r^2}
    \right].
\end{equation}
Here $\ell=0,1,2,\ldots$, and the three cases considered in this work correspond to $\epsilon=1$ for scalar perturbations $(s=0)$, $\epsilon=0$ for electromagnetic perturbations $(s=1)$, and $\epsilon=-3$ for vector-type gravitational perturbations $(s=2)$. Notice that in the present case $F(r)$ is given by the second equation in  \eqref{LE}. By means of the substitution $x=r/r_h$, the above equation can be recast in the equivalent form
\begin{equation}\label{ourODE}
  F(x)\frac{d}{dx}\left(F(x)\frac{d\psi_{\omega\ell\epsilon}}{dx}\right)+\left[r_h^2\omega^2-V_\epsilon(x)\right]\psi_{\omega\ell\epsilon}(x)=0, \qquad F(x)=1-\frac{2Mr_h^2 x^2}{r_h^3 x^3 +\alpha(r_h x+\gamma M)}.
\end{equation} 
Here $V_\epsilon(x)=r_h^2U_\epsilon(r_h x)$ is the dimensionless effective potential, namely
\begin{equation}
  V_\epsilon(x)=F(x)\left[\frac{\epsilon}{x}\frac{dF}{dx}+\frac{\ell(\ell+1)}{x^2}\right].
\end{equation}
In the following analysis, we focus on computing the QNMs for the spectral problem stated in \eqref{ourODE}. Before deriving the explicit asymptotic form of the radial solutions, let us briefly recall the physical meaning of the QNM boundary conditions. We adopt the time dependence $e^{-i\omega t}$ and write $\omega=\omega_R+i\omega_I$, with $\omega_I<0$ corresponding to a mode that is damped in time. In terms of the tortoise coordinate $r_*$, defined by \eqref{tort}, the effective potential vanishes both at the event horizon and at spatial infinity. Consequently, in these two asymptotic regions, the master equation reduces to a free-wave equation, and the radial function behaves locally as
\begin{equation}
  \psi_{\omega\ell\epsilon}\sim e^{\mp i\omega r_*}.
\end{equation}
The interpretation of these two signs follows from the advanced and retarded null coordinates $v=t+r_*$ and $u=t-r_*$. Indeed, the full perturbing field behaves as $e^{-i\omega t}e^{-i\omega r_*}=e^{-i\omega v}$ for the ingoing wave, whereas $e^{-i\omega t}e^{+i\omega r_*}=e^{-i\omega u}$ for the outgoing wave. Therefore, at the future event horizon, where $r_*\to -\infty$, the physically admissible QNM solution is purely ingoing, since no signal can propagate outward from inside the black hole. Conversely, at spatial infinity, where $r_*\to +\infty$, one imposes a purely outgoing wave, corresponding to radiation emitted away from the black hole, with no incoming radiation from infinity. These boundary conditions characterise QNMs as the free oscillations of an open dissipative system. They replace the usual normalizability condition of a conservative eigenvalue problem. For a generic complex value of $\omega$, a solution satisfying the ingoing condition at the horizon will behave at spatial infinity as a linear combination of incoming and outgoing waves. Requiring the incoming coefficient at infinity to vanish imposes a nontrivial spectral condition. This condition is satisfied only for isolated complex frequencies $\omega$, which constitute the QNM spectrum. In the numerical implementation below, the known ingoing and outgoing asymptotic factors are extracted explicitly from $\psi_{\omega\ell\epsilon}$, so that the remaining radial function is regular on the compactified interval $[-1,1]$. The allowed QNM frequencies are then precisely those values of $\omega$ for which such a regular solution exists.

\subsection{The non-extreme case}

This scenario focuses on mass parameters $M>M_c$. We begin by observing that the event horizon corresponds to a simple zero of the function $F(x)$ defined in \eqref{ourODE}. Furthermore, imposing the condition $F(1) = 0$, which is equivalent to $\alpha M\gamma = r_h[r_h(2M-r_h)-\alpha]$,  enables us to recast $F(x)$ in the form
\begin{equation}
  F(x) = 1 - \frac{2Mr_h^2 x^2}{r_h^3 x^3+\alpha r_h(x-1)+r_h^2(2M-r_h)}.
\end{equation}
To establish the QNM boundary conditions at the event horizon and at infinity, we first need to determine the asymptotic behaviour of the radial solution $\psi_{\omega\ell\epsilon}$ as $x \to 1^{+}$ and as $x \to +\infty$. We can then extract the QNM boundary conditions from this asymptotic data. Concerning the asymptotic behaviour as $x\to 1^+$, it is convenient to reformulate \eqref{ourODE} in the form
\begin{equation}\label{ODEZ}
\frac{d^2\psi_{\omega\ell\epsilon}}{dx^2}+p(x)\frac{d\psi_{\omega\ell\epsilon}}{dx}+q(x)\psi_{\omega\ell\epsilon}(x)=0,\quad
p(x)=\frac{F^{'}(x)}{F(x)},\quad
q(x)=\frac{x_h^2\omega^2-V_\epsilon(x)}{F^2(x)}.   
\end{equation}
Since $p$ and $q$ have poles of order one and two at $x = 1$, respectively, this point is classified as a regular singular point of \eqref{ODEZ}, according to Frobenius theory \cite{Ince1956}. Hence, we can construct solutions of the form
\begin{equation}
\psi_{\omega\ell\epsilon}(x)=(x-1)^\rho\sum_{\kappa=0}^\infty a_\kappa(x-1)^\kappa.
\end{equation}
The leading behavior at $x=1$ is represented by the term $(x-1)^\rho$ where $\rho$ is determined by the indicial equation
\begin{equation}\label{indicial}
\rho(\rho-1)+P_0\rho+Q_0=0
\end{equation}
with
\begin{equation}
P_0=\lim_{x\to 1}(x-1)p(x)=1,\qquad
Q_0=\lim_{x\to 1}(x-1)^2 q(x)=\left(\frac{r_h \omega}{F^{'}(1)}\right)^2.
\end{equation}
The roots of \eqref{indicial} are $\rho_\pm = \pm i r_h\omega/F^{'}(1)$ and the correct QNM boundary condition at $x=1$ reads
\begin{equation}\label{QNMBCz1}
\psi_{\omega\ell\epsilon}\underset{{x\to 1^+}}{\longrightarrow} (x-1)^{-i r_h a\omega},\quad a=\frac{1}{F^{'}(1)}=\frac{2Mr_h}{3r_h^2-4Mr_h+\alpha}.
\end{equation}
From the above expression, we immediately see that $a$ is always positive since $\alpha>0$. Regarding the asymptotic behaviour as $x\to+\infty$, we start by observing that
\begin{equation}
p(x) = \sum_{\kappa=0}^\infty\frac{\mathfrak{f}_\kappa}{x^k} = \mathcal{O}\left(\frac{1}{x^2}\right), \qquad
q(x) = \sum_{\kappa=0}^\infty\frac{\mathfrak{g}_\kappa}{z^k}=r_h^2\omega^2+\frac{4Mr_h\omega^2}{x}+\mathcal{O}\left(\frac{1}{x^2}\right).
\end{equation}
Consequently, the asymptotic behaviour of the solutions to equation \eqref{ODEZ} can be deduced using the method outlined in \cite{Olver1994MAA}. Given that at least one of the coefficients $\mathfrak{f}_0$, $\mathfrak{g}_0$, $\mathfrak{g}_1$ is nonzero, a formal solution to \eqref{ODEZ} is represented by \cite{Olver1994MAA}
\begin{equation}\label{olvers}
\psi^{(j)}_{\omega\ell\epsilon}(x) = x^{\mu_j}e^{\lambda_j x}\sum_{\kappa=0}^\infty\frac{a_{\kappa,j}}{x^\kappa}, \qquad j \in \{1,2\},
\end{equation}
where $\lambda_1$, $\lambda_2$, $\mu_1$ and $\mu_2$ are the roots of the characteristic equations
\begin{equation}\label{chareqns}
   \lambda^2+\mathfrak{f}_0\lambda+\mathfrak{g}_0=0,\quad
   \mu_j=-\frac{\mathfrak{f}_1\lambda_j+\mathfrak{g}_1}{\mathfrak{f}_0+2\lambda_j}.
\end{equation}
A straightforward computation shows that $\lambda_\pm = \pm ir_h\omega$ and $\mu_\pm = \pm 2iM\omega$. As a result, the QNM boundary condition at space-like infinity can be expressed as
\begin{equation}\label{QNMBCzinf}
    \psi_{\omega\ell\epsilon}\underset{{x\to +\infty}}{\longrightarrow} x^{2iM\omega}e^{i r_h\omega x}.
\end{equation}
At this point, we transform the radial function $\psi_{\omega\ell\epsilon}(x)$ into a new radial function $\Phi_{\omega\ell\epsilon}(x)$ such that the QNM boundary conditions are automatically implemented and  $\Phi_{\omega\ell\epsilon}(x)$ is regular at $x = 1$ and at space-like infinity. To this aim, we consider the transformation
\begin{equation}\label{Ansatz}
\psi_{\omega\ell\epsilon}(x) = x^{i(2M+ar_h)\omega}(x-1)^{-iar_h\omega}e^{ir_h\omega(x-1)} \Phi_{\omega\ell\epsilon}(x).
\end{equation}
If we substitute \eqref{Ansatz} into \eqref{ourODE}, we end up with the following ordinary differential equation for the radial eigenfunctions, namely
\begin{equation}\label{ODEznone}
    P_2(x)\Phi^{''}_{\omega\ell\epsilon}(x) + P_1(x)\Phi^{'}_{\omega\ell\epsilon}(x) + P_0(x)\Phi_{\omega\ell\epsilon}(x) = 0
\end{equation}
with
\begin{eqnarray}
P_2(x)&=&x^2(x-1)^2 F^2(x),\\
P_1(x)&=&x(x-1)F(x)\left\{x(x-1)F^{'}(x)+i\omega F(x)\left[2r_h x^2+2x(2M-r_h)-2ar_h-4M\right]\right\},\\
P_0(x)&=&-\omega^2 Q_+(x)Q_{-}(x)+i\omega F(x)L(x)-x^2(x-1)^2 V_\epsilon(x),\\
Q_\pm(x)&=&F(x)[(x-1)(r_h x+2M)-ar_h]\pm r_h x(x-1),\\
L(x)&=&x(x-1)[r_h x^2+(2M-r_h)x-2M-ar_h]F^{'}(x)-F(x)[2Mx^2-(2x-1)(2M+ar_h)].
\end{eqnarray}
Let us now introduce the transformation $x=2/(1-y)$ mapping the point at infinity and the event horizon to $y = 1$ and $y = -1$, respectively. Furthermore, a dot denotes differentiation with respect to the new variable $y$. Then, equation \eqref{ODEznone} becomes
\begin{equation}\label{ODEynone}
    S_2(y)\ddot{\Phi}_{\omega\ell\epsilon}(y) + S_1(y)\dot{\Phi}_{\omega\ell\epsilon}(y) + S_0(y)\Phi_{\omega\ell\epsilon}(y) = 0,
\end{equation}
where
\begin{eqnarray}
  S_2(y) &=&(1+y)^2 F^2(y), \label{S2onone} \\
  S_1(y) &=& i\omega\frac{1+y}{(1-y)^2}F^2(y)\left[8r_h+4(2M-r_h)(1-y)-2(2M+ar_h)(1-y)^2\right]\nonumber\\
  &&-2\frac{(1+y)^2}{1-y}F^2(y)+(1+y)^2 F(y)\dot{F}(y), \label{S1onone}\\
  S_0(y) &=& \omega^2\Sigma_2(y)+i\omega\Sigma_1(y)+\Sigma_0(y) \label{S0onone}
\end{eqnarray}
with
\begin{eqnarray}
\Sigma_2(y) &=& \frac{4r_h^2(1+y)^2}{(1-y)^4}-\frac{F^2(y)}{(1-y)^4}\left\{2M(1-y^2)-r_h[ay^2-2y(1+a)+a-2]\right\}^2,\\
\Sigma_1(y) &=&F(y)\left\{
\frac{1+y}{(1-y)^2}[4r_h+2(2M-r_h)(1-y)-(2M+ar_h)(1-y)^2]\dot{F}(y)\right.\nonumber\\
&&-\left.\frac{F(y)}{(1-y)^2}[8M-(2M+ar_h)(3+y)(1-y)]
\right\},\\
\Sigma_0(y)&=&-\frac{4(1+y)^2}{(1-y)^4}V_\epsilon(y).
\end{eqnarray}
Notice that we must also require that $\Phi_{\omega\ell\epsilon}(y)$ is regular at $y=\pm 1$. As a result of the transformation introduced above, we have 
\begin{equation}\label{fv}
F(y)=1-\frac{8Mr_h^2(1-y)}{8r_h^3+2\alpha r_h(1-y)^2-r_h(r_h^2-2Mr_h+\alpha)(1-y)^3}, \quad 
  V_\epsilon(y) = \frac{(1-y)^2}{4}F(y)\left[\epsilon (1-y)\dot{F}(y)+\ell(\ell+1)\right].
\end{equation}
\begin{table}
\caption{Classification of the points $y=\pm 1$ for the relevant functions defined by   (\ref{S2onone})--(\ref{S0onone}), and (\ref{fv}),. The abbreviations $z$ ord $n$ and $p$ ord $m$ stand for zero of order $n$ and pole of order $m$, respectively.}
\begin{center}
\begin{tabular}{ | c | c | c | c | c | c | c | c }
\hline
$y$  & $F(y)$  & $V_\epsilon(y)$ & $S_2(y)$ & $S_{1}(y)$ & $S_{0}(y)$\\ \hline
$-1$ & z \mbox{ord} 1 & z \mbox{ord} 1 & z \mbox{ord} 4& z \mbox{ord} 3 & z \mbox{ord} 3 \\ \hline
$+1$ & $+1$  & z \mbox{ord} 2 & $4$ & p \mbox{ord} 2 & p \mbox{ord} 2\\ \hline
\end{tabular}
\label{tableEinsnone}
\end{center}
\end{table}
Table~\ref{tableEinsnone} indicates that the coefficients of the differential equation \eqref{ODEynone} share a common zero of order $2$ at $y = -1$ while $y = 1$ is a pole of order $4$ for the coefficient $S_0$. Hence, in order to apply the spectral method, we need to multiply \eqref{ODEynone} by $(1-y)^2/(1+y)^3$. As a result, we end up with the following differential equation
\begin{equation}\label{ODEhynone}
M_2(y)\ddot{\Phi}_{\omega\ell\epsilon}(y) + M_1(y)\dot{\Phi}_{\omega\ell\epsilon}(y) + M_0(y)\Phi_{\omega\ell\epsilon}(y) = 0,
\end{equation}
where
\begin{equation}\label{S210honone}
M_2(y) = \frac{(1-y)^2}{1+y} F^2(y), \qquad
M_1(y) = i\omega N_1(y)+N_0(y), \qquad
M_0(y) = \omega^2 C_2(y)+i\omega C_1(y)+C_0(y)
\end{equation}
with
\begin{eqnarray}
N_1(y) &=&\frac{2F^2(y)}{(1+y)^2}\left[4r_h+2(2M-r_h)(1-y)-(ar_h+2M)(1-y)^2\right],\label{N1}\\
N_0(y) &=&\frac{1-y}{1+y}\left[(1-y)\dot{F}(y)-2F(y)\right],\label{N0}\\
C_2(y) &=& \frac{4r_h^2}{(1+y)(1-y)^2}-\frac{F^2(y)}{(1+y)^3(1-y)^2}\left\{2M(1-y^2)-r_h[ay^2-2y(1+a)+a-2]\right\}^2,\label{C2}\\
C_1(y) &=&\frac{F(y)}{(1+y)^3}\left\{(1+y)\dot{F}(y)\left[4r_h+2(2M-r_h)(1-y)-(ar_h+2M)(1-y)^2\right]\right.\nonumber\\
&&\left.-F(y)\left[8M-(3+y)(1-y)(ar_h+2M)\right]\right\},\label{C1}\\
C_0(y) &=& -\frac{4V_\epsilon(y)}{(1+y)(1-y)^2}.\label{C0}
\end{eqnarray}
It can be easily checked with Maple that
\begin{eqnarray}
    &&\lim_{y\to 1^{-}}M_2(y)=0=\lim_{y\to -1^{+}}M_2(y),\\
    &&\lim_{y\to 1^{-}}M_1(y)=2ir_h\omega,\quad
    \lim_{y\to -1^{+}}M_1(y)=i\omega\Lambda_1+\Lambda_0,\\
    &&\lim_{y\to 1^{-}}M_0(y)=A_2\omega^2 +A_0,\quad
     \lim_{y\to -1^{+}}M_0(y)=B_2\omega^2+i\omega B_1+B_0,
\end{eqnarray}
where
\begin{eqnarray}
\Lambda_1 &=&\frac{4Mr_h-3r_h^2-\alpha}{M},\quad
\Lambda_0 = \left(\frac{\Lambda_1}{2r_h}\right)^2,\quad
A_2=\frac{2M(r_h^3+6Mr_h^2-8M^2 r_h+2M\alpha)}{3r_h^2-4Mr_h+\alpha},\quad
A_0=-\frac{\ell(\ell+1)}{2},\label{coef1}\\
B_2&=&\frac{9r_h^5-9Mr_h^4-6(3M^2-\alpha)r_h^3+M(16M^2+\alpha)r_h^2-\alpha(8M^2-\alpha)r_h+M\alpha^2}{M(3r_h^2-4Mr_h+\alpha)},\label{coef2}\\
B_1&=&\frac{3r_h^2-4Mr_h+\alpha}{4Mr_h^2}B_2,\quad
B_0=-\frac{3r_h^2-4Mr_h+\alpha}{8M^2 r_h^2}\left\{3\epsilon r_h^2+2M[\ell(\ell+1)-2\epsilon]+\alpha\epsilon\right\}.\label{coef3}
\end{eqnarray}
In the final step leading to the application of the spectral method, we recast the differential equation \eqref{ODEhynone} into the following form
\begin{equation}\label{TSCH}
  L_0\left[\Phi_{\omega\ell\epsilon}, \dot{\Phi}_{\omega\ell\epsilon}, \ddot{\Phi}_{\omega\ell\epsilon}\right] +  i L_1\left[\Phi_{\omega\ell\epsilon}, \dot{\Phi}_{\omega\ell\epsilon}, \ddot{\Phi}_{\omega\ell\epsilon}\right]\omega +  L_2\left[\Phi_{\omega\ell\epsilon}, \dot{\Phi}_{\omega\ell\epsilon}, \ddot{\Phi}_{\omega\ell\epsilon}\right]\omega^2 = 0
\end{equation}
with
\begin{eqnarray}
L_0\left[\Phi_{\omega\ell\epsilon}, \dot{\Phi}_{\omega\ell\epsilon}, \ddot{\Phi}_{\omega\ell\epsilon}\right] &=& L_{00}(y)\Phi_{\omega\ell\epsilon} + L_{01}(y)\dot{\Phi}_{\omega\ell\epsilon} + L_{02}(y)\ddot{\Phi}_{\omega\ell\epsilon},\label{L0none}\\
L_1\left[\Phi_{\omega\ell\epsilon}, \dot{\Phi}_{\omega\ell\epsilon}, \ddot{\Phi}_{\omega\ell\epsilon}\right] &=&L_{10}(y)\Phi_{\omega\ell\epsilon} + L_{11}(y)\dot{\Phi}_{\omega\ell\epsilon} + L_{12}(y)\ddot{\Phi}_{\omega\ell\epsilon}, \label{L1none}\\
L_2\left[\Phi_{\omega\ell\epsilon}, \dot{\Phi}_{\omega\ell\epsilon}, \ddot{\Phi}_{\omega\ell\epsilon}\right] &=&L_{20}(y)\Phi_{\omega\ell\epsilon} + L_{21}(y)\dot{\Phi}_{\omega\ell\epsilon} + L_{22}(y)\ddot{\Phi}_{\omega\ell\epsilon}.\label{L2none}
\end{eqnarray}
Moreover, in Table~\ref{tableZweinone}, we have summarized the $L_{ij}$ appearing in (\ref{L0none})--(\ref{L2none}) and their limiting values at $y = \pm 1$.

\begin{table}
\caption{Definitions of the coefficients $L_{ij}$ and their corresponding behaviours at the endpoints of the interval $-1 \leq y \leq 1$. The symbols appearing in this table have been defined in (\ref{coef1})-(\ref{coef3}).}
\begin{center}
\begin{tabular}{ | c | c | c | c | c | c | c | c }
\hline
$(i,j)$  & $\displaystyle{\lim_{y\to -1^+}}L_{ij}$  & $L_{ij}$ & $\displaystyle{\lim_{y\to 1^-}}L_{ij}$  \\ \hline
$(0,0)$ &  $B_0$          & $C_0$                  & $A_0$\\ \hline
$(0,1)$ &  $\Lambda_0$    & $N_0$                  & $0$\\ \hline
$(0,2)$ &  $0$            & $M_2$                  & $0$\\ \hline 
$(1,0)$ &  $B_1$          & $C_1$                  & $0$\\ \hline 
$(1,1)$ &  $\Lambda_1$    & $N_1$                  & $2r_h$\\ \hline 
$(1,2)$ &  $0$            & $0$                    & $0$\\ \hline 
$(2,0)$ &  $B_2$          & $C_2$                  & $A_2$\\ \hline
$(2,1)$ &  $0$            & $0$                    & $0$\\ \hline
$(2,2)$ &  $0$            & $0$                    & $0$\\ \hline
\end{tabular}
\label{tableZweinone}
\end{center}
\end{table} 

\subsection{The extreme case}

In this configuration, the metric function $g_{00}$ (or equivalently $F(r)$) develops a double zero at $r = r_e$ when the mass parameter $M$ reaches a critical value $M_c$. Imposing the conditions $F(r_e) = 0$ and $F'(r_e) = 0$, we find numerically that $M_c = 3.502741812\ldots$ and $r_e = 4.484183919\ldots$. Solving these two equations yields the following relations between the critical mass $M_c$, the extremal radius $r_e$, and the parameters $\alpha$ and $\gamma$
\begin{equation}
  M_c = \frac{3r_e^2 + \alpha}{4r_e}, \quad \alpha \gamma M_c = \frac{r_e^3 - \alpha r_e}{2}.
\end{equation}
Substituting these expressions into the general form of $F(r)$, as given in equation \eqref{LE}, and using the running coupling relation in equation \eqref{run_c}, we obtain an explicit form of $F(r)$ in which the double root condition at $r = r_e$ is automatically satisfied
\begin{equation}\label{Fr}
  F(r) = 1 - \frac{(\alpha + 3r_e^2)r^2}{r_e(2r^3 + 2\alpha r + r_e^3 - \alpha r_e)}.
\end{equation}
At this stage, it is convenient to introduce the dimensionless radial coordinate $\xi = r/r_e$, which simplifies the analysis of the extremal geometry. Under this rescaling, equation \eqref{Fr} transforms into the following form
\begin{equation}\label{fze}
  F(\xi) = \frac{(2r_e^2 \xi + r_e^2 - \alpha)(\xi - 1)^2}{2r_e^2 \xi^3 + 2\alpha \xi + r_e^2 - \alpha}.
\end{equation}
In this new coordinate system, the radial part of the massless Klein–Gordon equation retains the structure given in equation \eqref{ODEZ}, with the substitution $x \rightarrow \xi$ and $r_h\rightarrow r_e$. To derive the QNM boundary conditions at the event horizon and at spatial infinity, we first analyse the asymptotic behaviour of the radial solution $\psi_{\omega\ell\epsilon}$ in the limits $\xi \to 1^{+}$ and $\xi \to +\infty$. This asymptotic analysis enables us to determine the appropriate QNM boundary conditions for purely ingoing waves at the horizon and purely outgoing waves at infinity. Returning to equation\eqref{ODEZ}, a straightforward expansion around $\xi = 1$ reveals the leading-order behavior of the functions $p(\xi)$ and $q(\xi)$ near the horizon
\begin{equation}
  p(\xi)=\frac{2}{\xi-1}+\mathcal{O}(1), \quad q(\xi)=\frac{A}{(\xi-1)^4}+\frac{B}{(\xi-1)^3}+\frac{C}{(\xi-1)^2}+\frac{D}{\xi-1}+\mathcal{O}(1),
\end{equation}
where the coefficients of the leading singular terms are given by
\begin{equation}
  A = \frac{r_e^2(\alpha+3r_e^2)^2}{(\alpha-3r_e^2)^2}\omega^2, \quad B = \frac{4r_e^2(\alpha+3r_e^2)^2(\alpha-2r_e^2)}{(\alpha-3r_e^2)^3}\omega^2.
\end{equation}
The expressions for the subleading coefficients $C$ and $D$ are omitted, as they are not relevant for the forthcoming analysis. Because $q(\xi)$ has a fourth-order pole at $\xi = 1$, it follows that $\xi = 1$ is an irregular singularity, rendering Frobenius's theory inapplicable in this scenario. On the other hand, since for $k = 1$ we have
\begin{equation}
    (\xi-1)^{k+1}p(\xi)=\mathcal{O}(\xi-1),\quad 
    (\xi-1)^{2k+2}q(\xi)=A+\mathcal{O}(\xi-1)
\end{equation}
with $A \neq 0$, then, according to \cite{Bender1999}, $\xi = 1$ is an irregular singular point of rank one. Consequently, the leading behaviour of the solutions to equation \eqref{ODEZ} in a neighbourhood of the event horizon can be deduced using the method outlined in \cite{Olver1994MAA}. To this purpose, we start by observing that by means of the transformation $\tau = (\xi-1)^{-1}$ mapping the event horizon at infinity and infinity to zero, \eqref{ODEZ} becomes
\begin{eqnarray}
&&\frac{d^2\psi_{\omega\ell\epsilon}}{d\tau^2}+\mathfrak{C}(\tau)\frac{d\psi_{\omega\ell\epsilon}}{d\tau}+\mathfrak{D}(\tau)\psi_{\omega\ell\epsilon}(\tau)=0,\label{ODEZe}\\
&&\mathfrak{C}(\tau)=\mathcal{O}\left(\frac{1}{\tau^2}\right),\quad
\mathfrak{D}(\tau)=A+\frac{B}{\tau}+\mathcal{O}\left(\frac{1}{\tau^2}\right).
\end{eqnarray}
Since at least one of the coefficients $A$, and $B$ is nonzero, a formal solution to \eqref{ODEZe} is given by \cite{Olver1994MAA}
\begin{equation}
  \psi^{(\pm)}_{\omega\ell\epsilon}(\tau)=\tau^{\mu_\pm}e^{\lambda_\pm \tau}\sum_{\kappa=0}^\infty\frac{\mathfrak{a}_{\kappa,\pm}}{\tau^\kappa},
\end{equation}
where $\lambda_\pm$, and $\mu_\pm$ are the roots of the characteristic equations
\begin{equation}
   \lambda_\pm^2+A=0,\quad
   \mu_\pm=-\frac{B}{2\lambda_\pm}.
\end{equation}
A straightforward computation shows that
\begin{equation}\label{lmu}
  \lambda_\pm=\pm\frac{i r_e(3r_e^2+\alpha)\omega}{3r_e^2-\alpha}, \quad \mu_\pm=\pm\frac{2i r_e(3r_e^2+\alpha)(2r_e^2-\alpha)\omega}{(3r_e^2-\alpha)^2}.
\end{equation}
At this point, it is important to observe that a radial field exhibiting purely ingoing behaviour near the event horizon ($\xi \to 1^+$) transforms, under the coordinate change $\tau = (\xi - 1)^{-1}$, into an outward-propagating mode as $\tau \to +\infty^{-}$. This correspondence supports the selection of the positive sign in the expressions for the characteristic exponents. Accordingly, the appropriate QNM boundary condition at the event horizon $\xi = 1$ takes the form
\begin{equation}\label{QNMBCe1}
  \psi_{\omega\ell\epsilon} \underset{{\xi \to 1^+}}{\longrightarrow} (\xi - 1)^{-\mu_+} \exp\left(\frac{\lambda_+}{\xi - 1}\right),
\end{equation}
where the parameters $\mu_+$ and $\lambda_+$ are defined in equation \eqref{lmu}. By means of the transformation $\eta = 1/\xi$, it is not difficult to verify that the point at infinity is again an irregular singular point of rank one. Therefore, in the extremal case, the asymptotic behaviour of the solutions to equation \eqref{ODEZ} can be derived according to the method outlined in \cite{Olver1994MAA}. To this purpose, we  observe that
\begin{equation}
  p(\xi)=\mathcal{O}\left(\frac{1}{\xi^2}\right), \quad q(\xi)=r_e^2\omega^2+\frac{(3r_e^2+\alpha)\omega^2}{\xi}+\mathcal{O}\left(\frac{1}{\xi^2}\right).
\end{equation}
With the help of \eqref{chareqns}, we immediately find that the QNM boundary condition at space-like infinity can be expressed as
 \begin{equation}\label{QNMBCzinfe}
    \psi_{\omega\ell\epsilon}\underset{{\xi\to +\infty}}{\longrightarrow} \xi^{i\widehat{a}\omega}e^{i r_e\omega\xi},\quad
    \widehat{a}=\frac{3r_e^2+\alpha}{2r_e}.
    \end{equation}
It is convenient to transform the radial function $\psi_{\omega\ell\epsilon}(\xi)$ into a new radial function $\Phi_{\omega\ell\epsilon}(\xi)$ such that the QNM boundary conditions are automatically implemented and  $\Phi_{\omega\ell s}(\xi)$ is regular at $\xi = 1$ and at space-like infinity. To this aim, we consider the transformation
\begin{equation}\label{Ansatze}
    \psi_{\omega\ell\epsilon}(\xi)=\xi^{i\widehat{a}\omega+\mu_+}(\xi-1)^{-\mu_+}e^{ir_e \omega(\xi-1)+\frac{\lambda_+}{\xi-1}} \Phi_{\omega\ell\epsilon}(\xi).
\end{equation}
If we rewrite it in a more compact form, namely
\begin{eqnarray}
\psi_{\omega\ell\epsilon}(\xi)&=&\xi^{ia\omega}(\xi-1)^{-ib\omega}e^{ir_e\omega\eta(\xi)} \Phi_{\omega\ell\epsilon}(\xi),\quad
\eta(\xi)=\xi-1+\frac{3r_e^2+\alpha}{(3r_e^2-\alpha)(\xi-1)}\\
a&=&\frac{(3r_e^2+\alpha)(17r_e^4-10\alpha r_e^2+\alpha^2)}{2r_e(3r_e^2-\alpha)^2},\quad
b=\frac{2r_e(3r_e^2+\alpha)(2r_e^2-\alpha)}{(3r_e^2-\alpha)^2},\quad
a-b=\widehat{a}
\end{eqnarray}
and we replace it into \eqref{ourODE} with $x$ and $r_h$ replaced by $\xi$ and $r_e$, respectively, we end up with the differential equation
\begin{equation}\label{ODEzext}
  P_{2e}(\xi)\Phi^{''}_{\omega\ell\epsilon}(\xi)+P_{1e}(\xi)\Phi^{'}_{\omega\ell\epsilon}(\xi)+P_{0e}(\xi)\Phi_{\omega\ell\epsilon}(\xi)=0
\end{equation}
with
\begin{eqnarray}
P_{2e}(\xi)&=&\xi^2(\xi-1)^2 F^2(\xi),\\
P_{1e}(\xi)&=&\xi(\xi-1)F(\xi)\left\{\xi(\xi-1)F^{'}(\xi)+i\omega F(\xi)\left[2r_e\xi(\xi-1)\eta^{'}(\xi)+2(\xi\widehat{a}+a)\right]\right\},\\
P_{0e}(\xi)&=&-\mathfrak{Q}_+(\xi)\mathfrak{Q}_-(\xi)\omega^2+i\omega F(\xi)\mathfrak{L}(\xi)-\xi^2(\xi-1)^2 V_\epsilon(\xi),\\
\mathfrak{Q}_\pm(\xi)&=&r_e\xi(\xi-1)F(\xi)\eta^{'}(\xi)+F(\xi)(\xi\widehat{a}-a)\pm r_e\xi(\xi-1),\\
\mathfrak{L}(\xi)&=&r_e\xi^2(\xi-1)^2\left[F(\xi)\eta^{''}(\xi)+F^{'}(\xi)\eta^{'}(\xi)\right]+\xi(\xi-1)(\widehat{a}\xi-a)F^{'}(\xi)-(\widehat{a}\xi^2-2a\xi+a)F(\xi).
\end{eqnarray}
As already done in the nonextremal case, we introduce the transformation $\xi = 2/(1-y)$. Furthermore, a dot denotes differentiation with respect to the new variable $y$. Then, equation \eqref{ODEzext} becomes
\begin{equation}\label{ODEye}
    S_{2e}(y)\ddot{\Phi}_{\omega\ell\epsilon}(y)+S_{1e}(y)\dot{\Phi}_{\omega\ell\epsilon}(y)+S_{0e}(y)\Phi_{\omega\ell\epsilon}(y)=0,
\end{equation}
where
\begin{eqnarray}
S_{2e}(y)&=&(1+y)^2 F^2(y),\label{S2oe}\\
S_{1e}(y)&=&2i\omega\frac{1+y}{1-y}F^2(y)\left[r_e(1-y^2)\dot{\eta}(y)+ay+a-2b\right]+(1+y)^2F(y)\left[\dot{F}(y)-\frac{2F(y)}{1-y}\right],\label{S1oe}\\
S_{0e}(y)&=&\omega^2\Sigma_{2e}(y)+i\omega\Sigma_{1e}(y)+\Sigma_{0e}(y)\label{S0oe}
\end{eqnarray}
with
\begin{eqnarray}
\Sigma_{2e}(y)&=&\frac{4r_e^2(1+y)^2}{(1-y)^4}-\frac{F^2(y)}{(1-y)^2}\left[r_e(1-y^2)\dot{\eta}(y)+ay+a-2b\right]^2,\\
\Sigma_{1e}(y)&=&r_e(1+y)^2 F^2(y)\ddot{\eta}(y)+r_e\frac{(1+y)^2}{1-y}F(y)\dot{\eta}(y)\left[(1-y)\dot{F}(y)-2F(y)\right]+\nonumber\\
&&\frac{1+y}{1-y}(ay+a-2b)F(y)\dot{F}(y)-\frac{ay^2+2ay+a-4b}{(1-y)^2}F^2(y),\\
\Sigma_{0e}(y)&=&-\frac{4(1+y)^2}{(1-y)^4}V_\epsilon(y).
\end{eqnarray}
and the requirement that $\Phi_{\omega\ell\epsilon}(y)$ is regular at $y = \pm 1$. As a result of the transformation introduced above, we have 
\begin{equation}\label{fve}
F(y)=\frac{4r_e^2+(r_e^2-\alpha)(1-y)}{16 r_e^2+4\alpha(1-y)^2+(r_e^2-\alpha)(1-y)^3}(1+y)^2, \qquad 
\eta(y) = \frac{1+y}{1-y}+\frac{3r_e^2+\alpha}{3r_e^2-\alpha}\cdot\frac{1-y}{1+y},
\end{equation}
while $V_\epsilon(y)$ is formally given by \eqref{fv}.
\begin{table}
\caption{Classification of the points $y = \pm 1$ for the relevant functions entering in (\ref{S2oe}), (\ref{S1oe}), and (\ref{S0oe}). The abbreviations $z$ ord $n$ and $p$ ord $m$ stand for zero of order $n$ and pole of order $m$, respectively.}
\begin{center}
\begin{tabular}{ | l | l | l | l |l |l | l | l}
\hline
$y$  & $F(y)$  & $V_\epsilon(y)$ & $\eta(y)$ & $S_{2e}(y)$ & $S_{1e}(y)$ & $S_{0e}(y)$\\ \hline
$-1$ & z \mbox{ord} 2 & z \mbox{ord} 2 & p \mbox{ord} 1 & z \mbox{ord} 6& z \mbox{ord} 4 & z \mbox{ord} 4 \\ \hline
$+1$ & $+4$  & z \mbox{ord} 2 & p \mbox{ord} 1 & $+1$ & p \mbox{ord} 2 & p \mbox{ord} 2\\ \hline
\end{tabular}
\label{table3}
\end{center}
\end{table}
Table~\ref{table3} indicates that the coefficients of the differential equation \eqref{ODEye} share a common zero of order $4$ at $y = -1$ while $y = 1$ is a pole of order $2$ for the coefficients $S_{1e}(y)$ and $S_{0e}(y)$. Hence, to apply the spectral method, we need to multiply \eqref{ODEye} by $(1-y)^2/(1+y)^4$. As a result, we end up with the following differential equation
\begin{equation}\label{ODEhynonee}
    M_{2e}(y)\ddot{\Phi}_{\omega\ell\epsilon}(y)+M_{1e}(y)\dot{\Phi}_{\omega\ell\epsilon}(y)+M_{0e}(y)\Phi_{\omega\ell\epsilon}(y)=0,
\end{equation}
where
\begin{equation}\label{S210hononee}
  M_{2e}(y)=\left(\frac{1-y}{1+y}\right)^2 F^2(y),\quad
  M_{1e}(y)=i\omega N_{1e}(y)+N_{0e}(y),\quad
  M_{0e}(y)=\omega^2 C_{2e}(y)+i\omega C_{1e}(y)+C_{0e}(y)
\end{equation}
with
\begin{eqnarray}
N_{1e}(y)&=&\frac{2(1-y)}{(1+y)^3}F^2(y)\left[r_e(1-y^2)\dot{\eta}(y)+ay+a-2b\right],\quad
N_{0e}(y)=\left(\frac{1-y}{1+y}\right)^2F(y)\left[\dot{F}(y)-\frac{2F(y)}{1-y}\right],\label{N0e}\\
C_{2e}(y)&=&\frac{4r_e^2}{(1-y^2)^2}-\frac{F^2(y)}{(1+y)^4}\left[r_e(1-y^2)\dot{\eta}(y)+ay+a-2b\right]^2,\label{C2e}\\
C_{1e}(y)&=&r_e\left(\frac{1-y}{1+y}\right)^2 F^2(y)\ddot{\eta}(y)+r_e\frac{1-y}{(1+y)^2}F(y)\dot{\eta}(y)\left[(1-y)\dot{F}(y)-2F(y)\right]+\nonumber\\
&&\frac{1-y}{(1+y)^3}(ay+a-2b)F(y)\dot{F}(y)-\frac{ay^2+2ay+a-4b}{(1+y)^4}F^2(y),\label{C1e}\\
    C_{0e}(y)&=&-\frac{4V_\epsilon(y)}{(1-y^2)^2}.\label{C0e}
\end{eqnarray}
It can be easily checked with Maple that
\begin{eqnarray}
    &&\lim_{y\to 1^{-}}M_{2e}(y)=0=\lim_{y\to -1^{+}}M_{2e}(y),\quad \lim_{y\to 1^{-}}M_{1e}(y)=ir_e\omega,\quad
    \lim_{y\to -1^{+}}M_{1e}(y)=i\omega\Lambda_{1e},\\
    &&\lim_{y\to 1^{-}}M_{0e}(y)=A_{2e}\omega^2 +A_{0e},\quad
     \lim_{y\to -1^{+}}M_{0e}(y)=B_{2e}\omega^2+B_{0e},
\end{eqnarray}
where
\begin{eqnarray}
\Lambda_{1e}&=&\frac{r_e(\alpha-3r_e^2)}{\alpha+3r_e^2},\quad
A_{2e}=\frac{165r_e^8-8\alpha r_e^6-30\alpha^2 r_e^4+\alpha^4}{8r_e^2(3r_e^2-\alpha)^2},\quad A_{0e}=-\frac{\ell(\ell+1)}{4},\label{Acoefnonee}\\
B_{2e}&=&\frac{267r_e^8-208\alpha r_e^6+30\alpha^2 r_e^4+8\alpha^3 r_e^2-\alpha^4}{4(3r_e^2+\alpha)(3r_e^2-\alpha)^2},\quad
B_{0e}=-\frac{(3r_e^2-\alpha)}{4(3r_e^2+\alpha)}\ell(\ell+1).\label{B0e}
\end{eqnarray}
Finally, in order to apply the spectral method, we rewrite the differential equation \eqref{ODEhynonee} into the following form
\begin{equation}\label{TSCHe}
\widehat{L}^{(e)}_0\left[\Phi_{\omega\ell\epsilon},\dot{\Phi}_{\omega\ell\epsilon},\ddot{\Phi}_{\omega\ell\epsilon}\right]+ i\widehat{L}^{(e)}_1\left[\Phi_{\omega\ell\epsilon},\dot{\Phi}_{\omega\ell\epsilon},\ddot{\Phi}_{\omega\ell\epsilon}\right]\omega+ \widehat{L}_2^{(e)}\left[\Phi_{\omega\ell\epsilon},\dot{\Phi}_{\omega\ell\epsilon},\ddot{\Phi}_{\omega\ell\epsilon}\right]\omega^2=0
\end{equation}
with
\begin{eqnarray}
\widehat{L}^{(e)}_0\left[\Phi_{\omega\ell\epsilon},\dot{\Phi}_{\omega\ell\epsilon},\ddot{\Phi}_{\omega\ell\epsilon}\right]&=&\widehat{L}^{(e)}_{00}(y)\Phi_{\omega\ell\epsilon}+\widehat{L}^{(e)}_{01}(y)\dot{\Phi}_{\omega\ell\epsilon}+\widehat{L}^{(e)}_{02}(y)\ddot{\Phi}_{\omega\ell\epsilon},\label{L0nonee}\\
\widehat{L}^{(e)}_1\left[\Phi_{\omega\ell\epsilon},\dot{\Phi}_{\omega\ell\epsilon},\ddot{\Phi}_{\omega\ell\epsilon}\right]&=&\widehat{L}^{(e)}_{10}(y)\Phi_{\omega\ell\epsilon}+\widehat{L}^{(e)}_{11}(y)\dot{\Phi}_{\omega\ell\epsilon}+\widehat{L}^{(e)}_{12}(y)\ddot{\Phi}_{\omega\ell\epsilon},\label{L1nonee}\\
\widehat{L}^{(e)}_2\left[\Phi_{\omega\ell\epsilon},\dot{\Phi}_{\omega\ell\epsilon},\ddot{\Phi}_{\omega\ell\epsilon}\right]&=&\widehat{L}^{(e)}_{20}(y)\Phi_{\omega\ell\epsilon}+\widehat{L}^{(e)}_{21}(y)\dot{\Phi}_{\omega\ell\epsilon}+\widehat{L}^{(e)}_{22}(y)\ddot{\Phi}_{\omega\ell\epsilon}.\label{L2nonee}
\end{eqnarray}
Moreover, in Table~\ref{table4}, we have summarized the $\widehat{L}^{(e)}_{ij}$ appearing in (\ref{L0nonee})--(\ref{L2nonee}) and their limiting values at $y = \pm 1$.

\begin{table}
\caption{Definitions of the coefficients $\widehat{L}^{(e)}_{ij}$ and their corresponding behaviours at the endpoints of the interval $-1\leq y\leq 1$. The symbols appearing in this table have been defined in (\ref{S210hononee})-(\ref{B0e}).}
\begin{center}
\begin{tabular}{ | l | l | l | l |l |l | l | l}
\hline
$(i,j)$  & $\displaystyle{\lim_{y\to -1^+}}\widehat{L}^{(e)}_{ij}$  & $\widehat{L}^{(e)}_{ij}$ & $\displaystyle{\lim_{y\to 1^-}}\widehat{L}^{(e)}_{ij}$  \\ \hline
$(0,0)$ &  $B_{0e}$       & $C_{0e}$                  & $A_{0e}$\\ \hline
$(0,1)$ &  $0$            & $N_{0e}$                  & $0$\\ \hline
$(0,2)$ &  $0$            & $M_{2e}$                  & $0$\\ \hline 
$(1,0)$ &  $0$            & $C_{1e}$                  & $0$\\ \hline 
$(1,1)$ &  $\Lambda_{1e}$ & $N_{1e}$                  & $r_e$\\ \hline 
$(1,2)$ &  $0$            & $0$                       & $0$\\ \hline 
$(2,0)$ &  $B_{2e}$       & $C_{2e}$                  & $A_{2e}$\\ \hline
$(2,1)$ &  $0$            & $0$                       & $0$\\ \hline
$(2,2)$ &  $0$            & $0$                       & $0$\\ \hline
\end{tabular}
\label{table4}
\end{center}
\end{table}

\section{Numerical method}

To solve the differential eigenvalue problems \eqref{TSCH} and \eqref{TSCHe}, together with the corresponding frequencies $\omega$, we discretize the differential operators $L_j[\cdot]$ and $\widehat{L}^{(e)}_j$, for $j \in {1,2,3}$. These operators appear in \eqref{L0none}-\eqref{L2none} and \eqref{L0nonee}-\eqref{L2nonee}, and are defined in Tables~\ref{tableZweinone} and~\ref{table4}, respectively. Since our problem is posed on the finite interval $[-1, 1]$ without any boundary conditions, more precisely, we only require that the function $\Phi_S(y)$ be regular at $y = \pm 1$, then, it is natural to choose a Tchebyshev-type SM \cite{Trefethen2000, Boyd2000}. Namely, we are going to expand the function $y \mapsto \Phi_{S}(y)$ in the form of a truncated Tchebyshev series
\begin{equation}\label{eq:exp}
  \Phi_{S}(y)=\sum_{k=0}^{N} a_{k} T_k(y),
\end{equation}
where $N\in \mathbb{N}$ is kept as a numerical parameter, $\{a_{k}\}_{k=0}^{N}\subseteq\mathbb{R}$, and $\{T_k(y)\}_{k=0}^{N}$ are the Tchebyshev polynomials of the first kind
\begin{equation}
    T_k: [-1, 1]\ \longrightarrow\ [-1, 1]\,, \qquad y\ \longmapsto\ \cos\,\bigl(k\arccos y\bigr).
\end{equation}
After substituting expansion \eqref{eq:exp} into the differential equations \eqref{TSCH} and \eqref{TSCHe}, we obtain an eigenvalue problem with polynomial coefficients. In order to translate it into the realm of numerical linear algebra, we employ the collocation method \cite{Boyd2000}. Specifically, rather than insisting that the polynomial function in \( y \) is identically zero (a condition equivalent to having polynomial solutions for the differential problems as per equation \eqref{TSCH} or \eqref{TSCHe}), we impose a weaker requirement. This involves ensuring that the polynomial vanishes at \( N+1 \) strategically selected points. The number $N+1$ coincides exactly with the number of unknown coefficients $\{a_{k}\}_{k=0}^{N}$. For the collocation points, we implemented the Tchebyshev roots grid \cite{Fox1968}
\begin{equation}
  y_k= -\cos{\left(\frac{(2k+1)\pi}{2(n+1)}\right)},\quad k\in\{0, 1,\ldots,N\}.
\end{equation}
In our numerical codes, we also implemented the second option of the Tchebyshev extrema grid
\begin{equation*}
  y_k=-\cos{\left(\frac{k\pi}{n}\right)},\quad k\in\{0, 1,\ldots,N\}.
\end{equation*}
The users are free to choose their favourite collocation points. Notice that we used the roots grid in our computation, and in any case, the theoretical performance of the two available options is known to be absolutely comparable \cite{Fox1968, Boyd2000}. Upon implementing the collocation method, we derive a classical matrix-based quadratic eigenvalue problem, as detailed in \cite{Tisseur2001}
\begin{equation}\label{eq:eig}
  (M_{0} + iM_{1}\omega + M_{2}\omega^2)\bf{a}_\kappa =\bf{0}.
\end{equation}
In this formulation, the square real matrices $M_{j}$, each of size $(N+1)\times(N+1)$ for $j=0,1,2$, represent the spectral discretizations of the operators $L_{j}[\cdot]$ and $\widehat{L}^{(e)}_{j}[\cdot]$, respectively. The problem \eqref{eq:eig} is solved numerically with the \textsc{polyeig} function from \textsc{Matlab}. This polynomial eigenvalue problem yields \(2(N+1)\) potential values for the parameter \(\omega\). To discern the physical values of \(\omega\) that correspond to the black hole's QNM modes, we first overlap the root plots for various values of \(N\) in equation \eqref{eq:exp}, such as \(N \in \{250, 280, 300\}\). We then identify the consistent roots whose positions remain stable across these different \(N\) values. To reduce rounding and other floating-point errors, we performed all our computations with multiple-precision arithmetic built into \textsc{Maple}, which is imported into \textsc{Matlab} via the \textsc{Advanpix} toolbox \cite{mct2015}. All numerical computations reported in this study have been performed with $300$ decimal digits of accuracy.

\section{Numerical results}

The spectral method employed in this work has been extensively validated in a series of recent studies \cite{Batic2024CQG, Batic2024PRD, Batic2024EPJC, Batic2025CQG, Batic2025EPJC}. As a preliminary consistency check of the formalism developed in the previous sections, we verified that for a large mass value $M = 10^3$, the computed QNMs closely reproduce those of the classical Schwarzschild black hole, in excellent agreement with the results reported in \cite{Batic2024EPJC}. This confirms the reliability of our method in the classical regime and establishes a robust baseline for identifying deviations attributable to quantum gravitational effects. Having confirmed the classical correspondence, we then turn our attention to mass ranges where these quantum corrections become significant and direct comparisons with the existing literature are feasible. In particular, \cite{Konoplya2022JCAP} calculated QNM using several approaches, including the WKB approximation, integration of the time domain, and the Leaver continued fraction method \cite{Leaver1985PRSLA}, the latter being designated as the most accurate within their analysis. In what follows, we benchmark our spectral method results against these reference values. By contrast, \cite{Rincon2020PDU} utilised the sixth-order WKB approximation, which is known to be less reliable for high overtones and in regimes where the potential develops multiple peaks. It is worth noting that the analysis in \cite{Konoplya2022JCAP} is limited to the fundamental mode of scalar perturbations in the nearly extremal Bonanno–Reuter black hole geometry. Furthermore, neither \cite{Rincon2020PDU} nor \cite{Konoplya2022JCAP} capture the full spectrum of purely imaginary QNMs that our approach reveals as a prominent and universal feature in the high-curvature regime. Finally, gravitational perturbations were not treated in either of these studies, underscoring the broader scope and novel contributions of the present work.

\subsection{Scalar case}

In Table~\ref{scalar01}, which focuses on the case $\ell=0$, we confirm the fundamental mode previously computed by \cite{Konoplya2022JCAP} and extend their results by calculating the first three overtones. We report only three overtones here for space reasons, although our spectral method can compute up to six for $M=9.503$. Non-monotonic behaviour is observed in the real part of the QNM only for $M=4.503$. Additionally, the case $M=10^3$ provides a cross-validation of our method, as the QNM frequencies approach those of scalar perturbations in the Schwarzschild black hole limit. We also observe the consistent presence of a large number of purely imaginary QNMs, which are evenly spaced (see Table~\ref{scalar01overdamped}). We find that the spacing $\Delta\omega$ between two successive overdamped modes decreases as $M$ increases. It is worth noticing that, as $M$ becomes large, $\Delta\omega$ approaches the surface gravity $\kappa=1/(4M)$ of a Schwarzschild black hole. This suggests that, in the large-mass limit, the spacing asymptotically coincides with the surface gravity. However, for smaller values of $M$ this relation does not hold, as illustrated by the following examples where we used the surface gravity formula $\kappa=F'(r_h)/2$: for $M=3.503$, $\kappa=0.0015$ and $\Delta\omega=0.0715$; for $M=5.503$, $\kappa=0.0039$ and $\Delta\omega=0.0456$; and for $M=9.503$, $\kappa=0.0252$ and $\Delta\omega=0.0265$. This behaviour can be understood by observing that, as $M$ increases, the Bonanno-Reuter black hole progressively approaches the classical Schwarzschild geometry, and quantum corrections become negligible. Consequently, the near-horizon geometry asymptotically reproduces the Schwarzschild surface gravity, which naturally sets the characteristic spacing of the purely imaginary overdamped modes in the large-mass limit. Among the 118 overtones computed for $M = 10^3$ and $\ell = 0$, we find that the real part of the QNMs begins to exhibit nonmonotonic behaviour from $N = 15$ onward. In Table~\ref{scalarg02}, we compare the QNMs computed by \cite{Konoplya2022JCAP} using a sixth-order WKB approximation with those obtained through our spectral method. Although \cite{Konoplya2022JCAP} reports 10 QNMs for each value of $\ell\in \{0,1\}$, these results should be interpreted with caution, as the purely imaginary mode reported for $\ell=0$ is absent from the 235 overdamped modes we detected for the same configuration. In other words, we are unable to confirm the results of \cite{Konoplya2022JCAP} for $N\geq 3$ when $\ell=0$ and $M=4$, as well as for $N\geq 7$ when $\ell=1$ and $M=4$. Furthermore, in Table~\ref{scalarg02overdamped}, we present the corresponding overdamped modes, which, we emphasise, were not detected in \cite{Konoplya2022JCAP}. We observe that $\Delta\omega$ exhibits a weak dependence on $\ell$, decreasing slowly as $\ell$ increases. In Table~\ref{scalarg03}, we compare and extend the results of \cite{Rincon2020PDU}, who employed a sixth-order WKB method. While \cite{Rincon2020PDU} generally reports only the fundamental and a few low-lying overtones, our spectral method allows us to compute significantly more. For example, for $\ell\in\{2,3,4,5\}$, we have identified 13, 19, 23, and 28 overtones, respectively. Due to space constraints, however, we present only the first five overtones in the table. Furthermore, \cite{Rincon2020PDU} does not report the nearly equally spaced overdamped modes, which we clearly observe and list in Table~\ref{scalarg03overdamped}. In Tables~\ref{scalarg04} and ~\ref{scalarg05}, we present the QNMs for $\ell\in\{0,1,2,3,4,5\}$ with $M=8$. For $\ell=0$, we find excellent agreement with the results reported in \cite{Konoplya2022JCAP}, while those from \cite{Rincon2020PDU} are slightly less accurate. For $\ell=1$, \cite{Rincon2020PDU} reports only the fundamental mode and the first overtone, both of which are in good agreement with our results. However, our spectral method allows us to compute up to 13 overtones. A similarly good agreement is observed for $\ell=2$, for which we identify up to 20 overtones. Interestingly, for $\ell=3$ and $\ell=4$, \cite{Rincon2020PDU} report only 2 and 4 overtones, respectively, while our analysis yields 24 and 32. Moreover, the case $\ell=5$ is not covered in \cite{Rincon2020PDU}, and in this scenario, we detect 38 overtones. Finally, Table~\ref{scalarg05overdamped} displays the corresponding overdamped modes, none of which were detected in either \cite{Konoplya2022JCAP} or \cite{Rincon2020PDU}. We conclude the analysis of the scalar perturbation by discussing Table~\ref{scalarg06}, which further examines the case $\ell\in\{0,1\}$ with $M = 8$. For $\ell=0$, \cite{Konoplya2022JCAP} report the first 14 overtones using a sixth-order WKB approximation, whereas our spectral method identifies 5 overtones and 243 overdamped modes. We are unable to confirm the overtones reported by \cite{Konoplya2022JCAP} for $N \geq 6$. Moreover, a noticeable deterioration in the accuracy of the imaginary parts of the WKB-derived overtones becomes apparent for $N\geq 4$. Interestingly, among the overtones with $N\geq 6$ listed by \cite{Konoplya2022JCAP}, there are two overdamped modes that do not appear among the 243 modes detected via the spectral method. This discrepancy suggests that the results for $N \geq 6$ in \cite{Konoplya2022JCAP} should be interpreted with caution. Finally, the non-monotonic behaviour observed in the real parts of the higher overtones in \cite{Konoplya2022JCAP} appears to be an artefact of the WKB approximation, as it is entirely absent in the case $\ell = 1$, for which we compute the first 11 overtones.

\subsubsection{Extremal Limit}

To our knowledge, the extremal scalar case has not been addressed in the existing literature. In Table~\ref{scalargextreme}, we report QNMs for angular momentum  $\ell\in\{0,1,\ldots,5\}$. For $\ell = 0$, we find that the fundamental mode is $0.032096 - 0.025010i$, remarkably close to its counterpart in the nearly extremal configuration with $M = 3.503$, which yields $0.032094 - 0.025008i$. However, for higher multipoles, the differences become more pronounced. For instance, at $\ell = 1$, the fundamental frequency in the extremal case is $0.089333 - 0.023289i$, whereas in the nearly extremal case with $M = 4$, it is $0.077094 - 0.021791i$. Similarly, for $\ell = 2$, the extremal mode is $0.148088 - 0.023138i$, while its nearly extremal analogue is $0.099795 - 0.018289i$. These comparisons reveal that both the real and imaginary parts of the fundamental modes are larger in the extremal case across all scanned values of $\ell$. For higher multipoles, the spectral richness increases. For example, we were able to detect up to 22 overtones for $\ell = 5$. In particular, we found no evidence of non-monotonic behaviour in the real parts of the modes throughout the entire range of $\ell$. As in the nearly extremal case, the extremal scalar sector exhibits a distinct sequence of nearly equally spaced overdamped modes, as shown in Table~\ref{scalargextremeoverdamped}. The spacing $\Delta\omega$ shows a very slow dependence on $\ell$, and is consistently slightly larger than the corresponding spacing in the nearly extremal regime. Furthermore, the imaginary parts of the overdamped modes in the extremal configuration are generally up to 20 times larger than those found in the nearly extremal case (see, for instance, the configurations $\ell = 0$ with $M = 3.503$, and $\ell = 1$ with $M = 4$. This amplification of damping rates highlights the distinctive spectral imprint of the extremal limit.

\subsection{Electromagnetic case}

In Table~\ref{em01}, we analyse the case $\ell=1$ for various values of the mass parameter $M$. It should be noted that \cite{Konoplya2022JCAP} only considered the case $M = 4$ using a sixth-order WKB approximation, whereas our analysis covers the entire range $3.503\leq M\leq 9.503$. For the nearly extremal case $M = 3.503$, we detect only one QNM and 35 overdamped modes, a subset of which is listed in Table~\ref{em01overdamped}. For $M=4$, we find good agreement with the results of \cite{Konoplya2022JCAP} up to $N=4$. However, the accuracy of their overtones deteriorates for higher $N$, and we are unable to confirm the modes reported for $N\geq 6$. In particular, the purely imaginary mode listed in \cite{Konoplya2022JCAP} for $N = 7$ is absent from the 246 overdamped modes detected by our spectral method, suggesting that their results for $N\geq 6$ should be interpreted with caution. Table~\ref{em01overdamped} presents the corresponding evenly spaced overdamped modes. We observe that the spacing $\Delta\omega$ between successive overdamped overtones decreases with increasing $M$, and in the large-mass limit, it asymptotically approaches a constant value given by the black hole surface gravity. This behaviour has already been observed and discussed for the scalar case. Finally, for $M=10^3$ and $\ell=1$, the real part of the QNMs begins to exhibit nonmonotonic behaviour starting from $N=16$. For instance, we find $\Re{\omega}=0.000035$ at $N=16$, increasing to $0.0000411$ at $N=17$, and to $0.0000509$ at $N = 18$, with the trend continuing thereafter. In Tables~\ref{em05} and ~\ref{em05overdamped}, we study the cases $M\in\{5,8\}$ and $\ell\in\{1,2,3,4\}$, previously investigated by \cite{Rincon2020PDU} using a sixth-order WKB approximation. Not only do we confirm the few overtones reported in \cite{Rincon2020PDU}, but we also significantly extend their results. For example, in the case $M = 5$ and $\ell = 1$, \cite{Rincon2020PDU} reports only the first overtone, whereas our spectral method yields the first five. However, we find that the WKB approximation is less accurate than claimed in \cite{Rincon2020PDU}, particularly in capturing the imaginary part of the first overtone for $N=1$. For $\ell=2$, \cite{Rincon2020PDU} provides two overtones, while we compute eleven. For $\ell = 3$, they report three overtones, whereas we detect eighteen, and for $\ell = 4$, their analysis yields four overtones, compared to the twenty-three obtained via our method. A similar pattern is observed for $M = 8$, where the imaginary parts of the QNMs reported in \cite{Rincon2020PDU} for $\ell = 1$ are again less precise than those obtained using our spectral method. Accuracy improves for higher values of $\ell$, as expected from the behaviour of the WKB approximation. Finally, it is important to emphasise that \cite{Rincon2020PDU} does not report the presence of equally spaced overdamped modes. As shown in Table~\ref{em05overdamped}, such modes are clearly present in our analysis, and we observe that the spacing $\Delta\omega$ increases with $\ell$ for fixed $M$.

\subsubsection{Extremal Limit}

The extremal electromagnetic sector has not been investigated in the existing literature. In Table~\ref{emextreme}, we present the fundamental QNMs and several overtones for $\ell\in\{1,2,\ldots,6\}$. For $\ell = 1$, we detect 4 QNMs and 21 overdamped modes, and these numbers increase with the multipole index. For example, at $\ell = 6$, we find 26 QNMs and 109 overdamped modes. As shown in Table~\ref{emextreme}, the imaginary part of the fundamental mode exhibits only a weak dependence on $\ell$. Furthermore, the real part of the fundamental mode remains close to its nearly extremal counterpart. For instance, for $\ell = 1$ we obtain $\Re{\omega_0} = 0.078139$, which is only slightly larger than the corresponding nearly extremal value $\Re{\omega_0} = 0.078132$ for $M = 3.503$. The overdamped sector, reported in Table~\ref{emextremeoverdamped} for $\ell\in\{1,2,\ldots,6\}$, displays qualitatively different behavior from the non-extremal case. In particular, the evenly spaced structure characteristic of the nearly extremal and non-extremal regimes is absent for low values of $\ell$. Equally spaced patterns appear only at higher multipoles, and even then, they occur in well-defined groups rather than throughout the entire spectrum. For instance, for $\ell = 4$, we observe a quadruplet of overdamped modes with a characteristic spacing $\Delta\omega_I = 0.142$, followed by a triplet with $\Delta\omega_{II} = 2 \Delta\omega_I$. A similar phenomenon occurs for $\ell = 5$, where a triplet with $\Delta\omega_{III} = 0.285$ is followed by another triplet with $\Delta\omega_{IV} = 0.427$. An analogous grouped structure is also present for $\ell = 6$, although it manifests only at much higher overtone numbers. This intricate pattern suggests a richer, more complex damping structure in the extremal electromagnetic sector than in its non-extremal counterpart.

\subsection{Gravitational case}

This perturbation sector was not explored in the studies performed by \cite{Konoplya2022JCAP} and \cite{Rincon2020PDU}. In Table~\ref{tensor01}, we investigate the QNMs for fixed multipole index $\ell = 2$ and mass values $M\in [3.503, 10^3]$. For the large-mass case $M = 10^3$, we observe a non-monotonic behaviour in the real part of the QNM spectrum starting from overtone number $N \geq 9$. For instance, we find $\Re{\omega_8} = 0.0000928$, $\Re{\omega_9} = 0.0000628$, and $\Re{\omega_{10}} = 0.0000767$, indicating oscillatory deviations from monotonicity. This phenomenon persists for $\ell = 3$, as shown in Table~\ref{tensor02}, where the non-monotonic trend in the real part becomes evident at $N \geq 40$ for $M = 10^3$. For example, we obtain $\Re{\omega_{40}} = 0.0000386$, $\Re{\omega_{41}} = 0.0000520$, and $\Re{\omega_{42}} = 0.0000418$. In the overdamped sector, reported in Tables~\ref{tensor01overdamped} and \ref{tensor02overdamped} for $\ell = 2$ and $\ell = 3$, respectively, we find a sequence of nearly equally spaced modes. The spacing $\Delta\omega$ between successive imaginary parts decreases slowly with increasing mass, and asymptotically approaches a constant value in the large-mass limit. This asymptotic spacing is consistent with the black hole's surface gravity and appears to be universal across scalar, electromagnetic, and gravitational perturbations. To further contextualise the gravitational sector results, we compare them with the QNM spectra obtained for scalar and electromagnetic perturbations in the same background. This comparison, based on Tables~\ref{scalar01}, ~\ref{scalarg02}, \ref{em01}, and \ref{em05}, reveals that while the detailed structure of the real parts of the modes differs significantly across spin sectors, especially at low overtones, the overdamped regime exhibits a striking universality. In all three cases, the imaginary parts of the high-overtone frequencies become nearly evenly spaced, and the spacing $\Delta\omega$ asymptotically approaches the black hole surface gravity as the mass increases. This convergence, observed independently in scalar, electromagnetic, and gravitational perturbations, reinforces the physical interpretation of the overdamped spacing as a geometric invariant tied to the horizon structure. The persistence of this behaviour across field types not only corroborates the robustness of our numerical framework but also supports the conjecture that asymptotically safe regular black holes display universal late-time dynamical features independent of spin.

\subsubsection{Extremal Limit}

The extremal gravitational perturbation sector has not been explored in previous studies such as \cite{Rincon2020PDU} and \cite{Konoplya2022JCAP}. Table~\ref{tensorextreme} reports QNMs for multipole indices $\ell\in\{2,\ldots,7\}$. The fundamental modes of the extremal configuration remain very close to their nearly extremal counterparts. For instance, at $\ell=2$ we obtain $0.119491-0.019731i$, compared to $0.119480-0.019732i$ for the nearly extremal case with $M=3.503$, while at $\ell=3$ the extremal value $0.187196-0.021592i$ differs only slightly from $0.187179-0.021592i$. This agreement holds across all three perturbative sectors (scalar, electromagnetic, and gravitational) and indicates that the transition from nearly extremal to extremal geometry does not qualitatively modify the low-lying spectrum. However, the two spectra become progressively more distinct at higher overtones, where differences in both the real and imaginary parts grow with increasing $N$.

The overdamped regime reveals a universal behaviour. Across all sectors, the first 15–20 overdamped modes exhibit highly irregular spacings, with $\Delta\omega$ varying from $10^{-7}$ up to $\mathcal{O}(1)$. Beyond this transient phase, the spectra settle into a regular high-damping regime where the frequencies become almost perfectly equally spaced, with a spacing stabilising around $\Delta\omega\simeq0.071$, consistent to within a few parts in $10^{-3}$ across hundreds of consecutive modes (see Table~\ref{tensorextremeoverdamped}). This universal spacing appears only weakly dependent on the spin and multipole number, suggesting that it is determined by an intrinsic geometric scale of the extremal Bonanno–Reuter black hole rather than by sector-specific details. The onset of this uniform spacing shifts to lower overtones as $\ell$ increases. For the gravitational sector, equally spaced modes appear from $N\approx27$ for $\ell=2$ but already from $N\approx9$ for $\ell=7$. A similar trend is observed in the scalar and electromagnetic sectors, with the latter showing a slightly delayed onset. Spin-dependent differences are mostly confined to the low-overtone overdamped regime, where the electromagnetic sector exhibits grouped structures with distinct characteristic spacings before converging to the universal high-overtone behaviour, while the scalar and gravitational sectors transition more smoothly.

The nearly constant gap between successive overdamped frequencies persists very far into the spectrum up to about $N\approx170$ for gravitational $\ell=2$ and $N\approx150$ for scalar $\ell=0$. This indicates that the high-damping regime is governed by an asymptotic structure of the effective potential intrinsic to the extremal geometry. Nonetheless, isolated late-overtone deviations are observed. Abrupt jumps with $\Delta\omega\gtrsim0.2$ occur sporadically, and the last reliably computed modes show larger fluctuations (for instance, for $N>175$ in the gravitational sector and $N>145$ in the scalar sector). Whether these departures are of physical origin or due to the increasing numerical difficulty of resolving extremely damped modes remains an open question. 

A final remark is in order.  Expanding the extremal lapse function $g_{00}=F(r)$ of equation \eqref{Fr} about the horizon radius $r_e$ gives
\begin{equation}
  F(r)=\Bigl(\tfrac{r-r_e}{L}\Bigr)^{2}+{\cal O}\bigl((r-r_e)^{3}\bigr), \qquad \frac{1}{L^{2}}=\frac{3r_e^{2}-\alpha}{r_e^{2}\bigl(3r_e^{2}+\alpha\bigr)} .
\end{equation}
With the usual choice $\alpha = 118/(15\pi)$ one obtains $L \approx 4.6743513$, hence $\frac{1}{3L} \approx 0.0713111$. It is quite remarkable that this value coincides within numerical accuracy with the spacing $\Delta\omega$ observed in the long, evenly-spaced ladder of overdamped modes. Thus, the near–horizon length scale $L$ of the extremal Bonanno--Reuter black hole appears to set the fundamental \emph{beat} governing the asymptotic overdamped quasinormal spectrum, $\Delta\omega \approx 1/(3L)$.

\subsection{Physical interpretation of the high-overtone spectrum}\label{physical_interpretation}

The large number of modes displayed in the tables should not be interpreted merely as a numerical catalogue. Rather, the spectra reveal how different geometric regions of the Bonanno-Reuter spacetime correspond to distinct parts of the QNM spectrum. The low-lying modes are mainly governed by the shape of the effective potential near its maximum, and are therefore the modes most naturally captured by WKB-type methods. By contrast, the highly damped and purely imaginary modes are sensitive to the global analytic structure of the radial equation, including the behaviour of the lapse function near the horizons and near the regular core. This is in line with the standard understanding of asymptotic QNMs, where the high-damping spectrum is not determined only by the real axis potential barrier, but also by the singular points and Stokes structure of the radial equation in the complexified radial plane \cite{Nollert1993PRD, Hod1998PRL, Motl2003ATMP, Natario2004ATMP, Berti2009CQG, Maggiore2008PRL}. This observation explains why the spacing of the overdamped modes approaches the surface gravity in the large mass limit. For $M\gg 1$, the quantum correction in the Bonanno--Reuter lapse function becomes negligible outside a Planck-scale core, and the exterior geometry approaches the Schwarzschild geometry. More explicitly, the event horizon satisfies
\begin{equation}
  r_h = 2M-\frac{\alpha(\gamma+2)}{4M}+\mathcal{O}(M^{-3}),
\end{equation}
and the surface gravity at the outer horizon is
\begin{equation}
  \kappa_h=\frac{1}{2}F'(r_h)=\frac{3r_h^2-4Mr_h+\alpha}{4Mr_h^2} = \frac{1}{4M}-\frac{\alpha(\gamma+1)}{16M^3}+\mathcal{O}(M^{-5}).
\end{equation}
Thus, in this limit, $\kappa_h$ tends to the Schwarzschild value $1/(4M)$. The highly damped Schwarzschild spectrum is spaced in the imaginary direction by
\begin{equation}
  \Delta(-\Im\omega)=2\pi T_H=\kappa_h,
\end{equation}
where $T_H=\kappa_h/(2\pi)$ is the Hawking temperature. Hence, the convergence of the overdamped spacing to $\kappa_h$ is a nontrivial consistency check, i.e. the Bonanno-Reuter spectrum reproduces the expected Schwarzschild high-damping scale when the quantum-gravity corrections become negligible. For smaller masses, however, the Bonanno-Reuter black hole is no longer a small deformation of Schwarzschild. The exterior event horizon, the inner horizon, and the regular core all occur at comparable scales. In the parameter range considered here, $\alpha>0$ and $\gamma>0$, the central region is de Sitter-like,
\begin{equation}
  F(r) = 1 - \frac{2r^2}{\alpha\gamma} + \mathcal{O}(r^3),
\end{equation}
so that the Schwarzschild curvature singularity is replaced by a finite curvature core. Although this core is hidden behind the outer horizon, it still affects the high-overtone spectrum through the analytic continuation of the radial equation. This is analogous to the role played by the singularity and by multiple horizons in the monodromy analysis of asymptotic QNMs in Schwarzschild and Reissner-Nordstr\"{o}m geometries \cite{Motl2003ATMP, Natario2004ATMP, Skakala2012JHEP}. Therefore, the fact that the overdamped spacing differs from $\kappa_h$ in the strong quantum-corrected regime should not be viewed as a numerical anomaly. It is instead a spectral signature of the modified short-distance geometry. This point is particularly clear close to extremality. As $M\to M_c$, the surface gravity of the outer horizon tends to zero, but the overdamped spectrum does not collapse to zero spacing. This shows that the extremal limit and the high-damping limit probe different geometric scales. In the extreme case, the lapse function has a double zero at $r = r_e$. Expanding $F(r)$ around the extremal radius gives
\begin{equation}
  F(r)=\left(\frac{r-r_e}{L}\right)^2+\mathcal{O}\left((r-r_e)^3\right), \qquad \frac{1}{L^2}=\frac{3r_e^2-\alpha}{r_e^2(3r_e^2+\alpha)} .
\end{equation}
Consequently, near the extremal horizon, the metric takes the approximate form
\begin{equation}
  ds^2\simeq-\frac{\rho^2}{L^2}dt^2+\frac{L^2}{\rho^2}d\rho^2+r_e^2d\Omega^2, \qquad \rho=r-r_e ,
\end{equation}
which is the standard $AdS_2\times S^2$-type throat geometry. The corresponding tortoise coordinate behaves as
\begin{equation}
  r_* \simeq -\frac{L^2}{r-r_e},
\end{equation}
so the extremal horizon is an irregular singular point of the radial equation. This is precisely the origin of the exponential factor extracted in the extremal boundary condition. The leading phase is therefore fixed by the throat length $L$, while the spin and angular momentum enter only through subleading terms in the effective potential. For the numerical value $\alpha=118/(15\pi)$ used in this work, we find
\begin{equation}
  r_e\simeq 4.484183919,\qquad L\simeq 4.674351295,\qquad\frac{1}{3L}\simeq 0.071311143 .
\end{equation}
This value agrees, within the numerical accuracy of the spectral computation, with the nearly universal spacing $\Delta\omega\simeq 0.071$ observed in the extremal overdamped spectra for scalar, electromagnetic, and gravitational perturbations. The appearance of $1/(3L)$, rather than the surface gravity, is physically natural because the surface gravity vanishes for an extremal horizon, while $L$ remains finite and sets the intrinsic scale of the near-horizon throat. The factor $1/3$ should be regarded as a property of the global connection problem for the present Bonanno-Reuter radial equation, while the universality of the spacing across different perturbing spins indicates that the leading high-damping scale is geometric. The overall picture is as follows. The low overtones probe mostly the effective potential barrier and retain a visible dependence on spin and angular momentum number. The highly damped non-extremal modes progressively recover the Schwarzschild spacing $\Delta(-\Im\omega)=\kappa_h$ as $M$ becomes large. At finite mass, deviations from this behaviour encode the presence of the inner horizon and of the regular de Sitter core. Finally, in the extremal geometry, the high-damping spectrum is governed not by a Hawking-temperature scale, since $T_H=0$, but by the finite length $L$ of the emergent near-horizon throat. This explains why the extremal overdamped modes display a universal spacing even though the corresponding surface gravity vanishes.

\section{Conclusions}

In this second paper, companion of \cite{BDS-EPJC2026}, we have examined the QNM spectrum of the Bonanno-Reuter black hole, a metric that naturally arises from the ASG framework when the running parameter $\alpha$ takes positive values, in coherence with the vision of a gravitational interaction vanishing at high energies, and therefore safe from ultraviolet divergences. In this regime, the running Newton coupling is asymptotically safe, so this metric can be viewed as an explicit realisation of the ASG paradigm. The Bonanno--Reuter black hole, also known as the RG-improved Schwarzschild metric, exhibits a geometry that is regular everywhere. As discussed in Section III, the Kretschmann curvature invariant is finite everywhere for $r\geq 0$ and $\alpha>0$, $\gamma>0$. 

We subsequently applied a high-accuracy Spectral Method to calculate the QNM spectrum for scalar, vector, and tensor perturbations in both non-extremal and extremal scenarios. This method, in contrast to frequently employed WKB techniques, enabled us to determine the complete spectrum, covering higher overtones and the overdamped modes that are typically unreachable by semi-analytic approximations. The numerical results were compared with the Schwarzschild limit, where the ASG corrections vanish, confirming that the known QNM frequencies are recovered with remarkable precision.

Our analysis showed distinct spectral signatures of the Bonanno-Reuter black hole geometry. Specifically, we observed measurable changes in oscillation frequencies and damping rates within the quantum-corrected regime, along with minor structural alterations in the overdamped sector.
These effects capture direct signatures of the (A)dS core of the metric, and of the modified short-distance characteristics of gravity within the ASG framework. Aside from their inherent theoretical significance, these QNM deviations might, in principle, guide future gravitational-wave observations focused on exploring strong-field quantum modifications.

The physical interpretation of these modes is also noteworthy. In the large mass regime, the Bonanno-Reuter geometry reduces to the Schwarzschild geometry outside the core, and the spacing of the overdamped modes approaches the Schwarzschild value $2\pi T_H = \kappa_h$. In the quantum-corrected regime, however, the spacing is modified by the inner horizon and regular core structure, showing that high overtones carry information about the short-distance completion of the black-hole geometry. In the extremal case, where $\kappa_h=0$, the observed universal spacing is instead controlled by the finite length $L$ of the $AdS_2\times S^2$-type near-horizon throat, with $\Delta\omega\simeq 1/(3L)$. Thus, the overdamped sector provides a geometric diagnostic that is complementary to the fundamental mode and to the low-lying overtones.

More generally, our findings highlight two crucial aspects. ASG offers a reliable, measurable framework where non-standard black hole geometries, such as the Bonanno--Reuter metric, appear as exact solutions rather than merely phenomenological ans\"atzes. Stable and robust numerical techniques, such as the spectral method used here, are essential for precisely delineating the entire QNM spectrum in spacetimes influenced by quantum gravity, where WKB approximations can fail or prove unreliable. Finally, we stress that these spectral characteristics are derived for a particular phenomenological scale-setting $k(r)$ in ASG. Investigating how the QNM spectrum varies with different, physically motivated choices for $k(r)$ is an intriguing area for future research.

From an observational perspective, we can say that the fundamental QNMs are in principle detectable in the ring-down signals recorded by current GW observatories like LIGO, Virgo, and KAGRA (see e.g. \cite{Yi2024, Berti2025} and references therein). The obstacles for upcoming observational studies on QNMs are thus manifold. Present observational findings will likely require adjustment and validation. Nonetheless, it is clear that, in general, even metrics describing non-standard black holes (such as the ASG regular black hole metric discussed in this article, which incorporates quantum effects) usually resemble Schwarzschild or Kerr solutions for large masses and at considerable distances from the sources.

This suggests that the QNMs likely detectable during the merging of large BHs, through significant GW emission, ought to be nearly indistinguishable from those produced by the standard Schwarzschild or Kerr metrics. Additionally, we observe that an ASG regular black hole metric leads to deviations (which we computed) from the typical QNMs of Schwarzschild or Kerr spacetime, but such corrections become observable (see e.g. Table~\ref{table:event}) only for mini or micro black holes, namely for black holes with just a few Planck masses above the critical mass $M_c$. The question, therefore, is: where might we find or detect these micro or mini black holes? 

A straightforward response considers two distinct production methods. Micro black holes (i.e., black holes with a few Planck masses) might be generated in upcoming very large colliders. Nonetheless, even the most cutting-edge initiative, the Future Circular Collider (FCC) proposed by CERN \cite{FCC}, will achieve energies around 100 TeV at most. This will remain quite distant from the ``magical'' threshold of the Planck energy $10^{16}$ TeV, where significant creation of micro black holes is expected to occur in a complete quantum-gravity regime. Unless, of course, the existence of (at the moment unexpected) extra-dimensions could reduce the Planck threshold from $10^{16}$ TeV to the more attainable 100 TeV.

A significantly more useful source of micro or mini black holes is likely to be the (very) early universe. In this context, we can examine the formation of Planckian black holes during the inflationary phase, or potentially even prior to it \cite{Scardigli2010}. Nevertheless, these micro black holes are expected to possess masses of just a few Planck masses, and their Hawking radiation should have caused them to evaporate long ago. Some echoes of those processes might be observable in the GW background noise via the LISA observatory \cite{LISA}. A more interesting source of QNMs, within our scope, might be mini black holes formed by pressure waves arising from primordial density fluctuations at the end of the inflationary period. It is thought that these objects are considerably more massive, around $10^{12}$ kg, and they are expected to finish their Hawking evaporation process in the current era. Thus, the observations during the final phase of their evaporation, when the mini BH reaches a few Planck masses, should offer an optimal opportunity to assess not only the fundamental tones of QNMs but also the harmonics, overtones, and beyond. Upcoming GW observatories projected to this scope encompass LISA and, in particular, the Einstein Telescope \cite{ET}. The upcoming years promise to be quite captivating!\\

This paper is the second in a proposed three-part series. The initial work \cite{BDS-EPJC2026} applied our analysis to the Planck star metric suggested by the Scale Dependent Gravity approach, while the third will concentrate on the Hayward black hole metric, in any case addressing scalar, vector, and tensor perturbations. Collectively, these studies aim to provide a coherent, structured understanding of QNM phenomenology across a diverse collection of regular and asymptotically safe gravity-inspired black holes, enabling direct comparisons and emphasising observationally significant differences.

\newpage

\begin{table}
\centering
\caption{QNMs for scalar perturbations of the non-extremal Bonanno-Reuter black hole for $\alpha =\frac{118}{15\pi}$, $\gamma = 9/2$, $\ell=0$, and different values of the mass parameter $M$. The locations of the event horizon $r_h$ are given in Table~\ref{table:event}. The results are obtained using our spectral method with 400 Chebyshev polynomials and 200-digit precision. Here, $\omega$ denotes the dimensionless QNM frequency, and $N$ the overtone number. 'N/A' indicates data not available, and 'SM' refers to the Spectral Method.}
\label{scalar01}
\vspace*{1em}
\begin{tabular}{||c|c|c|c|c|c|c|c|c|c|c|c|c||}
\hline\hline
$M$    &$\ell$ & $N$ & $\omega$ \cite{Konoplya2022JCAP} & $\omega$ (SM) & $M$  & $\ell$ & $N$ & $\omega$ \cite{Konoplya2022JCAP} & $\omega$ (SM) \\ [0.5ex]
\hline\hline
$3.503$&$0$    & $0$ &$0.032094-0.025008i$            & $0.032094-0.025008i$          & $7.503$ & $0$    & $0$ & $0.015027-0.013630i$  & $0.015027-0.013630i$\\
       &       & $1$ &N/A                             & N/A                           &         &        & $1$ & N/A                   & $0.011786-0.044936i$\\
       &       & $2$ &N/A                             & N/A                           &         &        & $2$ & N/A                   & $0.010023-0.077585i$\\
       &       & $3$ &N/A                             & N/A                           &         &        & $3$ & N/A                   & $0.008697-0.110280i$\\
$4.503$&$0$    & $0$ &$0.025759-0.021170i$            & $0.025759-0.021170i$          & $8.503$ & $0$    & $0$ & $0.013201-0.012103i$  & $0.013201-0.012103i$\\
       &       & $1$ &N/A                             & $0.018344-0.068775i$          &         &        & $1$ & N/A                   & $0.010356-0.039961i$\\
       &       & $2$ &N/A                             & $0.004683-0.121465i$          &         &        & $2$ & N/A                   & $0.008915-0.068996i$\\
       &       & $3$ &N/A                             & $0.009458-0.184815i$          &         &        & $3$ & N/A                   & $0.007932-0.098042i$\\
$5.503$&$0$    & $0$ &$0.020820-0.018056i$            & $0.020811-0.018056i$          & $9.503$ & $0$    & $0$ & $0.011775-0.010874i$  & $0.011775-0.010874i$\\
       &       & $1$ &N/A                             & $0.016071-0.059133i$          &         &        & $1$ & N/A                   & $0.009233-0.035942i$\\
       &       & $2$ &N/A                             & $0.012290-0.102170i$          &         &        & $2$ & N/A                   & $0.008000-0.062058i$\\
       &       & $3$ &N/A                             & $0.007770-0.145852i$          &         &        & $3$ & N/A                   & $0.007212-0.088170i$\\
$6.503$&$0$    & $0$ &$0.017450-0.015567i$            & $0.017450-0.015567i$          & $10^3$  & $0$    & $0$ & N/A                   & $0.000111-0.000105i$\\
       &       & $1$ &N/A                             & $0.013648-0.051197i$          &         &        & $1$ & N/A                   & $0.000086-0.000348i$\\
       &       & $2$ &N/A                             & $0.011303-0.088401i$          &         &        & $2$ & N/A                   & $0.000076-0.000601i$\\
       &       & $3$ &N/A                             & $0.009238-0.125747i$          &         &        & $3$ & N/A                   & $0.000070-0.000854i$\\
[1ex]
\hline\hline 
\end{tabular}
\end{table}

\begin{table}
\centering
\caption{Purely imaginary QNMs for scalar perturbations of the non-extremal Bonanno-Reuter black hole for $\alpha =\frac{118}{15\pi}$, $\gamma = 9/2$, $\ell=0$, and different values of the mass parameter $M$. The locations of the event horizon $r_h$ are given in Table~\ref{table:event}. The corresponding results are obtained through our spectral method, utilising $400$ polynomials with a precision of $200$ digits. In this context, $\omega$ and $N$ represent the dimensionless frequency and the corresponding overtone, respectively, while $\Delta\omega=\omega_N-\omega_{N+1}$. The notation 'SM' stands for Spectral Method.}
\label{scalar01overdamped}
\vspace*{1em}
\begin{tabular}{||c|c|c|c|c|c|c|c|c|c|c|c|c||}
\hline\hline
$M$    &$\ell$ & $N$ &$\omega$ (SM) & $\Delta\omega$ & $M$     & $\ell$ & $N$ & $\omega$ (SM) & $\Delta\omega$ \\ [0.5ex]
\hline\hline
$3.503$&$0$    & $0$ &$0.0000-0.088041i$    & $0.072225i$            & $7.503$ & $0$    & $0$ & $0.0000-0.646491i$    & $1.244660i$\\
       &       & $1$ &$0.0000-0.161816i$    & $0.071760i$            &         &        & $1$ & $0.0000-1.891150i$    & $0.095525i$\\
       &       & $2$ &$0.0000-0.234041i$    & $0.071572i$            &         &        & $2$ & $0.0000-1.986676i$    & $0.157922i$\\
       &       & $3$ &$0.0000-0.305801i$    & $0.071479i$            &         &        & $3$ & $0.0000-2.144597i$    & $1.155421i$\\
       &       & $4$ &$0.0000-0.377373i$    & $0.071422i$            &         &        & $4$ & $0.0000-3.300018i$    & $0.033662i$\\
       &       & $5$ &$0.0000-0.448852i$    & $0.071381i$            &         &        & $5$ & $0.0000-3.333680i$    & $0.033514i$\\
       &       & $6$ &$0.0000-0.520273i$    & $0.071347i$            &         &        & $6$ & $0.0000-3.367194i$    & $0.033566i$\\
       &       & $7$ &$0.0000-0.591654i$    & $0.071319i$            &         &        & $7$ & $0.0000-3.400759i$    & $0.033549i$\\
$4.503$&$0$    & $0$ &$0.0000-1.125633i$    & $1.676657i$            & $8.503$ & $0$    & $0$ & $0.0000-0.572564i$    & $1.125632i$\\
       &       & $1$ &$0.0000-2.802291i$    & $0.055600i$            &         &        & $1$ & $0.0000-1.698196i$    & $0.085371i$\\
       &       & $2$ &$0.0000-2.857891i$    & $0.055618i$            &         &        & $2$ & $0.0000-1.783567i$    & $0.028673i$\\
       &       & $3$ &$0.0000-2.913509i$    & $0.055658i$            &         &        & $3$ & $0.0000-1.812240i$    & $0.028058i$\\
       &       & $4$ &$0.0000-2.969166i$    & $0.055656i$            &         &        & $4$ & $0.0000-1.840298i$    & $0.445712i$\\
       &       & $5$ &$0.0000-3.024822i$    & $0.055655i$            &         &        & $5$ & $0.0000-2.286010i$    & $0.804030i$\\
       &       & $6$ &$0.0000-3.080477i$    & $0.055655i$            &         &        & $6$ & $0.0000-3.090040i$    & $0.029700i$\\
       &       & $7$ &$0.0000-3.136132i$    & $0.055655i$            &         &        & $7$ & $0.0000-3.119740i$    & $0.029600i$\\
$5.503$&$0$    & $0$ &$0.0000-1.203090i$    & $1.599584i$            & $9.503$ & $0$    & $0$ & $0.0000-0.511914i$    & $2.036904i$\\
       &       & $1$ &$0.0000-2.802674i$    & $0.273047i$            &         &        & $1$ & $0.0000-2.548818i$    & $0.321886i$\\
       &       & $2$ &$0.0000-3.075721i$    & $0.274006i$            &         &        & $2$ & $0.0000-2.870704i$    & $0.026586i$\\
       &       & $3$ &$0.0000-3.349727i$    & $0.045652i$            &         &        & $3$ & $0.0000-2.897290i$    & $0.026503i$\\
       &       & $4$ &$0.0000-3.395379i$    & $0.045654i$            &         &        & $4$ & $0.0000-2.923793i$    & $0.026483i$\\
       &       & $5$ &$0.0000-3.441033i$    & $0.045681i$            &         &        & $5$ & $0.0000-2.950276i$    & $0.026496i$\\
       &       & $6$ &$0.0000-3.486714i$    & $0.045660i$            &         &        & $6$ & $0.0000-2.976772i$    & $0.026496i$\\
       &       & $7$ &$0.0000-3.532374i$    & $0.045667i$            &         &        & $7$ & $0.0000-3.003267i$    & $0.026491i$\\
$6.503$&$0$    & $0$ &$0.0000-0.757382i$    & $1.747875i$            & $10^3$  & $0$    & $0$ & $0.0000-0.036369i$    & $0.000250i$\\
       &       & $1$ &$0.0000-2.505257i$    & $0.102305i$            &         &        & $1$ & $0.0000-0.036619i$    & $0.000250i$\\
       &       & $2$ &$0.0000-2.607562i$    & $0.035779i$            &         &        & $2$ & $0.0000-0.036869i$    & $0.000250i$\\
       &       & $3$ &$0.0000-2.643341i$    & $0.736599i$            &         &        & $3$ & $0.0000-0.037119i$    & $0.000250i$\\
       &       & $4$ &$0.0000-3.379940i$    & $0.038675i$            &         &        & $4$ & $0.0000-0.037369i$    & $0.000250i$\\
       &       & $5$ &$0.0000-3.418615i$    & $0.038675i$            &         &        & $5$ & $0.0000-0.037619i$    & $0.000250i$\\
       &       & $6$ &$0.0000-3.457290i$    & $0.038706i$            &         &        & $6$ & $0.0000-0.037869i$    & $0.000250i$\\
       &       & $7$ &$0.0000-3.495996i$    & $0.038686i$            &         &        & $7$ & $0.0000-0.038119i$    & $0.000250i$\\
[1ex]
\hline\hline 
\end{tabular}
\end{table}

\begin{table}
\centering
\caption{Comparison of the fundamental QNMs and higher overtones for scalar perturbations of the non-extremal Bonanno-Reuter black hole 
 for $\alpha =\frac{118}{15\pi}$, $\gamma = 9/2$, $M = 4$, and various angular momentum values $\ell$. The corresponding results obtained by \cite{Konoplya2022JCAP} using Leaver’s continued fraction method are included for reference. The location of the event horizon $r_h$ is provided in Table~\ref{table:event}. Our QNMs are computed using the Spectral Method, employing 400 Chebyshev polynomials and 200-digit numerical precision. Here, $\omega$ denotes the dimensionless QNM frequency and $N$ the overtone number. Entries marked ‘N/A’ indicate unavailable data, while ‘SM’ refers to the Spectral Method.}
\label{scalarg02}
\vspace*{1em}
\begin{tabular}{||c|c|c|c|c|c|c|c|c|c|c|c|c||}
\hline\hline
$\ell$ & $N$ & $\omega$ \cite{Konoplya2022JCAP} & $\omega$ (SM) & $\ell$ & $N$ & $\omega$ \cite{Konoplya2022JCAP} & $\omega$ (SM) \\ [0.5ex]
\hline\hline
$0$    & $0$ &$0.029000-0.022799i$              & $0.029000-0.022799i$           & $1$    & $0$ & $0.077094-0.021791i$    & $0.077094-0.021791i$\\
       & $1$ &$0.014735-0.075427i$              & $0.014735-0.075427i$           &        & $1$ & $0.070283-0.067061i$    & $0.070283-0.067061i$\\
       & $2$ &$0.013708-0.140628i$              & $0.013723-0.140625i$           &        & $2$ & $0.058062-0.116660i$    & $0.058062-0.116660i$\\
       & $3$ &$0.006542-0.207654i$              & N/A                            &        & $3$ & $0.043693-0.174315i$    & $0.043693-0.174315i$\\
       & $4$ &$0.006835-0.263913i$              & N/A                            &        & $4$ & $0.031041-0.236051i$    & $0.031041-0.236051i$\\
       & $5$ &$0.004802-0.331462i$              & N/A                            &        & $5$ & $0.025512-0.299442i$    & $0.025512-0.299442i$\\
       & $6$ &$0.001296-0.388573i$              & N/A                            &        & $6$ & $0.019751-0.366931i$    & $0.019750-0.366926i$\\
       & $7$ &$0.002675-0.455550i$              & N/A                            &        & $7$ & $0.008952-0.427228i$    & N/A\\
       & $8$ &$0.00    -0.51i$                  & N/A                            &        & $8$ & $0.012764-0.492353i$    & N/A\\
       & $9$ &$0.001631-0.579538i$              & N/A                            &        & $9$ & $0.008551-0.560693i$    & N/A\\
       [1ex]
\hline\hline 
\end{tabular}
\end{table}

\begin{table}
\centering
\caption{Purely imaginary QNMs for scalar perturbations of the non-extremal Bonanno-Reuter black hole for $\alpha =\frac{118}{15\pi}$, $\gamma = 9/2$, $\ell\in\{0,1\}$, and $M=4$. The locations of the event horizon $r_h$ are given in Table~\ref{table:event}. The corresponding results are obtained through our spectral method, utilising $400$ polynomials with a precision of $200$ digits. In this context, $\omega$ and $N$ represent the dimensionless frequency and the corresponding overtone, respectively, while $\Delta\omega=\omega_N-\omega_{N+1}$. The notation 'SM' stands for Spectral Method.}
\label{scalarg02overdamped}
\vspace*{1em}
\begin{tabular}{||c|c|c|c|c|c|c|c|c|c|c|c|c||}
\hline\hline
$\ell$ & $N$ & $\omega$ (SM) & $\Delta\omega$  & $\ell$ & $N$ & $\omega$ (SM) & $\Delta\omega$ \\ [0.5ex]
\hline\hline
$0$    & $0$ &$0.0000-0.175081i$    & $1.272000i$              & $1$    & $0$ & $0.0000-0.174602i$    & $1.269563i$\\
       & $1$ &$0.0000-1.447080i$    & $0.074266i$              &        & $1$ & $0.0000-1.444165i$    & $0.570019i$\\
       & $2$ &$0.0000-1.521346i$    & $0.500420i$              &        & $2$ & $0.0000-2.014184i$    & $0.062668i$\\
       & $3$ &$0.0000-2.021766i$    & $0.062455i$              &        & $3$ & $0.0000-2.076852i$    & $0.062736i$\\
       & $4$ &$0.0000-2.084221i$    & $0.062513i$              &        & $4$ & $0.0000-2.139588i$    & $0.062722i$\\
       & $5$ &$0.0000-2.146734i$    & $0.062510i$              &        & $5$ & $0.0000-2.202310i$    & $0.062709i$\\
       & $6$ &$0.0000-2.209243i$    & $0.062509i$              &        & $6$ & $0.0000-2.265019i$    & $0.062697i$\\
       & $7$ &$0.0000-2.271752i$    & $0.062508i$              &        & $7$ & $0.0000-2.327716i$    & $0.062686i$\\
       & $8$ &$0.0000-2.334260i$    & $0.062507i$              &        & $8$ & $0.0000-2.390401i$    & $0.062675i$\\
       & $9$ &$0.0000-2.396766i$    & $0.062506i$              &        & $9$ & $0.0000-2.453076i$    & $0.062666i$\\
       [1ex]
\hline\hline 
\end{tabular}
\end{table}

\begin{table}
\centering
\caption{Comparison of the fundamental QNMs and higher overtones for scalar perturbations of the non-extremal Bonanno-Reuter black hole 
 for $\alpha =\frac{118}{15\pi}$, $\gamma = 9/2$, $M = 5$, and various angular momentum values $\ell$. The corresponding results obtained by \cite{Rincon2020PDU} using the 6th order WKB approximation are included for reference. The location of the event horizon $r_h$ is provided in Table~\ref{table:event}. Our QNMs are computed using the Spectral Method, employing 400 Chebyshev polynomials and 200-digit numerical precision. Here, $\omega$ denotes the dimensionless QNM frequency and $N$ the overtone number. Entries marked ‘N/A’ indicate unavailable data, while ‘SM’ refers to the Spectral Method.}
\label{scalarg03}
\vspace*{1em}
\begin{tabular}{||c|c|c|c|c|c|c|c|c|c|c|c|c||}
\hline\hline
$\ell$ & $N$ & $\omega$ \cite{Rincon2020PDU}    & $\omega$ (SM) & $\ell$ & $N$ & $\omega$ \cite{Rincon2020PDU}            & $\omega$ (SM) \\ [0.5ex]
\hline\hline
$0$    & $0$ &$0.0241367-0.0181878i$            & $0.023057-0.019555i$           & $3$    & $0$ & $0.1393220-0.0182597i$  & $0.139322-0.018259i$\\
       & $1$ &N/A                               & $0.017425-0.063818i$           &        & $1$ & $0.1367950-0.0551772i$  & $0.136795-0.055171i$\\
       & $2$ &N/A                               & $0.011673-0.110470i$           &        & $2$ & $0.1320580-0.0932217i$  & $0.132058-0.093202i$\\
       & $3$ &N/A                               & N/A                            &        & $3$ & $0.1256980-0.1329170i$  & $0.125702-0.132861i$\\
       & $4$ &N/A                               & N/A                            &        & $4$ & N/A                     & $0.118436-0.174297i$\\
       & $5$ &N/A                               & N/A                            &        & $5$ & N/A                     & $0.110846-0.217315i$\\
       & $6$ &N/A                               & N/A                            &        & $6$ & N/A                     & $0.103240-0.261551i$\\
$1$    & $0$ &$0.0604828-0.0184862i$            & $0.060508-0.018408i$           & $4$    & $0$ & $0.1789250-0.0182468i$  & $0.178926-0.018247i$\\
       & $1$ &$0.0555291-0.0573753i$            & $0.055475-0.057094i$           &        & $1$ & $0.1769310-0.0549829i$  & $0.176931-0.054981i$\\
       & $2$ &N/A                               & $0.048309-0.099492i$           &        & $2$ & $0.1730940-0.0924225i$  & $0.173094-0.092418i$\\
       & $3$ &N/A                               & $0.041105-0.144380i$           &        & $3$ & $0.1677050-0.1309530i$  & $0.167707-0.130940i$\\
       & $4$ &N/A                               & $0.033531-0.190444i$           &        & $4$ & $0.1611590-0.1708380i$  & $0.161175-0.170794i$\\
       & $5$ &N/A                               & $0.023326-0.238123i$           &        & $5$ & N/A                     & $0.153935-0.212046i$\\
       & $6$ &N/A                               & $0.018407-0.297157i$           &        & $6$ & N/A                     & $0.146362-0.254589i$\\
$2$    & $0$ &$0.0997920-0.0182964i$            & $0.099795-0.018289i$           & $5$    & $0$ & N/A                     & $0.218562-0.018241i$\\
       & $1$ &$0.0963757-0.0556588i$            & $0.096377-0.055620i$           &        & $1$ & N/A                     & $0.216919-0.054885i$\\
       & $2$ &$0.0903912-0.0950511i$            & $0.090375-0.094943i$           &        & $2$ & N/A                     & $0.213715-0.092006i$\\
       & $3$ &N/A                               & $0.083099-0.136693i$           &        & $3$ & N/A                     & $0.209114-0.129888i$\\
       & $4$ &N/A                               & $0.075572-0.180438i$           &        & $4$ & N/A                     & $0.203353-0.168752i$\\
       & $5$ &N/A                               & $0.068091-0.225484i$           &        & $5$ & N/A                     & $0.196717-0.208734i$\\
       & $6$ &N/A                               & $0.060397-0.271391i$           &        & $6$ & N/A                     & $0.189501-0.249864i$\\
       [1ex]
\hline\hline 
\end{tabular}
\end{table}

\begin{table}
\centering
\caption{
Purely imaginary QNMs for scalar perturbations of the non-extremal Bonanno-Reuter black hole for $\alpha =\frac{118}{15\pi}$, $\gamma = 9/2$, $M=5$, and different values of $\ell$. The locations of the event horizon $r_h$ are given in Table~\ref{table:event}. The corresponding results are obtained through our spectral method, utilising $400$ polynomials with a precision of $200$ digits. In this context, $\omega$ and $N$ represent the dimensionless frequency and the corresponding overtone, respectively, while $\Delta\omega=\omega_N-\omega_{N+1}$. The notation 'SM' stands for Spectral Method.}
\label{scalarg03overdamped}
\vspace*{1em}
\begin{tabular}{||c|c|c|c|c|c|c|c|c|c|c|c|c||}
\hline\hline
$\ell$ & $N$ & $\omega$ (SM)    & $\Delta\omega$ & $\ell$ & $N$ & $\omega$ (SM)     & $\Delta\omega$ \\ [0.5ex]
\hline\hline
$0$    & $0$ &$0.000-0.127959i$         & $1.137445i$            & $3$    & $0$ & $0.000-1.032617i$         & $1.008511i$\\
       & $1$ &$0.000-1.265405i$         & $1.915128i$            &        & $1$ & $0.000-2.041127i$         & $0.203479i$\\
       & $2$ &$0.000-3.180533i$         & $0.050173i$            &        & $2$ & $0.000-2.244607i$         & $0.203596i$\\
       & $3$ &$0.000-3.230706i$         & $0.050191i$            &        & $3$ & $0.000-2.448202i$         & $0.203735i$\\
       & $4$ &$0.000-3.280897i$         & $0.050198i$            &        & $4$ & $0.000-2.651937i$         & $0.607309i$\\
       & $5$ &$0.000-3.331094i$         & $0.050207i$            &        & $5$ & $0.000-3.259246i$         & $0.050544i$\\
       & $6$ &$0.000-3.381301i$         & $0.050204i$            &        & $6$ & $0.000-3.309781i$         & $0.050543i$\\
$1$    & $0$ &$0.000-0.000804i$         & $1.009565i$            & $4$    & $0$ & $0.000-0.778948i$         & $1.598600i$\\
       & $1$ &$0.000-1.010369i$         & $2.216866i$            &        & $1$ & $0.000-2.377548i$         & $0.765217i$\\
       & $2$ &$0.000-3.227235i$         & $0.050241i$            &        & $2$ & $0.000-3.142765i$         & $0.050798i$\\
       & $3$ &$0.000-3.277476i$         & $0.050255i$            &        & $3$ & $0.000-3.193563i$         & $0.050833i$\\
       & $4$ &$0.000-3.327731i$         & $0.050259i$            &        & $4$ & $0.000-3.244396i$         & $0.050735i$\\
       & $5$ &$0.000-3.377990i$         & $0.050254i$            &        & $5$ & $0.000-3.295131i$         & $0.050769i$\\
       & $6$ &$0.000-3.428243i$         & $0.050252i$            &        & $6$ & $0.000-3.345900i$         & $0.050745i$\\
$2$    & $0$ &$0.000-0.000785i$         & $1.141300i$            & $5$    & $0$ & $0.000-0.896820i$         & $0.169170i$\\
       & $1$ &$0.000-1.142086i$         & $0.203944i$            &        & $1$ & $0.000-1.065991i$         & $2.161066i$\\
       & $2$ &$0.000-1.346030i$         & $1.873933i$            &        & $2$ & $0.000-3.227057i$         & $0.050998i$\\
       & $3$ &$0.000-3.219963i$         & $0.050402i$            &        & $3$ & $0.000-3.278054i$         & $0.050983i$\\
       & $4$ &$0.000-3.270365i$         & $0.050378i$            &        & $4$ & $0.000-3.329037i$         & $0.050955i$\\
       & $5$ &$0.000-3.320743i$         & $0.050369i$            &        & $5$ & $0.000-3.379991i$         & $0.050936i$\\
       & $6$ &$0.000-3.371112i$         & $0.050361i$            &        & $6$ & $0.000-3.430927i$         & $0.050915i$\\
       [1ex]
\hline\hline 
\end{tabular}
\end{table}

\begin{table}
\centering
\caption{Comparison of the fundamental QNMs and higher overtones for scalar perturbations of the non-extremal Bonanno-Reuter black hole 
 for $\alpha =\frac{118}{15\pi}$, $\gamma = 9/2$, $M = 8$, and various angular momentum values $\ell$. The corresponding results obtained by \cite{Rincon2020PDU} using the 6th order WKB approximation and by \cite{Konoplya2022JCAP} are included for reference. The location of the event horizon $r_h$ is provided in Table~\ref{table:event}. Our QNMs are computed using the Spectral Method, employing 400 Chebyshev polynomials and 200-digit numerical precision. Here, $\omega$ denotes the dimensionless QNM frequency and $N$ the overtone number. Entries marked ‘N/A’ indicate unavailable data, while ‘SM’ refers to the Spectral Method.}
\label{scalarg04}
\vspace*{1em}
\begin{tabular}{||c|c|c|c|c|c|c|c|c|c|c|c|c||}
\hline\hline
$\ell$ & $N$ & $\omega$ \cite{Rincon2020PDU} & $\omega$ \cite{Konoplya2022JCAP}  & $\omega$ (SM) \\ [0.5ex]
\hline\hline
$0$    & $0$ & $0.0141207-0.0125909i$        & $0.014060-0.012828i$              & $0.014056-0.012828i$ \\
       & $1$ & N/A                           & $0.011030-0.042326i$              & $0.011030-0.042326i$ \\
       & $2$ & N/A                           & $0.009447-0.073078i$              & $0.009447-0.073078i$ \\
       & $3$ & N/A                           & $0.008318-0.103855i$              & $0.008319-0.103855i$ \\
       & $4$ & N/A                           & $0.007282-0.134637i$              & $0.007282-0.134634i$ \\
       & $5$ & N/A                           & N/A                               & $0.006213-0.165493i$ \\ 
$1$    & $0$ & $0.0370570-0.0120052i$        & N/A                               & $0.037066-0.011974i$ \\
       & $1$ & $0.0337423-0.0375315i$        & N/A                               & $0.033718-0.037416i$ \\
       & $2$ & N/A                           & N/A                               & $0.029473-0.065731i$ \\
       & $3$ & N/A                           & N/A                               & $0.026066-0.095750i$ \\
       & $4$ & N/A                           & N/A                               & $0.023499-0.126327i$ \\
$2$    & $0$ & $0.0611573-0.0118763i$        & N/A                               & $0.061158-0.011874i$ \\
       & $1$ & $0.0588411-0.0362336i$        & N/A                               & $0.058840-0.036223i$ \\
       & $2$ & $0.0549027-0.0621947i$        & N/A                               & $0.054905-0.062157i$ \\
       & $3$ & N/A                           & N/A                               & $0.050474-0.089975i$ \\
       & $4$ & N/A                           & N/A                               & $0.046398-0.119228i$ \\
[1ex]
\hline\hline 
\end{tabular}
\end{table}

\begin{table}
\centering
\caption{Comparison of the fundamental QNMs and higher overtones for scalar perturbations of the non-extremal Bonanno-Reuter black hole 
 for $\alpha =\frac{118}{15\pi}$, $\gamma = 9/2$, $M = 8$, and various angular momentum values $\ell$. The corresponding results obtained by \cite{Rincon2020PDU} using the 6th order WKB approximation are included for reference. The location of the event horizon $r_h$ is provided in Table~\ref{table:event}. Our QNMs are computed using the Spectral Method, employing 400 Chebyshev polynomials and 200-digit numerical precision. Here, $\omega$ denotes the dimensionless QNM frequency and $N$ the overtone number. Entries marked ‘N/A’ indicate unavailable data, while ‘SM’ refers to the Spectral Method.}
\label{scalarg05}
\vspace*{1em}
\begin{tabular}{||c|c|c|c|c|c|c|c|c|c|c|c|c||}
\hline\hline
$\ell$ & $N$ & $\omega$ \cite{Rincon2020PDU}  & $\omega$ (SM) \\ [0.5ex]
\hline\hline
$3$    & $0$ &$0.0853888-0.0118467i$  & $0.085389-0.011846i$ \\
       & $1$ &$0.0836685-0.0358552i$  & $0.083668-0.035854i$ \\
       & $2$ &$0.0804891-0.0607583i$  & $0.080491-0.060754i$ \\
       & $3$ &$0.0763352-0.0869715i$  & $0.076359-0.086946i$ \\
       & $4$ &N/A                     & $0.071882-0.114518i$ \\
       & $5$ &N/A                     & $0.067552-0.143256i$ \\
       & $6$ &N/A                     & $0.063622-0.172820i$ \\
$4$    & $0$ &$0.1096640-0.0118352i$  & $0.109664-0.011835i$ \\
       & $1$ &$0.1083050-0.0356977i$  & $0.108305-0.035697i$ \\
       & $2$ &$0.1057100-0.0601196i$  & $0.105711-0.060119i$ \\
       & $3$ &$0.1021210-0.0854093i$  & $0.102128-0.085406i$ \\
       & $4$ &$0.0978641-0.1117760i$  & $0.097904-0.111750i$ \\ 
       & $5$ &N/A                     & $0.093415-0.139177i$ \\
       & $6$ &N/A                     & $0.088977-0.167570i$ \\
$5$    & $0$ &N/A                     & $0.133960-0.011829i$ \\
       & $1$ &N/A                     & $0.132839-0.035617i$ \\
       & $2$ &N/A                     & $0.130665-0.059786i$ \\
       & $3$ &N/A                     & $0.127574-0.084560i$ \\
       & $4$ &N/A                     & $0.123768-0.110118i$ \\ 
       & $5$ &N/A                     & $0.119492-0.136560i$ \\
       & $6$ &N/A                     & $0.115000-0.163893i$ \\
[1ex]
\hline\hline 
\end{tabular}
\end{table}

\begin{table}
\centering
\caption{
Purely imaginary QNMs for scalar perturbations of the non-extremal Bonanno-Reuter black hole for $\alpha =\frac{118}{15\pi}$, $\gamma = 9/2$, $M=8$, and different values of $\ell$. The locations of the event horizon $r_h$ are given in Table~\ref{table:event}. The corresponding results are obtained through our spectral method, utilising $400$ polynomials with a precision of $200$ digits. In this context, $\omega$ and $N$ represent the dimensionless frequency and the corresponding overtone, respectively, while $\Delta\omega=\omega_N-\omega_{N+1}$. The notation 'SM' stands for Spectral Method.}
\label{scalarg05overdamped}
\vspace*{1em}
\begin{tabular}{||c|c|c|c|c|c|c|c|c|c|c|c|c||}
\hline\hline
$\ell$ & $N$ & $\omega$ (SM)  & $\Delta\omega$ & $\ell$ & $N$ & $\omega$ (SM)  & $\Delta\omega$    \\ [0.5ex]
\hline\hline
$0$    & $0$ &$0.0000-2.587041i$      & $0.634363i$            & $3$    & $0$ & $0.0000-1.910808i$     & $0.062407i$\\
       & $1$ &$0.0000-3.221404i$      & $0.031352i$            &        & $1$ & $0.0000-1.973215i$     & $0.030428i$\\
       & $2$ &$0.0000-3.252756i$      & $0.031510i$            &        & $2$ & $0.0000-2.003643i$     & $0.318698i$\\
       & $3$ &$0.0000-3.284266i$      & $0.031461i$            &        & $3$ & $0.0000-2.322341i$     & $0.381182i$\\
       & $4$ &$0.0000-3.315726i$      & $0.031475i$            &        & $4$ & $0.0000-2.703522i$     & $0.411294i$\\ 
       & $5$ &$0.0000-3.347201i$      & $0.031471i$            &        & $5$ & $0.0000-3.114816i$     & $0.094792i$\\
       & $6$ &$0.0000-3.378672i$      & $0.031472i$            &        & $6$ & $0.0000-3.209608i$     & $0.031504i$\\
       & $7$ &$0.0000-3.410143i$      & $0.031472i$            &        & $7$ & $0.0000-3.241112i$     & $0.031627i$\\
       & $8$ &$0.0000-3.441615i$      & $0.031472i$            &        & $8$ & $0.0000-3.272739i$     & $0.031578i$\\
       & $9$ &$0.0000-3.473087i$      & $0.031472i$            &        & $9$ & $0.0000-3.304317i$     & $0.031591i$\\ 
$1$    & $0$ &$0.0000-2.616439i$      & $0.634470i$            & $4$    & $0$ & $0.0000-1.961257i$     & $0.032395i$\\
       & $1$ &$0.0000-3.250909i$      & $0.031538i$            &        & $1$ & $0.0000-1.993651i$     & $0.034648i$\\
       & $2$ &$0.0000-3.282446i$      & $0.031475i$            &        & $2$ & $0.0000-2.028299i$     & $1.141301i$\\
       & $3$ &$0.0000-3.313921i$      & $0.031493i$            &        & $3$ & $0.0000-3.169600i$     & $0.031720i$\\
       & $4$ &$0.0000-3.345414i$      & $0.031488i$            &        & $4$ & $0.0000-3.201319i$     & $0.031669i$\\ 
       & $5$ &$0.0000-3.376902i$      & $0.031489i$            &        & $5$ & $0.0000-3.232988i$     & $0.031672i$\\
       & $6$ &$0.0000-3.408391i$      & $0.031488i$            &        & $6$ & $0.0000-3.264661i$     & $0.031673i$\\
       & $7$ &$0.0000-3.439879i$      & $0.031488i$            &        & $7$ & $0.0000-3.296334i$     & $0.031663i$\\
       & $8$ &$0.0000-3.471367i$      & $0.031488i$            &        & $8$ & $0.0000-3.327997i$     & $0.031664i$\\
       & $9$ &$0.0000-3.502854i$      & $0.031487i$            &        & $9$ & $0.0000-3.359661i$     & $0.031658i$\\ 
$2$    & $0$ &$0.0000-1.349849i$      & $0.412291i$            & $5$    & $0$ & $0.0000-0.075361i$     & $3.147846i$\\
       & $1$ &$0.0000-1.762140i$      & $0.411725i$            &        & $1$ & $0.0000-3.223207i$     & $0.031841i$\\
       & $2$ &$0.0000-2.173865i$      & $1.073222i$            &        & $2$ & $0.0000-3.255048i$     & $0.031724i$\\
       & $3$ &$0.0000-3.247087i$      & $0.031587i$            &        & $3$ & $0.0000-3.286772i$     & $0.031754i$\\
       & $4$ &$0.0000-3.278673i$      & $0.031509i$            &        & $4$ & $0.0000-3.318526i$     & $0.031739i$\\
       & $5$ &$0.0000-3.310182i$      & $0.031532i$            &        & $5$ & $0.0000-3.350265i$     & $0.031737i$\\
       & $6$ &$0.0000-3.341715i$      & $0.031524i$            &        & $6$ & $0.0000-3.382002i$     & $0.031732i$\\
       & $7$ &$0.0000-3.373239i$      & $0.031525i$            &        & $7$ & $0.0000-3.413733i$     & $0.031727i$\\
       & $8$ &$0.0000-3.404764i$      & $0.031524i$            &        & $8$ & $0.0000-3.445460i$     & $0.031723i$\\
       & $9$ &$0.0000-3.436287i$      & $0.031523i$            &        & $9$ & $0.0000-3.477183i$     & $0.031718i$\\ 
[1ex]
\hline\hline 
\end{tabular}
\end{table}

\begin{table}
\centering
\caption{Comparison of the fundamental QNMs and higher overtones for scalar perturbations of the non-extremal Bonanno-Reuter black hole for 
 $\alpha =\frac{118}{15\pi}$, $\gamma = 9/2$, $M = 8$, and various angular momentum values $\ell$. The corresponding results obtained by \cite{Konoplya2022JCAP} using Leaver’s continued fraction method are included for reference. The location of the event horizon $r_h$ is provided in Table~\ref{table:event}. Our QNMs are computed using the Spectral Method, employing 400 Chebyshev polynomials and 200-digit numerical precision. Here, $\omega$ denotes the dimensionless QNM frequency and $N$ the overtone number. Entries marked ‘N/A’ indicate unavailable data, while ‘SM’ refers to the Spectral Method.}
\label{scalarg06}
\vspace*{1em}
\begin{tabular}{||c|c|c|c|c|c|c|c|c|c|c|c|c||}
\hline\hline
$\ell$ & $N$ & $\omega$ \cite{Konoplya2022JCAP} & $\omega$ (SM)        & $\ell$ & $N$ & $\omega$ \cite{Konoplya2022JCAP} & $\omega$ (SM) \\ [0.5ex]
\hline\hline
$0$    & $0$ &$0.014060-0.012828i$              & $0.014060-0.012828i$ & $1$    & $0$ & N/A             & $0.037066-0.011974i$\\
       & $1$ &$0.011030-0.042326i$              & $0.011030-0.042326i$ &        & $1$ & N/A             & $0.033718-0.037416i$\\
       & $2$ &$0.009447-0.073078i$              & $0.009447-0.073078i$ &        & $2$ & N/A             & $0.029473-0.065731i$\\
       & $3$ &$0.008318-0.103855i$              & $0.008319-0.103855i$ &        & $3$ & N/A             & $0.026066-0.095750i$\\
       & $4$ &$0.007282-0.134637i$              & $0.007282-0.134634i$ &        & $4$ & N/A             & $0.023499-0.126327i$\\
       & $5$ &$0.006213-0.165471i$              & $0.006213-0.165493i$ &        & $5$ & N/A             & $0.021435-0.157049i$\\
       & $6$ &$0.005055-0.196409i$              & N/A                  &        & $6$ & N/A             & $0.019643-0.187796i$\\
       & $7$ &$0.003780-0.227529i$              & N/A                  &        & $7$ & N/A             & $0.017981-0.218542i$\\
       & $8$ &$0.002403-0.258950i$              & N/A                  &        & $8$ & N/A             & $0.016355-0.249284i$\\
       & $9$ &$0.001041-0.290822i$              & N/A                  &        & $9$ & N/A             & $0.014690-0.280035i$\\
       & $10$&$0.000042-0.323183i$              & N/A                  &        & $10$& N/A             & $0.012916-0.310804i$\\
       & $11$&$0.000000-0.355670i$              & N/A                  &        & $11$& N/A             & $0.010938-0.341676i$\\
       & $12$&$0.000000-0.387832i$              & N/A                  &        & $12$& N/A             & $0.003681-0.601453i$\\
       & $13$&$0.000091-0.419620i$              & N/A                  &        & $13$& N/A             & N/A\\
       & $14$&$0.000279-0.451184i$              & N/A                  &        & $14$& N/A             & N/A\\
       & $15$&$0.000389-0.482640i$              & N/A                  &        & $15$& N/A             & N/A\\
       [1ex]
\hline\hline 
\end{tabular}
\end{table}

\begin{table}
\centering
\caption{QNMs for electromagnetic perturbations of the non-extremal Bonanno-Reuter black hole for $\alpha =\frac{118}{15\pi}$, $\gamma = 9/2$, $\ell=1$, and different values of the mass parameter $M$. The locations of the event horizon $r_h$ are given in Table~\ref{table:event}. The results are obtained using our spectral method with 400 Chebyshev polynomials and 200-digit precision. Here, $\omega$ denotes the dimensionless QNM frequency, and $N$ the overtone number. 'N/A' indicates data not available, and 'SM' refers to the Spectral Method.}
\label{em01}
\vspace*{1em}
\begin{tabular}{||c|c|c|c|c|c|c|c|c|c|c|c|c||}
\hline\hline
$M$    &$\ell$ & $N$ & $\omega$ \cite{Konoplya2022JCAP} & $\omega$ (SM)         & $M$  & $\ell$ & $N$ & $\omega$ \cite{Konoplya2022JCAP} & $\omega$ (SM) \\ [0.5ex]
\hline\hline
$3.503$&$1$    & $0$ &N/A                             & $0.078132-0.021572i$    & $7.503$ & $1$    & $0$ & N/A   & $0.033740-0.012074i$\\
       &       & $1$ &N/A                             & N/A                     &         &        & $1$ & N/A   & $0.029609-0.038126i$\\
       &       & $2$ &N/A                             & N/A                     &         &        & $2$ & N/A   & $0.024565-0.067775i$\\
       &       & $3$ &N/A                             & N/A                     &         &        & $3$ & N/A   & $0.020685-0.099329i$\\
       &       & $4$ &N/A                             & N/A                     &         &        & $4$ & N/A   & $0.017783-0.131415i$\\
       &       & $5$ &N/A                             & N/A                     &         &        & $5$ & N/A   & $0.015417-0.163616i$\\
       &       & $6$ &N/A                             & N/A                     &         &        & $6$ & N/A   & $0.013316-0.195825i$\\
       &       & $7$ &N/A                             & N/A                     &         &        & $7$ & N/A   & $0.011310-0.228029i$\\
$4.0$  &$1$    & $0$ &$0.066877-0.020549i$            & $0.066877-0.020549i$    & $8.0$   & $1$    & $0$ & N/A   & $0.031568-0.011356i$\\
       &       & $1$ &$0.059764-0.063499i$            & $0.059764-0.063499i$    &         &        & $1$ & N/A   & $0.027653-0.035885i$\\
       &       & $2$ &$0.047392-0.110619i$            & $0.047392-0.110619i$    &         &        & $2$ & N/A   & $0.022900-0.063841i$\\
       &       & $3$ &$0.030448-0.166115i$            & $0.030449-0.166115i$    &         &        & $3$ & N/A   & $0.019284-0.093599i$\\
       &       & $4$ &$0.019951-0.224263i$            & $0.019951-0.224263i$    &         &        & $4$ & N/A   & $0.016620-0.123854i$\\
       &       & $5$ &$0.012209-0.290487i$            & $0.012230-0.290461i$    &         &        & $5$ & N/A   & $0.014488-0.154211i$\\
       &       & $6$ &$0.000983-0.354117i$            & N/A                     &         &        & $6$ & N/A   & $0.012636-0.184572i$\\
       &       & $7$ &$0.000-0.420i$                  & N/A                     &         &        & $7$ & N/A   & $0.010919-0.214916i$\\
$4.503$&$1$    & $0$ &N/A                             & $0.058405-0.018980i$    & $8.503$ & $1$    & $0$ & N/A   & $0.029640-0.010709i$\\
       &       & $1$ &N/A                             & $0.052287-0.059106i$    &         &        & $1$ & N/A   & $0.025925-0.033860i$\\
       &       & $2$ &N/A                             & $0.043388-0.103494i$    &         &        & $2$ & N/A   & $0.021435-0.060278i$\\
       &       & $3$ &N/A                             & $0.033784-0.150621i$    &         &        & $3$ & N/A   & $0.018047-0.088401i$\\
       &       & $4$ &N/A                             & $0.021668-0.199381i$    &         &        & $4$ & N/A   & $0.015579-0.116991i$\\
       &       & $5$ &N/A                             & $0.011727-0.261899i$    &         &        & $5$ & N/A   & $0.013633-0.145674i$\\
       &       & $6$ &N/A                             & N/A                     &         &        & $6$ & N/A   & $0.011973-0.174358i$\\
       &       & $7$ &N/A                             & N/A                     &         &        & $7$ & N/A   & $0.010464-0.203019i$\\
$5.503$&$1$    & $0$ &N/A                             & $0.046833-0.016077i$    & $9.503$ & $1$    & $0$ & N/A   & $0.026440-0.009614i$\\
       &       & $1$ &N/A                             & $0.041582-0.050470i$    &         &        & $1$ & N/A   & $0.023071-0.030427i$\\
       &       & $2$ &N/A                             & $0.034791-0.089150i$    &         &        & $2$ & N/A   & $0.019025-0.054218i$\\
       &       & $3$ &N/A                             & $0.028953-0.130251i$    &         &        & $3$ & N/A   & $0.016009-0.079552i$\\
       &       & $4$ &N/A                             & $0.023875-0.172116i$    &         &        & $4$ & N/A   & $0.013845-0.105300i$\\
       &       & $5$ &N/A                             & $0.018875-0.214234i$    &         &        & $5$ & N/A   & $0.012174-0.131129i$\\
       &       & $6$ &N/A                             & $0.013250-0.256612i$    &         &        & $6$ & N/A   & $0.010784-0.156956i$\\
       &       & $7$ &N/A                             & N/A                     &         &        & $7$ & N/A   & $0.009557-0.182761i$\\
$6.503$&$1$    & $0$ &N/A                             & $0.039193-0.013814i$    & $10^3$  & $1$    & $0$ & N/A   & $0.000248-0.000093i$\\
       &       & $1$ &N/A                             & $0.034559-0.043531i$    &         &        & $1$ & N/A   & $0.000215-0.000294i$\\
       &       & $2$ &N/A                             & $0.028795-0.077207i$    &         &        & $2$ & N/A   & $0.000175-0.000525i$\\
       &       & $3$ &N/A                             & $0.024203-0.113029i$    &         &        & $3$ & N/A   & $0.000146-0.000772i$\\
       &       & $4$ &N/A                             & $0.020601-0.149473i$    &         &        & $4$ & N/A   & $0.000127-0.001023i$\\
       &       & $5$ &N/A                             & $0.017481-0.186069i$    &         &        & $5$ & N/A   & $0.000112-0.001274i$\\
       &       & $6$ &N/A                             & $0.014504-0.222708i$    &         &        & $6$ & N/A   & $0.000101-0.001525i$\\
       &       & $7$ &N/A                             & $0.011427-0.259407i$    &         &        & $7$ & N/A   & $0.000092-0.001776i$\\
[1ex]
\hline\hline 
\end{tabular}
\end{table}

\begin{table}
\centering
\caption{Purely imaginary QNMs for electromagnetic perturbations of the non-extremal Bonanno-Reuter black hole for $\alpha =\frac{118}{15\pi}$, $\gamma = 9/2$, $\ell=1$, and different values of the mass parameter $M$. The locations of the event horizon $r_h$ are given in Table~\ref{table:event}. The corresponding results are obtained through our spectral method, utilising $400$ polynomials with a precision of $200$ digits. In this context, $\omega$ and $N$ represent the dimensionless frequency and the corresponding overtone, respectively, while $\Delta\omega=\omega_N-\omega_{N+1}$. The notation 'SM' stands for Spectral Method.}
\label{em01overdamped}
\vspace*{1em}
\begin{tabular}{||c|c|c|c|c|c|c|c|c|c|c|c|c||}
\hline\hline
$M$    &$\ell$ & $N$ &$\omega$ (SM) & $\Delta\omega$ & $M$     & $\ell$ & $N$ & $\omega$ (SM) & $\Delta\omega$ \\ [0.5ex]
\hline\hline
$3.503$&$1$    & $0$ &$0.0000-0.325857i$    & $0.072324i$            & $7.503$ & $1$    & $0$ & $0.0000-2.917989i$    & $0.124182i$\\
       &       & $1$ &$0.0000-0.398181i$    & $0.071934i$            &         &        & $1$ & $0.0000-3.042172i$    & $0.335931i$\\
       &       & $2$ &$0.0000-0.470114i$    & $0.071705i$            &         &        & $2$ & $0.0000-3.378102i$    & $0.033441i$\\
       &       & $3$ &$0.0000-0.541820i$    & $0.071563i$            &         &        & $3$ & $0.0000-3.411543i$    & $0.033578i$\\
       &       & $4$ &$0.0000-0.613382i$    & $0.071470i$            &         &        & $4$ & $0.0000-3.445121i$    & $0.033557i$\\
       &       & $5$ &$0.0000-0.684852i$    & $0.071407i$            &         &        & $5$ & $0.0000-3.478678i$    & $0.033554i$\\
       &       & $6$ &$0.0000-0.756259i$    & $0.071361i$            &         &        & $6$ & $0.0000-3.512232i$    & $0.033559i$\\
       &       & $7$ &$0.0000-0.827620i$    & $0.071328i$            &         &        & $7$ & $0.0000-3.545791i$    & $0.033556i$\\
$4.0$  &$1$    & $0$ &$0.0000-1.719638i$    & $0.322067i$            & $8.0$   & $1$    & $0$ & $0.0000-2.851631i$    & $0.411267i$\\
       &       & $1$ &$0.0000-2.041705i$    & $0.062544i$            &         &        & $1$ & $0.0000-3.262897i$    & $0.031486i$\\
       &       & $2$ &$0.0000-2.104249i$    & $0.062528i$            &         &        & $2$ & $0.0000-3.294383i$    & $0.031442i$\\
       &       & $3$ &$0.0000-2.166778i$    & $0.062539i$            &         &        & $3$ & $0.0000-3.325825i$    & $0.031501i$\\
       &       & $4$ &$0.0000-2.229316i$    & $0.062536i$            &         &        & $4$ & $0.0000-3.357326i$    & $0.031460i$\\
       &       & $5$ &$0.0000-2.291852i$    & $0.062534i$            &         &        & $5$ & $0.0000-3.388786i$    & $0.031483i$\\
       &       & $6$ &$0.0000-2.354386i$    & $0.062532i$            &         &        & $6$ & $0.0000-3.420269i$    & $0.031472i$\\
       &       & $7$ &$0.0000-2.416918i$    & $0.062531i$            &         &        & $7$ & $0.0000-3.451740i$    & $0.031477i$\\       
$4.503$&$1$    & $0$ &$0.0000-2.460719i$    & $0.415123i$            & $8.503$ & $1$    & $0$ & $0.0000-3.010761i$    & $0.148087i$\\
       &       & $1$ &$0.0000-2.875843i$    & $0.055716i$            &         &        & $1$ & $0.0000-3.158848i$    & $0.029609i$\\
       &       & $2$ &$0.0000-2.931559i$    & $0.055663i$            &         &        & $2$ & $0.0000-3.188457i$    & $0.029573i$\\
       &       & $3$ &$0.0000-2.987222i$    & $0.055678i$            &         &        & $3$ & $0.0000-3.218030i$    & $0.029650i$\\
       &       & $4$ &$0.0000-3.042900i$    & $0.055670i$            &         &        & $4$ & $0.0000-3.247679i$    & $0.029594i$\\
       &       & $5$ &$0.0000-3.098569i$    & $0.055670i$            &         &        & $5$ & $0.0000-3.277273i$    & $0.029625i$\\
       &       & $6$ &$0.0000-3.154239i$    & $0.055669i$            &         &        & $6$ & $0.0000-3.306897i$    & $0.029610i$\\
       &       & $7$ &$0.0000-3.209908i$    & $0.055669i$            &         &        & $7$ & $0.0000-3.336507i$    & $0.029616i$\\
$5.503$&$1$    & $0$ &$0.0000-2.999403i$    & $0.502123i$            & $9.503$ & $1$    & $0$ & $0.0000-2.826325i$    & $0.158819i$\\
       &       & $1$ &$0.0000-3.501526i$    & $0.045692i$            &         &        & $1$ & $0.0000-2.985144i$    & $0.026588i$\\
       &       & $2$ &$0.0000-3.547218i$    & $0.045678i$            &         &        & $2$ & $0.0000-3.011732i$    & $0.026491i$\\
       &       & $3$ &$0.0000-3.592896i$    & $0.045672i$            &         &        & $3$ & $0.0000-3.038222i$    & $0.026483i$\\
       &       & $4$ &$0.0000-3.638568i$    & $0.045674i$            &         &        & $4$ & $0.0000-3.064705i$    & $0.026509i$\\
       &       & $5$ &$0.0000-3.684242i$    & $0.045673i$            &         &        & $5$ & $0.0000-3.091213i$    & $0.026489i$\\
       &       & $6$ &$0.0000-3.729915i$    & $0.045673i$            &         &        & $6$ & $0.0000-3.117702i$    & $0.026500i$\\
       &       & $7$ &$0.0000-3.775588i$    & $0.045673i$            &         &        & $7$ & $0.0000-3.144202i$    & $0.026495i$\\
$6.503$&$1$    & $0$ &$0.0000-2.957749i$    & $0.550667i$            & $10^3$  & $1$    & $0$ & $0.0000-0.039984i$    & $0.000250i$\\
       &       & $1$ &$0.0000-3.508417i$    & $0.038800i$            &         &        & $1$ & $0.0000-0.040234i$    & $0.000250i$\\
       &       & $2$ &$0.0000-3.547217i$    & $0.038690i$            &         &        & $2$ & $0.0000-0.040484i$    & $0.000250i$\\
       &       & $3$ &$0.0000-3.585907i$    & $0.038692i$            &         &        & $3$ & $0.0000-0.040735i$    & $0.000250i$\\
       &       & $4$ &$0.0000-3.624599i$    & $0.038702i$            &         &        & $4$ & $0.0000-0.040985i$    & $0.000250i$\\
       &       & $5$ &$0.0000-3.663300i$    & $0.038695i$            &         &        & $5$ & $0.0000-0.041235i$    & $0.000250i$\\
       &       & $6$ &$0.0000-3.701996i$    & $0.038698i$            &         &        & $6$ & $0.0000-0.041485i$    & $0.000250i$\\
       &       & $7$ &$0.0000-3.740693i$    & $0.038697i$            &         &        & $7$ & $0.0000-0.041735i$    & $0.000250i$\\
[1ex]
\hline\hline 
\end{tabular}
\end{table}

\begin{table}
\centering
\caption{Comparison of the fundamental QNMs and higher overtones for electromagnetic perturbations of the non-extremal Bonanno-Reuter black hole for $\alpha =\frac{118}{15\pi}$, $\gamma = 9/2$, and various values of the angular momentum $\ell$ and the mass parameter $M$. The corresponding results obtained by \cite{Rincon2020PDU} using the 6th order WKB approximation are included for reference. The location of the event horizon $r_h$ is provided in Table~\ref{table:event}. Our QNMs are computed using the Spectral Method, employing 400 Chebyshev polynomials and 200-digit numerical precision. Here, $\omega$ denotes the dimensionless QNM frequency and $N$ the overtone number. Entries marked ‘N/A’ indicate unavailable data, while ‘SM’ refers to the Spectral Method.}
\label{em05}
\vspace*{1em}
\begin{tabular}{||c|c|c|c|c|c|c|c|c|c|c|c|c||}
\hline\hline
$M$   &$\ell$ & $N$ &$\omega$ \cite{Rincon2020PDU}   & $\omega$ (SM)      &   $M$ &$\ell$ & $N$ &$\omega$ \cite{Rincon2020PDU}   & $\omega$ (SM)           \\ [0.5ex]
\hline\hline
$5.0$ &$1$    & $0$ &$0.0518962-0.0176598i$          & $0.051986-0.017459i$ & $8$ &$1$ &$0$ &$0.0315513-0.0114188i$  & $0.031568-0.011356i$\\
      &       & $1$ &$0.0464157-0.0555774i$          & $0.046358-0.054636i$ &     &    &$1$ &$0.0276887-0.0361299i$  & $0.027653-0.035885i$\\
      &       & $2$ &N/A                             & $0.038778-0.096174i$ &     &    &$2$ &N/A                     & $0.022900-0.063841i$\\
      &       & $3$ &N/A                             & $0.031733-0.140272i$ &     &    &$3$ &N/A                     & $0.019284-0.093599i$ \\
      &       & $4$ &N/A                             & $0.024899-0.185258i$ &     &    &$4$ &N/A                     & $0.016620-0.123854i$\\ 
      &       & $5$ &N/A                             & $0.017009-0.230734i$ &     &    &$5$ &N/A                     & $0.014488-0.154211i$\\
      &       & $6$ &N/A                             & N/A                  &     &    &$6$ &N/A                     & $0.012636-0.184572i$\\
      &       & $7$ &N/A                             & N/A                  &     &    &$7$ &N/A                     & $0.010919-0.214916i$\\
      &       & $8$ &N/A                             & N/A                  &     &    &$8$ &N/A                     & $0.009240-0.245232i$\\
      &$2$    & $0$ &$0.0947619-0.0179553i$          & $0.094765-0.017946i$ &     &$2$ &$0$ &$0.0579389-0.0116614i$  & $0.057939-0.011659i$\\
      &       & $1$ &$0.0912135-0.0546942i$          & $0.091216-0.054643i$ &     &    &$1$ &$0.0554914-0.0356318i$  & $0.055490-0.035619i$ \\        
      &       & $2$ &$0.0850381-0.0936163i$          & $0.085021-0.093469i$ &     &    &$2$ &$0.0513362-0.0613273i$  & $0.051346-0.061285i$ \\
      &       & $3$ &N/A                             & $0.077620-0.134868i$ &     &    &$3$ &N/A                     & $0.046730-0.088960i$ \\
      &       & $4$ &N/A                             & $0.070117-0.178320i$ &     &    &$4$ &N/A                     & $0.042551-0.118131i$ \\
      &       & $5$ &N/A                             & $0.062827-0.223035i$ &     &    &$5$ &N/A                     & $0.039057-0.148119i$ \\
      &       & $6$ &N/A                             & $0.055507-0.268490i$ &     &    &$6$ &N/A                     & $0.036157-0.178490i$ \\
      &       & $7$ &N/A                             & $0.047574-0.314559i$ &     &    &$7$ &N/A                     & $0.033696-0.209027i$ \\
      &       & $8$ &N/A                             & $0.037792-0.362045i$ &     &    &$8$ &N/A                     & $0.031544-0.239629i$ \\
      &$3$    & $0$ &$0.1357420-0.0180840i$          & $0.135743-0.018083i$ &     &$3$ &$0$ &$0.0831040-0.0117377i$  & $0.083104-0.011737i$ \\
      &       & $1$ &$0.1331670-0.0546641i$          & $0.133168-0.054657i$ &     &    &$1$ &$0.0813353-0.0355388i$  & $0.081335-0.035537i$ \\
      &       & $2$ &$0.1283470-0.0924114i$          & $0.128347-0.092389i$ &     &    &$2$ &$0.0780673-0.0602670i$  & $0.078070-0.060263i$ \\
      &       & $3$ &$0.1218910-0.1318680i$          & $0.121899-0.131805i$ &     &    &$3$ &$0.0738002-0.0863555i$  & $0.073832-0.086329i$ \\
      &       & $4$ &N/A                             & $0.114568-0.173053i$ &     &    &$4$ &N/A                     & $0.069259-0.113822i$\\ 
      &       & $5$ &N/A                             & $0.106967-0.215916i$ &     &    &$5$ &N/A                     & $0.064862-0.142512i$\\ 
      &       & $6$ &N/A                             & $0.099412-0.260000i$ &     &    &$6$ &N/A                     & $0.060894-0.172038i$\\  
      &       & $7$ &N/A                             & $0.091943-0.304940i$ &     &    &$7$ &N/A                     & $0.057405-0.202085i$\\ 
      &       & $8$ &N/A                             & $0.084414-0.350501i$ &     &    &$8$ &N/A                     & $0.054343-0.232436i$\\  
      &$4$    & $0$ &$0.1761450-0.0181403i$          & $0.176146-0.018140i$ &     &$4$ &$0$ &$0.1078920-0.0117693i$  & $0.107892-0.011769i$ \\
      &       & $1$ &$0.1741290-0.0546682i$          & $0.174129-0.054666i$ &     &    &$1$ &$0.1065100-0.0355041i$  & $0.106510-0.035504i$ \\
      &       & $2$ &$0.1702500-0.0919144i$          & $0.170250-0.091909i$ &     &    &$2$ &$0.1038710-0.0598101i$  & $0.103872-0.059809i$ \\
      &       & $3$ &$0.1648060-0.1302740i$          & $0.164810-0.130261i$ &     &    &$3$ &$0.1002220-0.0850032i$  & $0.100231-0.085001i$ \\
      &       & $4$ &$0.1582030-0.1700180i$          & $0.158224-0.169972i$ &     &    &$4$ &$0.0958965-0.1112980i$  & $0.095944-0.111271i$ \\
      &       & $5$ &N/A                             & $0.150944-0.211107i$ &     &    &$5$ &N/A                     & $0.091395-0.138649i$\\ 
      &       & $6$ &N/A                             & $0.143354-0.253551i$ &     &    &$6$ &N/A                     & $0.086911-0.167008i$\\  
      &       & $7$ &N/A                             & $0.135707-0.297087i$ &     &    &$7$ &N/A                     & $0.082696-0.196140i$\\ 
      &       & $8$ &N/A                             & $0.128109-0.341473i$ &     &    &$8$ &N/A                     & $0.078837-0.225830i$\\  
[1ex]
\hline\hline 
\end{tabular}
\end{table}

\begin{table}
\centering
\caption{
Purely imaginary QNMs for electromagnetic perturbations of the non-extremal Bonanno-Reuter black hole for $\alpha =\frac{118}{15\pi}$, $\gamma = 9/2$, and different values of $\ell$ and $M$. The locations of the event horizon $r_h$ are given in Table~\ref{table:event}. The corresponding results are obtained through our spectral method, utilising $400$ polynomials with a precision of $200$ digits. In this context, $\omega$ and $N$ represent the dimensionless frequency and the corresponding overtone, respectively, while $\Delta\omega=\omega_N-\omega_{N+1}$. The notation 'SM' stands for Spectral Method.}
\label{em05overdamped}
\vspace*{1em}
\begin{tabular}{||c|c|c|c|c|c|c|c|c|c|c|c|c||}
\hline\hline
$M$   & $\ell$ & $N$ & $\omega$ (SM)  & $\Delta\omega$ & $M$   & $\ell$ & $N$ & $\omega$ (SM)  & $\Delta\omega$    \\ [0.5ex]
\hline\hline
$5.0$ &$1$     & $0$ &$0.0000-2.816503i$      & $0.530919i$            & $8.0$ & $1$    & $0$ & $0.0000-2.851631i$     & $0.411267i$\\
      &        & $1$ &$0.0000-3.347422i$      & $0.050217i$            &       &        & $1$ & $0.0000-3.262897i$     & $0.031486i$\\
      &        & $2$ &$0.0000-3.397639i$      & $0.050212i$            &       &        & $2$ & $0.0000-3.294383i$     & $0.031442i$\\
      &        & $3$ &$0.0000-3.447851i$      & $0.050214i$            &       &        & $3$ & $0.0000-3.325825i$     & $0.031501i$\\
      &        & $4$ &$0.0000-3.498065i$      & $0.050214i$            &       &        & $4$ & $0.0000-3.357326i$     & $0.031460i$\\ 
      &        & $5$ &$0.0000-3.548279i$      & $0.050213i$            &       &        & $5$ & $0.0000-3.388786i$     & $0.031483i$\\
      &        & $6$ &$0.0000-3.598492i$      & $0.050213i$            &       &        & $6$ & $0.0000-3.420269i$     & $0.031472i$\\
      &        & $7$ &$0.0000-3.648705i$      & $0.050213i$            &       &        & $7$ & $0.0000-3.451740i$     & $0.031477i$\\
      &        & $8$ &$0.0000-3.698918i$      & $0.050213i$            &       &        & $8$ & $0.0000-3.483217i$     & $0.031474i$\\
      &        & $9$ &$0.0000-3.749131i$      & $0.050213i$            &       &        & $9$ & $0.0000-3.514691i$     & $0.031475i$\\
      &$2$     & $0$ &$0.0000-2.735055i$      & $0.557678i$            &       & $2$    & $0$ & $0.0000-2.822257i$     & $0.438272i$\\
      &        & $1$ &$0.0000-3.292733i$      & $0.050331i$            &       &        & $1$ & $0.0000-3.260529i$     & $0.031464i$\\
      &        & $2$ &$0.0000-3.343064i$      & $0.050276i$            &       &        & $2$ & $0.0000-3.291993i$     & $0.031498i$\\
      &        & $3$ &$0.0000-3.393340i$      & $0.050276i$            &       &        & $3$ & $0.0000-3.323491i$     & $0.031503i$\\
      &        & $4$ &$0.0000-3.443616i$      & $0.050275i$            &       &        & $4$ & $0.0000-3.354993i$     & $0.031493i$\\ 
      &        & $5$ &$0.0000-3.493891i$      & $0.050273i$            &       &        & $5$ & $0.0000-3.386486i$     & $0.031499i$\\
      &        & $6$ &$0.0000-3.544164i$      & $0.050271i$            &       &        & $6$ & $0.0000-3.417984i$     & $0.031495i$\\
      &        & $7$ &$0.0000-3.594435i$      & $0.050269i$            &       &        & $7$ & $0.0000-3.449479i$     & $0.031496i$\\
      &        & $8$ &$0.0000-3.644705i$      & $0.050268i$            &       &        & $8$ & $0.0000-3.480975i$     & $0.031495i$\\  
      &        & $9$ &$0.0000-3.694972i$      & $0.050266i$            &       &        & $9$ & $0.0000-3.512470i$     & $0.031495i$\\
      &$3$     & $0$ &$0.0000-2.124379i$      & $1.060934i$            &       & $3$    & $0$ & $0.0000-2.499365i$     & $0.757560i$\\
      &        & $1$ &$0.0000-3.185313i$      & $0.050358i$            &       &        & $1$ & $0.0000-3.256924i$     & $0.031467i$\\
      &        & $2$ &$0.0000-3.235671i$      & $0.050364i$            &       &        & $2$ & $0.0000-3.288391i$     & $0.031550i$\\
      &        & $3$ &$0.0000-3.286034i$      & $0.050414i$            &       &        & $3$ & $0.0000-3.319941i$     & $0.031528i$\\
      &        & $4$ &$0.0000-3.336448i$      & $0.050374i$            &       &        & $4$ & $0.0000-3.351469i$     & $0.031530i$\\ 
      &        & $5$ &$0.0000-3.386822i$      & $0.050375i$            &       &        & $5$ & $0.0000-3.382999i$     & $0.031530i$\\
      &        & $6$ &$0.0000-3.437197i$      & $0.050370i$            &       &        & $6$ & $0.0000-3.414529i$     & $0.031528i$\\
      &        & $7$ &$0.0000-3.487567i$      & $0.050365i$            &       &        & $7$ & $0.0000-3.446056i$     & $0.031527i$\\
      &        & $8$ &$0.0000-3.537932i$      & $0.050361i$            &       &        & $8$ & $0.0000-3.477584i$     & $0.031526i$\\  
      &        & $9$ &$0.0000-3.588293i$      & $0.050356i$            &       &        & $9$ & $0.0000-3.509110i$     & $0.031525i$\\
      &$4$     & $0$ &$0.0000-0.621997i$      & $2.604387i$            &       & $4$    & $0$ & $0.0000-0.075719i$     & $3.176306i$\\
      &        & $1$ &$0.0000-3.226384i$      & $0.050547i$            &       &        & $1$ & $0.0000-3.252025i$     & $0.031498i$\\
      &        & $2$ &$0.0000-3.276931i$      & $0.050538i$            &       &        & $2$ & $0.0000-3.283523i$     & $0.031609i$\\
      &        & $3$ &$0.0000-3.327468i$      & $0.050518i$            &       &        & $3$ & $0.0000-3.315132i$     & $0.031571i$\\
      &        & $4$ &$0.0000-3.377987i$      & $0.050514i$            &       &        & $4$ & $0.0000-3.346704i$     & $0.031579i$\\ 
      &        & $5$ &$0.0000-3.428501i$      & $0.050504i$            &       &        & $5$ & $0.0000-3.378283i$     & $0.031575i$\\
      &        & $6$ &$0.0000-3.479005i$      & $0.050495i$            &       &        & $6$ & $0.0000-3.409858i$     & $0.031573i$\\
      &        & $7$ &$0.0000-3.529500i$      & $0.050486i$            &       &        & $7$ & $0.0000-3.441431i$     & $0.031571i$\\
      &        & $8$ &$0.0000-3.579986i$      & $0.050478i$            &       &        & $8$ & $0.0000-3.473003i$     & $0.031570i$\\
      &        & $9$ &$0.0000-3.630464i$      & $0.050470i$            &       &        & $9$ & $0.0000-3.504572i$     & $0.031568i$\\
[1ex]
\hline\hline 
\end{tabular}
\end{table}

\begin{table}
\centering
\caption{QNMs for spin $2$ perturbations of the non-extremal Bonanno-Reuter black hole for $\alpha =\frac{118}{15\pi}$, $\gamma = 9/2$, $\ell=2$, and different values of the mass parameter $M$. The locations of the event horizon $r_h$ are given in Table~\ref{table:event}. The results are obtained using our spectral method with 400 Chebyshev polynomials and 200-digit precision. Here, $\omega$ denotes the dimensionless QNM frequency, and $N$ the overtone number. 'N/A' indicates data not available, and 'SM' refers to the Spectral Method.}
\label{tensor01}
\vspace*{1em}
\begin{tabular}{||c|c|c|c|c|c|c|c|c|c|c|c|c||}
\hline\hline
$M$    &$\ell$ & $N$ & $\omega$ (SM) & $M$     & $\ell$ & $N$ & $\omega$ (SM) \\ [0.5ex]
\hline\hline
$3.503$&$2$    & $0$ & $0.119480-0.019732i$           & $7.503$ & $2$    & $0$ & $0.050798-0.011541i$\\
       &       & $1$ & $0.113936-0.060248i$           &         &        & $1$ & $0.047686-0.035466i$\\
       &       & $2$ & N/A                            &         &        & $2$ & $0.042480-0.061692i$\\
       &       & $3$ & N/A                            &         &        & $3$ & $0.036949-0.090601i$\\
       &       & $4$ & N/A                            &         &        & $4$ & $0.032311-0.121351i$\\
       &       & $5$ & N/A                            &         &        & $5$ & $0.028775-0.153016i$\\
       &       & $6$ & N/A                            &         &        & $6$ & $0.026143-0.185097i$\\
       &       & $7$ & N/A                            &         &        & $7$ & $0.024186-0.217374i$\\
       &       & $8$ & N/A                            &         &        & $8$ & $0.022712-0.249751i$\\
$4.503$&$2$    & $0$ & $0.088172-0.017915i$           & $8.503$ & $2$    & $0$ & $0.044620-0.010250i$\\
       &       & $1$ & $0.084271-0.054775i$           &         &        & $1$ & $0.041781-0.031515i$\\
       &       & $2$ & $0.077561-0.094414i$           &         &        & $2$ & $0.037023-0.054868i$\\
       &       & $3$ & $0.069789-0.137451i$           &         &        & $3$ & $0.031957-0.080651i$\\
       &       & $4$ & $0.062120-0.183215i$           &         &        & $4$ & $0.027688-0.108078i$\\
       &       & $5$ & $0.054557-0.230756i$           &         &        & $5$ & $0.024398-0.136302i$\\
       &       & $6$ & $0.046238-0.279645i$           &         &        & $6$ & $0.021920-0.164877i$\\
       &       & $7$ & $0.035164-0.331368i$           &         &        & $7$ & $0.020068-0.193608i$\\
       &       & $8$ & $0.029836-0.391186i$           &         &        & $8$ & $0.018690-0.222416i$\\
$5.503$&$2$    & $0$ & $0.070567-0.015279i$           & $9.503$ & $2$    & $0$ & $0.039800-0.009212i$\\
       &       & $1$ & $0.066868-0.046857i$           &         &        & $1$ & $0.037201-0.028331i$\\
       &       & $2$ & $0.060674-0.081190i$           &         &        & $2$ & $0.032837-0.049355i$\\
       &       & $3$ & $0.054005-0.118791i$           &         &        & $3$ & $0.028181-0.072591i$\\
       &       & $4$ & $0.048267-0.158776i$           &         &        & $4$ & $0.024233-0.097312i$\\
       &       & $5$ & $0.043694-0.200069i$           &         &        & $5$ & $0.021151-0.122742i$\\
       &       & $6$ & $0.039981-0.242025i$           &         &        & $6$ & $0.018786-0.148475i$\\
       &       & $7$ & $0.036755-0.284331i$           &         &        & $7$ & $0.016983-0.174339i$\\
       &       & $8$ & $0.033690-0.326844i$           &         &        & $8$ & $0.015624-0.200261i$\\
$6.503$&$2$    & $0$ & $0.059021-0.013176i$           & $10^3$  & $2$    & $0$ & $0.000374-0.000089i$\\
       &       & $1$ & $0.055611-0.040462i$           &         &        & $1$ & $0.000347-0.000274i$\\
       &       & $2$ & $0.049915-0.070283i$           &         &        & $2$ & $0.000301-0.000478i$\\
       &       & $3$ & $0.043858-0.103077i$           &         &        & $3$ & $0.000252-0.000705i$\\
       &       & $4$ & $0.038776-0.137956i$           &         &        & $4$ & $0.000208-0.000947i$\\
       &       & $5$ & $0.034897-0.173912i$           &         &        & $5$ & $0.000169-0.001196i$\\
       &       & $6$ & $0.031983-0.210379i$           &         &        & $6$ & $0.000133-0.001448i$\\
       &       & $7$ & $0.029745-0.247100i$           &         &        & $7$ & $0.000093-0.001704i$\\
       &       & $8$ & $0.027946-0.283956i$           &         &        & $8$ & $0.000063-0.002303i$\\
[1ex]
\hline\hline 
\end{tabular}
\end{table}

\begin{table}
\centering
\caption{Purely imaginary QNMs for gravitational perturbations of the non-extremal Bonanno-Reuter black hole for $\alpha =\frac{118}{15\pi}$, $\gamma = 9/2$, $\ell=2$, and different values of the mass parameter $M$. The locations of the event horizon $r_h$ are given in Table~\ref{table:event}. The corresponding results are obtained through our spectral method, utilising $400$ polynomials with a precision of $200$ digits. In this context, $\omega$ and $N$ represent the dimensionless frequency and the corresponding overtone, respectively, while $\Delta\omega=\omega_N-\omega_{N+1}$. The notation 'SM' stands for Spectral Method.}
\label{tensor01overdamped}
\vspace*{1em}
\begin{tabular}{||c|c|c|c|c|c|c|c|c|c|c|c|c||}
\hline\hline
$M$    &$\ell$ & $N$ &$\omega$ (SM) & $\Delta\omega$ & $M$     & $\ell$ & $N$ & $\omega$ (SM) & $\Delta\omega$ \\ [0.5ex]
\hline\hline
$3.503$&$2$    & $0$ &$0.0000-0.107010i$    & $0.083939i$            & $7.503$ & $2$    & $0$ & $0.0000-2.157554i$    & $1.108554i$\\
       &       & $1$ &$0.0000-0.190949i$    & $0.078407i$            &         &        & $1$ & $0.0000-3.266108i$    & $0.033651i$\\
       &       & $2$ &$0.0000-0.269356i$    & $0.076241i$            &         &        & $2$ & $0.0000-3.299760i$    & $0.033542i$\\
       &       & $3$ &$0.0000-0.345597i$    & $0.075138i$            &         &        & $3$ & $0.0000-3.333301i$    & $0.033577i$\\
       &       & $4$ &$0.0000-0.420735i$    & $0.074451i$            &         &        & $4$ & $0.0000-3.366878i$    & $0.033566i$\\
       &       & $5$ &$0.0000-0.495186i$    & $0.073952i$            &         &        & $5$ & $0.0000-3.400445i$    & $0.033569i$\\
       &       & $6$ &$0.0000-0.569138i$    & $0.073555i$            &         &        & $6$ & $0.0000-3.434013i$    & $0.033568i$\\
       &       & $7$ &$0.0000-0.642692i$    & $0.073224i$            &         &        & $7$ & $0.0000-3.467581i$    & $0.033568i$\\
       &       & $8$ &$0.0000-0.715916i$    & $0.072944i$            &         &        & $8$ & $0.0000-3.501148i$    & $0.033567i$\\
       &       & $9$ &$0.0000-0.788860i$    & $0.072707i$            &         &        & $9$ & $0.0000-3.534715i$    & $0.033567i$\\
$4.503$&$2$    & $0$ &$0.0000-0.500969i$    & $2.299188i$            & $8.503$ & $2$    & $0$ & $0.0000-1.904318i$    & $1.126214i$\\
       &       & $1$ &$0.0000-2.800158i$    & $0.055770i$            &         &        & $1$ & $0.0000-3.030533i$    & $0.029679i$\\
       &       & $2$ &$0.0000-2.855927i$    & $0.055735i$            &         &        & $2$ & $0.0000-3.060211i$    & $0.029571i$\\
       &       & $3$ &$0.0000-2.911663i$    & $0.055728i$            &         &        & $3$ & $0.0000-3.089782i$    & $0.029658i$\\
       &       & $4$ &$0.0000-2.967390i$    & $0.055724i$            &         &        & $4$ & $0.0000-3.119440i$    & $0.029606i$\\
       &       & $5$ &$0.0000-3.023115i$    & $0.055722i$            &         &        & $5$ & $0.0000-3.149046i$    & $0.029633i$\\
       &       & $6$ &$0.0000-3.078837i$    & $0.055719i$            &         &        & $6$ & $0.0000-3.178679i$    & $0.029620i$\\
       &       & $7$ &$0.0000-3.134556i$    & $0.055717i$            &         &        & $7$ & $0.0000-3.208299i$    & $0.029626i$\\
       &       & $8$ &$0.0000-3.190273i$    & $0.055715i$            &         &        & $8$ & $0.0000-3.237925i$    & $0.029623i$\\
       &       & $9$ &$0.0000-3.245987i$    & $0.055712i$            &         &        & $9$ & $0.0000-3.267547i$    & $0.029624i$\\
$5.503$&$2$    & $0$ &$0.0000-0.000912i$    & $3.302283i$            & $9.503$ & $2$    & $0$ & $0.0000-2.844036i$    & $0.026452i$\\
       &       & $1$ &$0.0000-3.303196i$    & $0.045649i$            &         &        & $1$ & $0.0000-2.870487i$    & $0.026520i$\\
       &       & $2$ &$0.0000-3.348845i$    & $0.045725i$            &         &        & $2$ & $0.0000-2.897007i$    & $0.026511i$\\
       &       & $3$ &$0.0000-3.394570i$    & $0.045701i$            &         &        & $3$ & $0.0000-2.923517i$    & $0.026497i$\\
       &       & $4$ &$0.0000-3.440271i$    & $0.045694i$            &         &        & $4$ & $0.0000-2.950015i$    & $0.026511i$\\
       &       & $5$ &$0.0000-3.485965i$    & $0.045698i$            &         &        & $5$ & $0.0000-2.976526i$    & $0.026501i$\\
       &       & $6$ &$0.0000-3.531663i$    & $0.045696i$            &         &        & $6$ & $0.0000-3.003027i$    & $0.026507i$\\
       &       & $7$ &$0.0000-3.577359i$    & $0.045695i$            &         &        & $7$ & $0.0000-3.029534i$    & $0.026504i$\\
       &       & $8$ &$0.0000-3.623055i$    & $0.045695i$            &         &        & $8$ & $0.0000-3.056037i$    & $0.026505i$\\
       &       & $9$ &$0.0000-3.668749i$    & $0.045694i$            &         &        & $9$ & $0.0000-3.082542i$    & $0.026504i$\\
$6.503$&$2$    & $0$ &$0.0000-0.096822i$    & $2.392207i$            & $10^3$  & $2$    & $0$ & $0.0000-0.061102i$    & $0.000250i$\\
       &       & $1$ &$0.0000-2.489029i$    & $0.038660i$            &         &        & $1$ & $0.0000-0.061352i$    & $0.000250i$\\
       &       & $2$ &$0.0000-2.527689i$    & $0.851696i$            &         &        & $2$ & $0.0000-0.061602i$    & $0.000250i$\\
       &       & $3$ &$0.0000-3.379385i$    & $0.038726i$            &         &        & $3$ & $0.0000-0.061852i$    & $0.000250i$\\
       &       & $4$ &$0.0000-3.418111i$    & $0.038717i$            &         &        & $4$ & $0.0000-0.062102i$    & $0.000250i$\\
       &       & $5$ &$0.0000-3.456829i$    & $0.038710i$            &         &        & $5$ & $0.0000-0.062352i$    & $0.000250i$\\
       &       & $6$ &$0.0000-3.495539i$    & $0.038714i$            &         &        & $6$ & $0.0000-0.062602i$    & $0.000250i$\\
       &       & $7$ &$0.0000-3.534253i$    & $0.038712i$            &         &        & $7$ & $0.0000-0.062853i$    & $0.000250i$\\
       &       & $8$ &$0.0000-3.572965i$    & $0.038712i$            &         &        & $8$ & $0.0000-0.063103i$    & $0.000250i$\\
       &       & $9$ &$0.0000-3.611676i$    & $0.038711i$            &         &        & $9$ & $0.0000-0.063353i$    & $0.000250i$\\
[1ex]
\hline\hline 
\end{tabular}
\end{table}

\begin{table}
\centering
\caption{QNMs for spin $2$ perturbations of the non-extremal Bonanno-Reuter black hole for $\alpha =\frac{118}{15\pi}$, $\gamma = 9/2$, $\ell=3$, and different values of the mass parameter $M$. The locations of the event horizon $r_h$ are given in Table~\ref{table:event}. The results are obtained using our spectral method with 400 Chebyshev polynomials and 200-digit precision. Here, $\omega$ denotes the dimensionless QNM frequency, and $N$ the overtone number. 'N/A' indicates data not available, and 'SM' refers to the Spectral Method.}
\label{tensor02}
\vspace*{1em}
\begin{tabular}{||c|c|c|c|c|c|c|c|c|c|c|c|c||}
\hline\hline
$M$    &$\ell$ & $N$ & $\omega$ (SM)                  & $M$      & $\ell$ & $N$ & $\omega$ (SM) \\ [0.5ex]
\hline\hline
$3.503$&$3$    & $0$ & $0.187179-0.021592i$           & $7.503$ & $3$    & $0$ & $0.081173-0.012083i$\\
       &       & $1$ & $0.182764-0.065217i$           &         &        & $1$ & $0.079145-0.036630i$\\
       &       & $2$ & $0.173955-0.110236i$           &         &        & $2$ & $0.075404-0.062267i$\\
       &       & $3$ & $0.698132-0.709050i$           &         &        & $3$ & $0.070578-0.089493i$\\
       &       & $4$ & N/A                            &         &        & $4$ & $0.065440-0.118399i$\\
       &       & $5$ & N/A                            &         &        & $5$ & $0.060592-0.148679i$\\
       &       & $6$ & N/A                            &         &        & $6$ & $0.056304-0.179889i$\\
       &       & $7$ & N/A                            &         &        & $7$ & $0.052605-0.211662i$\\
       &       & $8$ & N/A                            &         &        & $8$ & $0.049417-0.243761i$\\
$4.503$&$3$    & $0$ & $0.139712-0.018985i$           & $8.503$ & $3$    & $0$ & $0.071366-0.010720i$\\
       &       & $1$ & $0.136842-0.057401i$           &         &        & $1$ & $0.069536-0.032505i$\\
       &       & $2$ & $0.131445-0.097088i$           &         &        & $2$ & $0.066162-0.055281i$\\
       &       & $3$ & $0.124164-0.138633i$           &         &        & $3$ & $0.061818-0.079502i$\\
       &       & $4$ & $0.115770-0.182222i$           &         &        & $4$ & $0.057210-0.105246i$\\
       &       & $5$ & $0.106866-0.227653i$           &         &        & $5$ & $0.052886-0.132228i$\\
       &       & $6$ & $0.097696-0.274556i$           &         &        & $6$ & $0.049087-0.160040i$\\
       &       & $7$ & $0.088135-0.322670i$           &         &        & $7$ & $0.045831-0.188348i$\\
       &       & $8$ & $0.077754-0.372135i$           &         &        & $8$ & $0.043048-0.216938i$\\
$5.503$&$3$    & $0$ & $0.112336-0.016077i$           & $9.503$ & $3$    & $0$ & $0.063696-0.009627i$\\
       &       & $1$ & $0.109800-0.048685i$           &         &        & $1$ & $0.062033-0.029195i$\\
       &       & $2$ & $0.105099-0.082585i$           &         &        & $2$ & $0.058968-0.049667i$\\
       &       & $3$ & $0.098955-0.118367i$           &         &        & $3$ & $0.055025-0.071458i$\\
       &       & $4$ & $0.092250-0.156161i$           &         &        & $4$ & $0.050852-0.094636i$\\
       &       & $5$ & $0.085689-0.195656i$           &         &        & $5$ & $0.046948-0.118937i$\\
       &       & $6$ & $0.079612-0.236355i$           &         &        & $6$ & $0.043530-0.143985i$\\
       &       & $7$ & $0.074068-0.277828i$           &         &        & $7$ & $0.040611-0.169477i$\\
       &       & $8$ & $0.068959-0.319779i$           &         &        & $8$ & $0.038125-0.195218i$\\
$6.503$&$3$    & $0$ & $0.094181-0.013820i$           & $10^3$  & $3$    & $0$ & $0.000599-0.000093i$\\
       &       & $1$ & $0.091918-0.041879i$           &         &        & $1$ & $0.000583-0.000281i$\\
       &       & $2$ & $0.087740-0.071137i$           &         &        & $2$ & $0.000552-0.000479i$\\
       &       & $3$ & $0.082329-0.102140i$           &         &        & $3$ & $0.000512-0.000690i$\\
       &       & $4$ & $0.076526-0.134995i$           &         &        & $4$ & $0.000470-0.000916i$\\
       &       & $5$ & $0.070989-0.169383i$           &         &        & $5$ & $0.000431-0.001152i$\\
       &       & $6$ & $0.066028-0.204826i$           &         &        & $6$ & $0.000398-0.001396i$\\
       &       & $7$ & $0.061680-0.240919i$           &         &        & $7$ & $0.000369-0.001644i$\\
       &       & $8$ & $0.057864-0.277397i$           &         &        & $8$ & $0.000345-0.001894i$\\
[1ex]
\hline\hline 
\end{tabular}
\end{table}

\begin{table}
\centering
\caption{Purely imaginary QNMs for gravitational perturbations of the non-extremal Bonanno-Reuter black hole for $\alpha =\frac{118}{15\pi}$, $\gamma = 9/2$, $\ell=3$, and different values of the mass parameter $M$. The locations of the event horizon $r_h$ are given in Table~\ref{table:event}. The corresponding results are obtained through our spectral method, utilising $400$ polynomials with a precision of $200$ digits. In this context, $\omega$ and $N$ represent the dimensionless frequency and the corresponding overtone, respectively, while $\Delta\omega=\omega_N-\omega_{N+1}$. The notation 'SM' stands for Spectral Method.}
\label{tensor02overdamped}
\vspace*{1em}
\begin{tabular}{||c|c|c|c|c|c|c|c|c|c|c|c|c||}
\hline\hline
$M$    &$\ell$ & $N$ &$\omega$ (SM) & $\Delta\omega$ & $M$     & $\ell$ & $N$ & $\omega$ (SM) & $\Delta\omega$ \\ [0.5ex]
\hline\hline
$3.503$&$3$    & $0$ &$0.0000-0.207505i$    & $0.083783i$            & $7.503$ & $3$    & $0$ & $0.0000-0.081765i$    & $2.004529i$\\
       &       & $1$ &$0.0000-0.291288i$    & $0.080075i$            &         &        & $1$ & $0.0000-2.086294i$    & $1.176870i$\\
       &       & $2$ &$0.0000-0.371363i$    & $0.078098i$            &         &        & $2$ & $0.0000-3.263164i$    & $0.033654i$\\
       &       & $3$ &$0.0000-0.449461i$    & $0.076940i$            &         &        & $3$ & $0.0000-3.296819i$    & $0.033580i$\\
       &       & $4$ &$0.0000-0.526400i$    & $0.076205i$            &         &        & $4$ & $0.0000-3.330399i$    & $0.033603i$\\
       &       & $5$ &$0.0000-0.602605i$    & $0.075695i$            &         &        & $5$ & $0.0000-3.364002i$    & $0.033595i$\\
       &       & $6$ &$0.0000-0.678300i$    & $0.075299i$            &         &        & $6$ & $0.0000-3.397598i$    & $0.033596i$\\
       &       & $7$ &$0.0000-0.753599i$    & $0.074958i$            &         &        & $7$ & $0.0000-3.431194i$    & $0.033595i$\\
       &       & $8$ &$0.0000-0.828558i$    & $0.074642i$            &         &        & $8$ & $0.0000-3.464789i$    & $0.033594i$\\
       &       & $9$ &$0.0000-0.903199i$    & $0.074338i$            &         &        & $9$ & $0.0000-3.498383i$    & $0.033593i$\\
$4.503$&$3$    & $0$ &$0.0000-0.150864i$    & $2.642359i$            & $8.503$ & $3$    & $0$ & $0.0000-0.071437i$    & $1.767043i$\\
       &       & $1$ &$0.0000-2.793223i$    & $0.055938i$            &         &        & $1$ & $0.0000-1.838480i$    & $0.061828i$\\
       &       & $2$ &$0.0000-2.849161i$    & $0.055865i$            &         &        & $2$ & $0.0000-1.900308i$    & $0.413528i$\\
       &       & $3$ &$0.0000-2.905025i$    & $0.055856i$            &         &        & $3$ & $0.0000-2.313836i$    & $0.684430i$\\
       &       & $4$ &$0.0000-2.960881i$    & $0.055847i$            &         &        & $4$ & $0.0000-2.998266i$    & $0.029529i$\\
       &       & $5$ &$0.0000-3.016729i$    & $0.055840i$            &         &        & $5$ & $0.0000-3.027795i$    & $0.029770i$\\
       &       & $6$ &$0.0000-3.072569i$    & $0.055833i$            &         &        & $6$ & $0.0000-3.057566i$    & $0.029564i$\\
       &       & $7$ &$0.0000-3.128402i$    & $0.055827i$            &         &        & $7$ & $0.0000-3.087129i$    & $0.029694i$\\
       &       & $8$ &$0.0000-3.184228i$    & $0.055820i$            &         &        & $8$ & $0.0000-3.116824i$    & $0.029627i$\\
       &       & $9$ &$0.0000-3.240049i$    & $0.055814i$            &         &        & $9$ & $0.0000-3.146451i$    & $0.029659i$\\
$5.503$&$3$    & $0$ &$0.0000-0.625927i$    & $2.256811i$            & $9.503$ & $3$    & $0$ & $0.0000-0.063493i$    & $1.503068i$\\
       &       & $1$ &$0.0000-2.882738i$    & $0.141286i$            &         &        & $1$ & $0.0000-1.566561i$    & $1.168784i$\\
       &       & $2$ &$0.0000-3.024024i$    & $0.274661i$            &         &        & $2$ & $0.0000-2.735345i$    & $0.053220i$\\
       &       & $3$ &$0.0000-3.298684i$    & $0.045761i$            &         &        & $3$ & $0.0000-2.788565i$    & $0.052973i$\\
       &       & $4$ &$0.0000-3.344445i$    & $0.045796i$            &         &        & $4$ & $0.0000-2.841539i$    & $0.026532i$\\
       &       & $5$ &$0.0000-3.390240i$    & $0.045751i$            &         &        & $5$ & $0.0000-2.868070i$    & $0.026498i$\\
       &       & $6$ &$0.0000-3.435991i$    & $0.045753i$            &         &        & $6$ & $0.0000-2.894568i$    & $0.026555i$\\
       &       & $7$ &$0.0000-3.481744i$    & $0.045753i$            &         &        & $7$ & $0.0000-2.921123i$    & $0.026509i$\\
       &       & $8$ &$0.0000-3.527497i$    & $0.045749i$            &         &        & $8$ & $0.0000-2.947631i$    & $0.026537i$\\
       &       & $9$ &$0.0000-3.573246i$    & $0.045747i$            &         &        & $9$ & $0.0000-2.974168i$    & $0.026521i$\\
$6.503$&$3$    & $0$ &$0.0000-0.095831i$    & $2.426904i$            & $10^3$  & $3$    & $0$ & $0.0000-0.036059i$    & $0.000250i$\\
       &       & $1$ &$0.0000-2.522735i$    & $0.853147i$            &         &        & $1$ & $0.0000-0.036309i$    & $0.000250i$\\
       &       & $2$ &$0.0000-3.375882i$    & $0.038802i$            &         &        & $2$ & $0.0000-0.036559i$    & $0.000250i$\\
       &       & $3$ &$0.0000-3.414684i$    & $0.038743i$            &         &        & $3$ & $0.0000-0.036809i$    & $0.000250i$\\
       &       & $4$ &$0.0000-3.453427i$    & $0.038752i$            &         &        & $4$ & $0.0000-0.037060i$    & $0.000250i$\\
       &       & $5$ &$0.0000-3.492179i$    & $0.038750i$            &         &        & $5$ & $0.0000-0.037310i$    & $0.000250i$\\
       &       & $6$ &$0.0000-3.530928i$    & $0.038748i$            &         &        & $6$ & $0.0000-0.037560i$    & $0.000250i$\\
       &       & $7$ &$0.0000-3.569676i$    & $0.038747i$            &         &        & $7$ & $0.0000-0.037810i$    & $0.000250i$\\
       &       & $8$ &$0.0000-3.608423i$    & $0.038746i$            &         &        & $8$ & $0.0000-0.038061i$    & $0.000250i$\\
       &       & $9$ &$0.0000-3.647169i$    & $0.038745i$            &         &        & $9$ & $0.0000-0.038311i$    & $0.000250i$\\
[1ex]
\hline\hline 
\end{tabular}
\end{table}

\begin{table}
\centering
\caption{QNMs for scalar perturbations of the extremal Bonanno-Reuter black hole 
 for $\alpha =\frac{118}{15\pi}$, $\gamma = 9/2$, and various angular momentum values $\ell$. The location of the event horizon $r_e$ is provided in Table~\ref{table:event}. Our QNMs are computed using the Spectral Method, employing 400 Chebyshev polynomials and 200-digit numerical precision. Here, $\omega$ denotes the dimensionless QNM frequency and $N$ the overtone number. Entries marked ‘N/A’ indicate unavailable data, while ‘SM’ refers to the Spectral Method.}
\label{scalargextreme}
\vspace*{1em}
\begin{tabular}{||c|c|c|c|c|c|c|c|c|c|c|c|c||}
\hline\hline
$\ell$ & $N$ & $\omega$ (SM)   & $\ell$ & $N$ & $\omega$ (SM) \\ [0.5ex]
\hline\hline
$0$    & $0$ &$0.032096-0.025010i$              & $3$    & $0$ & $0.207033-0.023102i$\\
       & $1$ &$0.018528-0.088491i$              &        & $1$ & $0.202571-0.069734i$\\
       & $2$ &$0.011491-0.159887i$              &        & $2$ & $0.193707-0.117704i$\\
       & $3$ &N/A                               &        & $3$ & $0.180671-0.168072i$\\
       & $4$ &N/A                               &        & $4$ & $0.164078-0.222034i$\\
       & $5$ &N/A                               &        & $5$ & $0.145180-0.280652i$\\\
       & $6$ &N/A                               &        & $6$ & $0.125798-0.344302i$\\
       & $7$ &N/A                               &        & $7$ & $0.107668-0.412323i$\\
       & $8$ &N/A                               &        & $8$ & $0.091755-0.483416i$\\
$1$    & $0$ &$0.089333-0.023289i$              & $4$    & $0$ & $0.266037-0.023089i$\\
       & $1$ &$0.079433-0.072311i$              &        & $1$ & $0.262550-0.069524i$\\
       & $2$ &$0.062146-0.129283i$              &        & $2$ & $0.255600-0.116750i$\\
       & $3$ &$0.044863-0.196349i$              &        & $3$ & $0.245259-0.165361i$\\
       & $4$ &$0.032304-0.268601i$              &        & $4$ & $0.231716-0.216037i$\\
       & $5$ &$0.023583-0.342134i$              &        & $5$ & $0.215369-0.269515i$\\
       & $6$ &N/A                               &        & $6$ & $0.196926-0.326486i$\\
       & $7$ &N/A                               &        & $7$ & $0.177415-0.387379i$\\
       & $8$ &N/A                               &        & $8$ & $0.158011-0.452152i$\\
$2$    & $0$ &$0.148088-0.023138i$              & $5$    & $0$ & $0.325066-0.023083i$\\
       & $1$ &$0.141906-0.070264i$              &        & $1$ & $0.322207-0.069420i$\\
       & $2$ &$0.129802-0.120135i$              &        & $2$ & $0.316498-0.116279i$\\
       & $3$ &$0.112865-0.174989i$              &        & $3$ & $0.307973-0.164041i$\\
       & $4$ &$0.093766-0.236544i$              &        & $4$ & $0.296704-0.213129i$\\
       & $5$ &$0.075923-0.304306i$              &        & $5$ & $0.282846-0.264012i$\\
       & $6$ &$0.061172-0.375922i$              &        & $6$ & $0.266677-0.317189i$\\
       & $7$ &$0.049474-0.449301i$              &        & $7$ & $0.248649-0.373139i$\\
       & $8$ &$0.040157-0.523325i$              &        & $8$ & $0.229400-0.432220i$\\
       [1ex]
\hline\hline 
\end{tabular}
\end{table}

\begin{table}
\centering
\caption{Purely imaginary QNMs for scalar perturbations of the extremal Bonanno-Reuter black hole for $\alpha =\frac{118}{15\pi}$, $\gamma = 9/2$, and several values of the angular momentum $\ell$. The location of the event horizon $r_e$ is given in Table~\ref{table:event}. The corresponding results are obtained through our spectral method, utilising $400$ polynomials with a precision of $200$ digits. In this context, $\omega$ and $N$ represent the dimensionless frequency and the corresponding overtone, respectively, while $\Delta\omega=\omega_N-\omega_{N+1}$. The notation 'SM' stands for Spectral Method.}
\label{scalargextremeoverdamped}
\vspace*{1em}
\begin{tabular}{||c|c|c|c|c|c|c|c|c|c|c|c|c||}
\hline\hline
$\ell$ & $N$ & $\omega$ (SM) & $\Delta\omega$  & $\ell$ & $N$ & $\omega$ (SM) & $\Delta\omega$ \\ [0.5ex]
\hline\hline
$0$    & $0$ &$0.0000-0.259004i$    & $5.860426i$              & $3$    & $0$ & $0.0000-0.001129i$    & $0.000398i$\\
       & $1$ &$0.0000-6.119430i$    & $0.071379i$              &        & $1$ & $0.0000-0.001527i$    & $6.252907i$\\
       & $2$ &$0.0000-6.190809i$    & $0.071105i$              &        & $2$ & $0.0000-6.254434i$    & $0.071028i$\\
       & $3$ &$0.0000-6.261915i$    & $0.071158i$              &        & $3$ & $0.0000-6.325462i$    & $0.071558i$\\
       & $4$ &$0.0000-6.333072i$    & $0.071321i$              &        & $4$ & $0.0000-6.397020i$    & $0.070923i$\\
       & $5$ &$0.0000-6.404394i$    & $0.070948i$              &        & $5$ & $0.0000-6.467943i$    & $0.071617i$\\
       & $6$ &$0.0000-6.475341i$    & $0.071501i$              &        & $6$ & $0.0000-6.539560i$    & $0.070900i$\\
       & $7$ &$0.0000-6.546843i$    & $0.070808i$              &        & $7$ & $0.0000-6.610460i$    & $0.071570i$\\
       & $8$ &$0.0000-6.617650i$    & $0.071575i$              &        & $8$ & $0.0000-6.682029i$    & $0.070995i$\\
       & $9$ &$0.0000-6.689226i$    & $0.070805i$              &        & $9$ & $0.0000-6.753024i$    & $0.071408i$\\
$1$    & $0$ &$0.0000-0.002684i$    & $0.237898i$              & $4$    & $0$ & $0.0000-0.001063i$    & $0.000391i$\\
       & $1$ &$0.0000-0.240583i$    & $5.741699i$              &        & $1$ & $0.0000-0.001454i$    & $6.100565i$\\
       & $2$ &$0.0000-5.982281i$    & $0.071783i$              &        & $2$ & $0.0000-6.102019i$    & $0.071092i$\\
       & $3$ &$0.0000-6.054064i$    & $0.070590i$              &        & $3$ & $0.0000-6.173111i$    & $0.071668i$\\
       & $4$ &$0.0000-6.124654i$    & $0.071556i$              &        & $4$ & $0.0000-6.244779i$    & $0.071056i$\\
       & $5$ &$0.0000-6.196210i$    & $0.070830i$              &        & $5$ & $0.0000-6.315835i$    & $0.071665i$\\
       & $6$ &$0.0000-6.267040i$    & $0.071299i$              &        & $6$ & $0.0000-6.387500i$    & $0.071063i$\\
       & $7$ &$0.0000-6.338339i$    & $0.071095i$              &        & $7$ & $0.0000-6.458563i$    & $0.071606i$\\
       & $8$ &$0.0000-6.409434i$    & $0.071031i$              &        & $8$ & $0.0000-6.530169i$    & $0.071138i$\\
       & $9$ &$0.0000-6.480465i$    & $0.071350i$              &        & $9$ & $0.0000-6.601307i$    & $0.071476i$\\
$2$    & $0$ &$0.0000-0.237357i$    & $0.021647i$              & $5$    & $0$ & $0.0000-0.000671i$    & $0.000309i$\\
       & $1$ &$0.0000-0.259004i$    & $5.860426i$              &        & $1$ & $0.0000-0.000980i$    & $5.731456i$\\
       & $2$ &$0.0000-6.119430i$    & $0.071379i$              &        & $2$ & $0.0000-5.732435i$    & $0.071654i$\\
       & $3$ &$0.0000-6.190809i$    & $0.071105i$              &        & $3$ & $0.0000-5.804089i$    & $0.071378i$\\
       & $4$ &$0.0000-6.261915i$    & $0.071158i$              &        & $4$ & $0.0000-5.875468i$    & $0.071690i$\\
       & $5$ &$0.0000-6.333072i$    & $0.071321i$              &        & $5$ & $0.0000-5.947158i$    & $0.071328i$\\ 
       & $6$ &$0.0000-6.404394i$    & $0.070948i$              &        & $6$ & $0.0000-6.018485i$    & $0.071687i$\\
       & $7$ &$0.0000-6.475341i$    & $0.071501i$              &        & $7$ & $0.0000-6.090173i$    & $0.428752i$\\
       & $8$ &$0.0000-6.546843i$    & $0.070808i$              &        & $8$ & $0.0000-6.518925i$    & $0.071467i$\\
       & $9$ &$0.0000-6.617650i$    & $0.071575i$              &        & $9$ & $0.0000-6.590391i$    & $0.071371i$\\
       [1ex]
\hline\hline 
\end{tabular}
\end{table}

\begin{table}
\centering
\caption{QNMs for electromagnetic perturbations of the extremal Bonanno-Reuter black hole 
 for $\alpha =\frac{118}{15\pi}$, $\gamma = 9/2$, and various angular momentum values $\ell$. The location of the event horizon $r_e$ is provided in Table~\ref{table:event}. Our QNMs are computed using the Spectral Method, employing 400 Chebyshev polynomials and 200-digit numerical precision. Here, $\omega$ denotes the dimensionless QNM frequency and $N$ the overtone number. Entries marked ‘N/A’ indicate unavailable data, while ‘SM’ refers to the Spectral Method.}
\label{emextreme}
\vspace*{1em}
\begin{tabular}{||c|c|c|c|c|c|c|c|c|c|c|c|c||}
\hline\hline
$\ell$ & $N$ & $\omega$ (SM)   & $\ell$ & $N$ & $\omega$ (SM) \\ [0.5ex]
\hline\hline
$1$    & $0$ &$0.078139-0.021573i$              & $4$    & $0$ & $0.262306-0.022893i$\\
       & $1$ &$0.067611-0.067145i$              &        & $1$ & $0.258807-0.068935i$\\
       & $2$ &$0.048821-0.121145i$              &        & $2$ & $0.251828-0.115767i$\\
       & $3$ &$0.030047-0.186239i$              &        & $3$ & $0.241440-0.163987i$\\
       & $4$ &$0.016422-0.256686i$              &        & $4$ & $0.227826-0.214278i$\\
       & $5$ &N/A                               &        & $5$ & $0.211381-0.267389i$\\
       & $6$ &N/A                               &        & $6$ & $0.192819-0.324027i$\\
       & $7$ &N/A                               &        & $7$ & $0.173185-0.384633i$\\
       & $8$ &N/A                               &        & $8$ & $0.153678-0.449162i$\\
$2$    & $0$ &$0.141379-0.022510i$              & $5$    & $0$ & $0.322013-0.022951i$\\
       & $1$ &$0.135101-0.068375i$              &        & $1$ & $0.319147-0.069025i$\\
       & $2$ &$0.122767-0.116995i$              &        & $2$ & $0.313424-0.115620i$\\
       & $3$ &$0.105435-0.170696i$              &        & $3$ & $0.304875-0.163119i$\\
       & $4$ &$0.085878-0.231326i$              &        & $4$ & $0.293572-0.211945i$\\
       & $5$ &$0.067710-0.298339i$              &        & $5$ & $0.279666-0.262570i$\\
       & $6$ &$0.052796-0.369203i$              &        & $6$ & $0.263434-0.315501i$\\
       & $7$ &$0.041005-0.441748i$              &        & $7$ & $0.245331-0.371221i$\\
       & $8$ &$0.031592-0.514862i$              &        & $8$ & $0.226001-0.430097i$\\
$3$    & $0$ &$0.202238-0.022779i$              & $6$    & $0$ & $0.381525-0.022985i$\\
       & $1$ &$0.197746-0.068764i$              &        & $1$ & $0.379098-0.069077i$\\
       & $2$ &$0.188815-0.116086i$              &        & $2$ & $0.374249-0.115541i$\\
       & $3$ &$0.175661-0.165816i$              &        & $3$ & $0.366995-0.162640i$\\
       & $4$ &$0.158894-0.219177i$              &        & $4$ & $0.357369-0.210661i$\\
       & $5$ &$0.139785-0.277273i$              &        & $5$ & $0.345440-0.259923i$\\
       & $6$ &$0.120208-0.340494i$              &        & $6$ & $0.331333-0.310771i$\\
       & $7$ &$0.101954-0.408147i$              &        & $7$ & $0.315255-0.363566i$\\
       & $8$ &$0.085996-0.478880i$              &        & $8$ & $0.297517-0.418662i$\\
       [1ex]
\hline\hline 
\end{tabular}
\end{table}

\begin{table}
\centering
\caption{Purely imaginary QNMs for electromagnetic perturbations of the extremal Bonanno-Reuter black hole for $\alpha =\frac{118}{15\pi}$, $\gamma = 9/2$, and several values of the angular momentum $\ell$. The location of the event horizon $r_e$ is given in Table~\ref{table:event}. The corresponding results are obtained through our spectral method, utilising $400$ polynomials with a precision of $200$ digits. In this context, $\omega$ and $N$ represent the dimensionless frequency and the corresponding overtone, respectively, while $\Delta\omega=\omega_N-\omega_{N+1}$. The notation 'SM' stands for Spectral Method.}
\label{emextremeoverdamped}
\vspace*{1em}
\begin{tabular}{||c|c|c|c|c|c|c|c|c|c|c|c|c||}
\hline\hline
$\ell$ & $N$ & $\omega$ (SM) & $\Delta\omega$  & $\ell$ & $N$ & $\omega$ (SM) & $\Delta\omega$ \\ [0.5ex]
\hline\hline
$1$    & $0$ &$0.0000-0.000384i$    & $0.000024i$              & $4$    & $0$ & $0.0000-0.000746i$    & $0.000317i$\\
       & $1$ &$0.0000-0.000408i$    & $0.000169i$              &        & $1$ & $0.0000-0.001063i$    & $0.000391i$\\
       & $2$ &$0.0000-0.000577i$    & $0.000038i$              &        & $2$ & $0.0000-0.001454i$    & $7.151451i$\\
       & $3$ &$0.0000-0.000615i$    & $0.000210i$              &        & $3$ & $0.0000-7.152905i$    & $2.564406i$\\
       & $4$ &$0.0000-0.000825i$    & $0.000056i$              &        & $4$ & $0.0000-9.717310i$    & $0.498183i$\\
       & $5$ &$0.0000-0.000880i$    & $0.000332i$              &        & $5$ & $0.0000-10.21550i$    & $0.142339i$\\
       & $6$ &$0.0000-0.001213i$    & $0.000406i$              &        & $6$ & $0.0000-10.35783i$    & $0.142334i$\\
       & $7$ &$0.0000-0.001619i$    & $0.000488i$              &        & $7$ & $0.0000-10.50017i$    & $0.142431i$\\
       & $8$ &$0.0000-0.002107i$    & $0.000577i$              &        & $8$ & $0.0000-10.64260i$    & $0.284730i$\\
       & $9$ &$0.0000-0.002684i$    & $0.239100i$              &        & $9$ & $0.0000-10.92733i$    & $0.284704i$\\
$2$    & $0$ &$0.0000-0.000588i$    & $0.000219i$              & $5$    & $0$ & $0.0000-0.000659i$    & $0.000012i$\\
       & $1$ &$0.0000-0.000807i$    & $0.000044i$              &        & $1$ & $0.0000-0.000671i$    & $0.000309i$\\
       & $2$ &$0.0000-0.000850i$    & $0.000329i$              &        & $2$ & $0.0000-0.000980i$    & $7.073463i$\\
       & $3$ &$0.0000-0.001179i$    & $0.000403i$              &        & $3$ & $0.0000-7.074442i$    & $0.070704i$\\
       & $4$ &$0.0000-0.001582i$    & $0.000485i$              &        & $4$ & $0.0000-7.145146i$    & $0.285643i$\\
       & $5$ &$0.0000-0.002067i$    & $0.235873i$              &        & $5$ & $0.0000-7.430789i$    & $0.285047i$\\
       & $6$ &$0.0000-0.237940i$    & $0.021687i$              &        & $6$ & $0.0000-7.715835i$    & $0.784331i$\\
       & $7$ &$0.0000-0.259626i$    & $9.821099i$              &        & $7$ & $0.0000-8.500166i$    & $0.427547i$\\
       & $8$ &$0.0000-10.08073i$    & $4.481123i$              &        & $8$ & $0.0000-8.927713i$    & $0.427418i$\\
       & $9$ &$0.0000-14.56185i$    & $0.285080i$              &        & $9$ & $0.0000-9.355131i$    & $0.356329i$\\
$3$    & $0$ &$0.0000-0.000549i$    & $0.000225i$              & $6$    & $0$ & $0.0000-0.000581i$    & $0.000287i$\\
       & $1$ &$0.0000-0.000773i$    & $0.000032i$              &        & $1$ & $0.0000-0.000867i$    & $0.000012i$\\
       & $2$ &$0.0000-0.000806i$    & $0.000324i$              &        & $2$ & $0.0000-0.000879i$    & $6.420588i$\\
       & $3$ &$0.0000-0.001129i$    & $0.000398i$              &        & $3$ & $0.0000-6.421467i$    & $0.071118i$\\
       & $4$ &$0.0000-0.001527i$    & $10.07595i$              &        & $4$ & $0.0000-6.492585i$    & $0.071793i$\\
       & $5$ &$0.0000-10.07748i$    & $0.142392i$              &        & $5$ & $0.0000-6.564378i$    & $0.071289i$\\
       & $6$ &$0.0000-10.21987i$    & $0.142335i$              &        & $6$ & $0.0000-6.635667i$    & $0.071545i$\\
       & $7$ &$0.0000-10.36221i$    & $3.912665i$              &        & $7$ & $0.0000-6.707212i$    & $0.214571i$\\
       & $8$ &$0.0000-14.27487i$    & $0.284628i$              &        & $8$ & $0.0000-6.921783i$    & $0.071040i$\\
       & $9$ &$0.0000-14.55950i$    & $0.284788i$              &        & $9$ & $0.0000-6.992823i$    & $0.071874i$\\
       [1ex]
\hline\hline 
\end{tabular}
\end{table}

\begin{table}
\centering
\caption{QNMs for tensor perturbations of the extremal Bonanno-Reuter black hole 
 for $\alpha =\frac{118}{15\pi}$, $\gamma = 9/2$, and various angular momentum values $\ell$. The location of the event horizon $r_e$ is provided in Table~\ref{table:event}. Our QNMs are computed using the Spectral Method, employing 400 Chebyshev polynomials and 200-digit numerical precision. Here, $\omega$ denotes the dimensionless QNM frequency and $N$ the overtone number. Entries marked ‘N/A’ indicate unavailable data, while ‘SM’ refers to the Spectral Method.}
\label{tensorextreme}
\vspace*{1em}
\begin{tabular}{||c|c|c|c|c|c|c|c|c|c|c|c|c||}
\hline\hline
$\ell$ & $N$ & $\omega$ (SM)   & $\ell$ & $N$ & $\omega$ (SM) \\ [0.5ex]
\hline\hline
$2$    & $0$ &$0.119491-0.019731i$              & $5$    & $0$ & $0.312683-0.022522i$\\
       & $1$ &$0.113946-0.060247i$              &        & $1$ & $0.309812-0.067738i$\\
       & $2$ &$0.102777-0.104191i$              &        & $2$ & $0.304078-0.113478i$\\
       & $3$ &$0.086672-0.154486i$              &        & $3$ & $0.295505-0.160126i$\\
       & $4$ &$0.068579-0.213458i$              &        & $4$ & $0.284158-0.208114i$\\
       & $5$ &$0.052695-0.280108i$              &        & $5$ & $0.270182-0.257928i$\\
       & $6$ &$0.040777-0.350988i$              &        & $6$ & $0.253850-0.310090i$\\
       & $7$ &$0.032232-0.423530i$              &        & $7$ & $0.235622-0.365110i$\\
       & $8$ &$0.026033-0.496575i$              &        & $8$ & $0.216159-0.423377i$\\
$3$    & $0$ &$0.187196-0.021592i$              & $6$    & $0$ & $0.373670-0.022685i$\\
       & $1$ &$0.182779-0.065217i$              &        & $1$ & $0.371238-0.068176i$\\
       & $2$ &$0.173968-0.110238i$              &        & $2$ & $0.366378-0.114039i$\\
       & $3$ &$0.160926-0.157798i$              &        & $3$ & $0.359103-0.160539i$\\
       & $4$ &$0.144226-0.209252i$              &        & $4$ & $0.349446-0.207966i$\\
       & $5$ &$0.125189-0.265865i$              &        & $5$ & $0.337470-0.256641i$\\
       & $6$ &$0.105862-0.328106i$              &        & $6$ & $0.323297-0.306917i$\\
       & $7$ &$0.088192-0.395176i$              &        & $7$ & $0.307131-0.359165i$\\
       & $8$ &$0.073167-0.465548i$              &        & $8$ & $0.289288-0.413750i$\\
$4$    & $0$ &$0.250803-0.022227i$              & $7$    & $0$ & $0.434133-0.022784i$\\
       & $1$ &$0.247309-0.066938i$              &        & $1$ & $0.432025-0.068443i$\\
       & $2$ &$0.240334-0.112450i$              &        & $2$ & $0.427810-0.114380i$\\
       & $3$ &$0.229932-0.159372i$              &        & $3$ & $0.421497-0.160789i$\\
       & $4$ &$0.216271-0.208415i$              &        & $4$ & $0.413104-0.207880i$\\
       & $5$ &$0.199735-0.260372i$              &        & $5$ & $0.402664-0.255879i$\\
       & $6$ &$0.181054-0.316006i$              &        & $6$ & $0.390238-0.305035i$\\
       & $7$ &$0.161324-0.375802i$              &        & $7$ & $0.375928-0.355615i$\\
       & $8$ &$0.141814-0.439721i$              &        & $8$ & $0.359893-0.407898i$\\
       [1ex]
\hline\hline 
\end{tabular}
\end{table}

\begin{table}
\centering
\caption{Purely imaginary QNMs for tensor perturbations of the extremal Bonanno-Reuter black hole for $\alpha =\frac{118}{15\pi}$, $\gamma = 9/2$, and several values of the angular momentum $\ell$. The location of the event horizon $r_e$ is given in Table~\ref{table:event}. The corresponding results are obtained through our spectral method, utilising $400$ polynomials with a precision of $200$ digits. In this context, $\omega$ and $N$ represent the dimensionless frequency and the corresponding overtone, respectively, while $\Delta\omega=\omega_N-\omega_{N+1}$. The notation 'SM' stands for Spectral Method.}
\label{tensorextremeoverdamped}
\vspace*{1em}
\begin{tabular}{||c|c|c|c|c|c|c|c|c|c|c|c|c||}
\hline\hline
$\ell$ & $N$ & $\omega$ (SM) & $\Delta\omega$  & $\ell$ & $N$ & $\omega$ (SM) & $\Delta\omega$ \\ [0.5ex]
\hline\hline
$2$    & $0$ &$0.0000-0.002067i$    & $0.238010i$              & $5$    & $0$ & $0.0000-0.000429i$    & $0.000230i$\\
       & $1$ &$0.0000-0.240077i$    & $0.021367i$              &        & $1$ & $0.0000-0.000659i$    & $0.000012i$\\
       & $2$ &$0.0000-0.261444i$    & $5.229540i$              &        & $2$ & $0.0000-0.000671i$    & $0.000309i$\\
       & $3$ &$0.0000-5.490984i$    & $0.426753i$              &        & $3$ & $0.0000-0.000980i$    & $5.115273i$\\
       & $4$ &$0.0000-5.917736i$    & $0.071430i$              &        & $4$ & $0.0000-5.116253i$    & $0.642035i$\\
       & $5$ &$0.0000-5.989166i$    & $0.070837i$              &        & $5$ & $0.0000-5.758287i$    & $0.071392i$\\
       & $6$ &$0.0000-6.060003i$    & $0.071363i$              &        & $6$ & $0.0000-5.829679i$    & $0.071242i$\\
       & $7$ &$0.0000-6.131366i$    & $0.070906i$              &        & $7$ & $0.0000-5.900920i$    & $0.071374i$\\
       & $8$ &$0.0000-6.202272i$    & $0.071288i$              &        & $8$ & $0.0000-5.972294i$    & $0.071251i$\\
       & $9$ &$0.0000-6.273560i$    & $0.070985i$              &        & $9$ & $0.0000-6.043545i$    & $0.071335i$\\
$3$    & $0$ &$0.0000-0.000806i$    & $0.000324i$              & $6$    & $0$ & $0.0000-0.000581i$    & $0.000287i$\\
       & $1$ &$0.0000-0.001129i$    & $0.000398i$              &        & $1$ & $0.0000-0.000867i$    & $0.000012i$\\
       & $2$ &$0.0000-0.001527i$    & $5.556448i$              &        & $2$ & $0.0000-0.000879i$    & $4.389305i$\\
       & $3$ &$0.0000-5.557975i$    & $0.070853i$              &        & $3$ & $0.0000-4.390185i$    & $0.143384i$\\
       & $4$ &$0.0000-5.628829i$    & $0.071497i$              &        & $4$ & $0.0000-4.533568i$    & $0.286479i$\\
       & $5$ &$0.0000-5.700326i$    & $0.070848i$              &        & $5$ & $0.0000-4.820047i$    & $0.429238i$\\
       & $6$ &$0.0000-5.771173i$    & $0.071474i$              &        & $6$ & $0.0000-5.249285i$    & $0.071432i$\\
       & $7$ &$0.0000-5.842647i$    & $0.070880i$              &        & $7$ & $0.0000-5.320716i$    & $0.071504i$\\
       & $8$ &$0.0000-5.913528i$    & $0.071424i$              &        & $8$ & $0.0000-5.392220i$    & $0.285833i$\\
       & $9$ &$0.0000-5.984952i$    & $0.070932i$              &        & $9$ & $0.0000-5.678054i$    & $0.071458i$\\
$4$    & $0$ &$0.0000-0.000724i$    & $0.000022i$              & $7$    & $0$ & $0.0000-0.000759i$    & $0.000002i$\\
       & $1$ &$0.0000-0.000746i$    & $0.000317i$              &        & $1$ & $0.0000-0.000761i$    & $0.029611i$\\
       & $2$ &$0.0000-0.001063i$    & $0.000391i$              &        & $2$ & $0.0000-0.030373i$    & $0.233173i$\\
       & $3$ &$0.0000-0.001454i$    & $3.696282i$              &        & $3$ & $0.0000-0.263545i$    & $4.471661i$\\
       & $4$ &$0.0000-3.697736i$    & $1.925258i$              &        & $4$ & $0.0000-4.735207i$    & $0.071766i$\\
       & $5$ &$0.0000-5.622994i$    & $0.071433i$              &        & $5$ & $0.0000-4.806973i$    & $0.071897i$\\     
       & $6$ &$0.0000-5.694427i$    & $0.071035i$              &        & $6$ & $0.0000-4.878870i$    & $0.071692i$\\
       & $7$ &$0.0000-5.765462i$    & $0.071433i$              &        & $7$ & $0.0000-4.950561i$    & $0.071789i$\\
       & $8$ &$0.0000-5.836895i$    & $0.071041i$              &        & $8$ & $0.0000-5.022350i$    & $0.071761i$\\
       & $9$ &$0.0000-5.907936i$    & $0.071400i$              &        & $9$ & $0.0000-5.094112i$    & $0.071639i$\\
       [1ex]
\hline\hline 
\end{tabular}
\end{table}


\begin{thebibliography}{73}
\expandafter\ifx\csname natexlab\endcsname\relax\def\natexlab#1{#1}\fi
\expandafter\ifx\csname bibnamefont\endcsname\relax
  \def\bibnamefont#1{#1}\fi
\expandafter\ifx\csname bibfnamefont\endcsname\relax
  \def\bibfnamefont#1{#1}\fi
\expandafter\ifx\csname citenamefont\endcsname\relax
  \def\citenamefont#1{#1}\fi
\expandafter\ifx\csname url\endcsname\relax
  \def\url#1{\texttt{#1}}\fi
\expandafter\ifx\csname urlprefix\endcsname\relax\def\urlprefix{URL }\fi
\providecommand{\bibinfo}[2]{#2}
\providecommand{\eprint}[2][]{\url{#2}}

\bibitem[{\citenamefont{Goroff and Sagnotti}(1986)}]{Sagnotti1985}
\bibinfo{author}{\bibfnamefont{M.~H.} \bibnamefont{Goroff}} \bibnamefont{and} \bibinfo{author}{\bibfnamefont{A.}~\bibnamefont{Sagnotti}}, \bibinfo{journal}{Nucl. Phys. B} \textbf{\bibinfo{volume}{266}}, \bibinfo{pages}{709} (\bibinfo{year}{1986}).

\bibitem[{\citenamefont{Weinberg}(1976)}]{Weinberg1}
\bibinfo{author}{\bibfnamefont{S.}~\bibnamefont{Weinberg}}, in \emph{\bibinfo{booktitle}{{14th International School of Subnuclear Physics: Understanding the Fundamental Constitutents of Matter}}} (\bibinfo{year}{1976}).

\bibitem[{\citenamefont{Weinberg}(1980)}]{Weinberg2}
\bibinfo{author}{\bibfnamefont{S.}~\bibnamefont{Weinberg}}, \emph{\bibinfo{title}{{Ultaviolet divergences in quantum theory of gravitation}}} (\bibinfo{year}{1980}), pp. \bibinfo{pages}{790--831}.

\bibitem[{\citenamefont{Wetterich}(1993)}]{Wetterich93}
\bibinfo{author}{\bibfnamefont{C.}~\bibnamefont{Wetterich}}, \bibinfo{journal}{Phys. Lett. B} \textbf{\bibinfo{volume}{301}}, \bibinfo{pages}{90} (\bibinfo{year}{1993}), \eprint{1710.05815}.

\bibitem[{\citenamefont{Reuter}(1998)}]{Reuter98}
\bibinfo{author}{\bibfnamefont{M.}~\bibnamefont{Reuter}}, \bibinfo{journal}{Phys. Rev. D} \textbf{\bibinfo{volume}{57}}, \bibinfo{pages}{971} (\bibinfo{year}{1998}), \eprint{hep-th/9605030}.

\bibitem[{\citenamefont{Niedermaier and Reuter}(2006)}]{Niedermaier2006}
\bibinfo{author}{\bibfnamefont{M.}~\bibnamefont{Niedermaier}} \bibnamefont{and} \bibinfo{author}{\bibfnamefont{M.}~\bibnamefont{Reuter}}, \bibinfo{journal}{Living Rev. Rel.} \textbf{\bibinfo{volume}{9}}, \bibinfo{pages}{5} (\bibinfo{year}{2006}).

\bibitem[{\citenamefont{Koch et~al.}(2016)\citenamefont{Koch, Reyes, and Rinc{\'o}n}}]{Koch16}
\bibinfo{author}{\bibfnamefont{B.}~\bibnamefont{Koch}}, \bibinfo{author}{\bibfnamefont{I.~A.} \bibnamefont{Reyes}}, \bibnamefont{and} \bibinfo{author}{\bibfnamefont{{\'A}.}~\bibnamefont{Rinc{\'o}n}}, \bibinfo{journal}{Class. Quant. Grav.} \textbf{\bibinfo{volume}{33}}, \bibinfo{pages}{225010} (\bibinfo{year}{2016}), \eprint{1606.04123}.

\bibitem[{\citenamefont{Contreras et~al.}(2020)\citenamefont{Contreras, Rinc{\'o}n, Panotopoulos, Bargue{\~n}o, and Koch}}]{Ricon19}
\bibinfo{author}{\bibfnamefont{E.}~\bibnamefont{Contreras}}, \bibinfo{author}{\bibfnamefont{{\'A}.}~\bibnamefont{Rinc{\'o}n}}, \bibinfo{author}{\bibfnamefont{G.}~\bibnamefont{Panotopoulos}}, \bibinfo{author}{\bibfnamefont{P.}~\bibnamefont{Bargue{\~n}o}}, \bibnamefont{and} \bibinfo{author}{\bibfnamefont{B.}~\bibnamefont{Koch}}, \bibinfo{journal}{Phys. Rev. D} \textbf{\bibinfo{volume}{101}}, \bibinfo{pages}{064053} (\bibinfo{year}{2020}), \eprint{1906.06990}.

\bibitem[{\citenamefont{Hassannejad et~al.}(2025)\citenamefont{Hassannejad, Lambiase, Scardigli, and Shojai}}]{Hassannejad2025PRD}
\bibinfo{author}{\bibfnamefont{R.}~\bibnamefont{Hassannejad}}, \bibinfo{author}{\bibfnamefont{G.}~\bibnamefont{Lambiase}}, \bibinfo{author}{\bibfnamefont{F.}~\bibnamefont{Scardigli}}, \bibnamefont{and} \bibinfo{author}{\bibfnamefont{F.}~\bibnamefont{Shojai}}, \bibinfo{journal}{Phys. Rev. D} \textbf{\bibinfo{volume}{111}}, \bibinfo{pages}{064069} (\bibinfo{year}{2025}).

\bibitem[{\citenamefont{Bonanno and M.~Reuter}(2000)}]{Bonanno2000PRD}
\bibinfo{author}{\bibfnamefont{A.}~\bibnamefont{Bonanno}} \bibnamefont{and} \bibinfo{author}{\bibfnamefont{M.}~\bibnamefont{M.~Reuter}}, \bibinfo{journal}{Phys. Rev. D} \textbf{\bibinfo{volume}{62}}, \bibinfo{pages}{043008} (\bibinfo{year}{2000}).

\bibitem[{\citenamefont{Batic et~al.}(2026)\citenamefont{Batic, Dutykh, and Scardigli}}]{BDS-EPJC2026}
\bibinfo{author}{\bibfnamefont{D.}~\bibnamefont{Batic}}, \bibinfo{author}{\bibfnamefont{D.}~\bibnamefont{Dutykh}}, \bibnamefont{and} \bibinfo{author}{\bibfnamefont{F.}~\bibnamefont{Scardigli}}, \bibinfo{journal}{Eur. Phys. J. C} \textbf{\bibinfo{volume}{86}}, \bibinfo{pages}{165} (\bibinfo{year}{2026}), \eprint{2602.19833}.

\bibitem[{\citenamefont{Rincon and Panotopoulos}(2020)}]{Rincon2020PDU}
\bibinfo{author}{\bibfnamefont{A.}~\bibnamefont{Rincon}} \bibnamefont{and} \bibinfo{author}{\bibfnamefont{G.}~\bibnamefont{Panotopoulos}}, \bibinfo{journal}{Phys. Dark Universe} \textbf{\bibinfo{volume}{30}}, \bibinfo{pages}{100639} (\bibinfo{year}{2020}).

\bibitem[{\citenamefont{Konoplya et~al.}(2022)\citenamefont{Konoplya, Zinhailo, Kunz, Stuchlik, and Zhidenko}}]{Konoplya2022JCAP}
\bibinfo{author}{\bibfnamefont{R.~A.} \bibnamefont{Konoplya}}, \bibinfo{author}{\bibfnamefont{A.~F.} \bibnamefont{Zinhailo}}, \bibinfo{author}{\bibfnamefont{J.}~\bibnamefont{Kunz}}, \bibinfo{author}{\bibfnamefont{Z.}~\bibnamefont{Stuchlik}}, \bibnamefont{and} \bibinfo{author}{\bibfnamefont{A.}~\bibnamefont{Zhidenko}}, \bibinfo{journal}{JCAP} \textbf{\bibinfo{volume}{10}}, \bibinfo{pages}{091} (\bibinfo{year}{2022}).

\bibitem[{\citenamefont{Abbott and et~al.}(2016)}]{Abbott2016PRL}
\bibinfo{author}{\bibfnamefont{B.~P.} \bibnamefont{Abbott}} \bibnamefont{and} \bibinfo{author}{\bibnamefont{et~al.}}, \bibinfo{journal}{Phys. Rev. Lett.} \textbf{\bibinfo{volume}{116}}, \bibinfo{pages}{061102} (\bibinfo{year}{2016}).

\bibitem[{\citenamefont{Abbott et~al.}(2019)}]{Abbott2019}
\bibinfo{author}{\bibfnamefont{B.~P.} \bibnamefont{Abbott}} \bibnamefont{et~al.} (\bibinfo{collaboration}{LIGO Scientific, Virgo}), \bibinfo{journal}{Phys. Rev. X} \textbf{\bibinfo{volume}{9}}, \bibinfo{pages}{031040} (\bibinfo{year}{2019}), \eprint{1811.12907}.

\bibitem[{\citenamefont{Regge and Wheeler}(1957)}]{Regge1957PR}
\bibinfo{author}{\bibfnamefont{T.}~\bibnamefont{Regge}} \bibnamefont{and} \bibinfo{author}{\bibfnamefont{J.~A.} \bibnamefont{Wheeler}}, \bibinfo{journal}{Phys. Rev.} \textbf{\bibinfo{volume}{108}}, \bibinfo{pages}{1063} (\bibinfo{year}{1957}).

\bibitem[{\citenamefont{Teukolsky}(1972)}]{Teukolsky1972}
\bibinfo{author}{\bibfnamefont{S.~A.} \bibnamefont{Teukolsky}}, \bibinfo{journal}{Phys. Rev. Lett.} \textbf{\bibinfo{volume}{29}}, \bibinfo{pages}{1114} (\bibinfo{year}{1972}).

\bibitem[{\citenamefont{Ferrari and Gualtieri}(2008)}]{Ferrari2008}
\bibinfo{author}{\bibfnamefont{V.}~\bibnamefont{Ferrari}} \bibnamefont{and} \bibinfo{author}{\bibfnamefont{L.}~\bibnamefont{Gualtieri}}, \bibinfo{journal}{Gen. Rel. Grav.} \textbf{\bibinfo{volume}{40}}, \bibinfo{pages}{945} (\bibinfo{year}{2008}), \eprint{0709.0657}.

\bibitem[{\citenamefont{Cardoso and Lemos}(2003)}]{Cardoso2003}
\bibinfo{author}{\bibfnamefont{V.}~\bibnamefont{Cardoso}} \bibnamefont{and} \bibinfo{author}{\bibfnamefont{J.~P.~S.} \bibnamefont{Lemos}}, \bibinfo{journal}{Phys. Rev. D} \textbf{\bibinfo{volume}{67}}, \bibinfo{pages}{084020} (\bibinfo{year}{2003}), \eprint{gr-qc/0301078}.

\bibitem[{\citenamefont{Berti et~al.}(2009)\citenamefont{Berti, Cardoso, and Starinets}}]{Berti2009CQG}
\bibinfo{author}{\bibfnamefont{E.}~\bibnamefont{Berti}}, \bibinfo{author}{\bibfnamefont{V.}~\bibnamefont{Cardoso}}, \bibnamefont{and} \bibinfo{author}{\bibfnamefont{A.~O.} \bibnamefont{Starinets}}, \bibinfo{journal}{Class. Quantum Grav.} \textbf{\bibinfo{volume}{26}}, \bibinfo{pages}{163001} (\bibinfo{year}{2009}).

\bibitem[{\citenamefont{Konoplya and Zhidenko}(2011)}]{Konoplya2011RMP}
\bibinfo{author}{\bibfnamefont{R.~A.} \bibnamefont{Konoplya}} \bibnamefont{and} \bibinfo{author}{\bibfnamefont{A.}~\bibnamefont{Zhidenko}}, \bibinfo{journal}{Rev. Mod. Phys.} \textbf{\bibinfo{volume}{83}}, \bibinfo{pages}{793} (\bibinfo{year}{2011}).

\bibitem[{\citenamefont{Liu et~al.}(2012)\citenamefont{Liu, Yang, Zhai, and Li}}]{Liu2012}
\bibinfo{author}{\bibfnamefont{D.-J.} \bibnamefont{Liu}}, \bibinfo{author}{\bibfnamefont{B.}~\bibnamefont{Yang}}, \bibinfo{author}{\bibfnamefont{Y.-J.} \bibnamefont{Zhai}}, \bibnamefont{and} \bibinfo{author}{\bibfnamefont{X.-Z.} \bibnamefont{Li}}, \bibinfo{journal}{Class. Quant. Grav.} \textbf{\bibinfo{volume}{29}}, \bibinfo{pages}{145009} (\bibinfo{year}{2012}), \eprint{1205.4792}.

\bibitem[{\citenamefont{Franchini and V{\"o}lkel}(2024)}]{Franchini2023}
\bibinfo{author}{\bibfnamefont{N.}~\bibnamefont{Franchini}} \bibnamefont{and} \bibinfo{author}{\bibfnamefont{S.~H.} \bibnamefont{V{\"o}lkel}}, \emph{\bibinfo{title}{{Testing General Relativity with Black Hole Quasi-normal Modes}}} (\bibinfo{year}{2024}), \eprint{2305.01696}.

\bibitem[{\citenamefont{Giesler et~al.}(2019)\citenamefont{Giesler, Isi, Scheel, and Teukolsky}}]{Giesler2019PRX}
\bibinfo{author}{\bibfnamefont{M.}~\bibnamefont{Giesler}}, \bibinfo{author}{\bibfnamefont{M.}~\bibnamefont{Isi}}, \bibinfo{author}{\bibfnamefont{M.~A.} \bibnamefont{Scheel}}, \bibnamefont{and} \bibinfo{author}{\bibfnamefont{S.~A.} \bibnamefont{Teukolsky}}, \bibinfo{journal}{Phys. Rev. X} \textbf{\bibinfo{volume}{9}}, \bibinfo{pages}{041060} (\bibinfo{year}{2019}).

\bibitem[{\citenamefont{Spina et~al.}(2024)\citenamefont{Spina, Silveravalle, and Bonanno}}]{Spina24}
\bibinfo{author}{\bibfnamefont{A.}~\bibnamefont{Spina}}, \bibinfo{author}{\bibfnamefont{S.}~\bibnamefont{Silveravalle}}, \bibnamefont{and} \bibinfo{author}{\bibfnamefont{A.}~\bibnamefont{Bonanno}}, in \emph{\bibinfo{booktitle}{{17th Marcel Grossmann Meeting}}} (\bibinfo{year}{2024}), \eprint{2410.05936}.

\bibitem[{\citenamefont{Konoplya and Zhidenko}(2022)}]{Konoplya2022PRL}
\bibinfo{author}{\bibfnamefont{R.~A.} \bibnamefont{Konoplya}} \bibnamefont{and} \bibinfo{author}{\bibfnamefont{A.}~\bibnamefont{Zhidenko}}, \bibinfo{journal}{Phys. Rev. Lett.} \textbf{\bibinfo{volume}{128}}, \bibinfo{pages}{091104} (\bibinfo{year}{2022}).

\bibitem[{\citenamefont{Konoplya et~al.}(2023)\citenamefont{Konoplya, Stuchlik, Zhidenko, and Zinhailo}}]{KonoplyaPRD2023}
\bibinfo{author}{\bibfnamefont{R.~A.} \bibnamefont{Konoplya}}, \bibinfo{author}{\bibfnamefont{Z.}~\bibnamefont{Stuchlik}}, \bibinfo{author}{\bibfnamefont{A.}~\bibnamefont{Zhidenko}}, \bibnamefont{and} \bibinfo{author}{\bibfnamefont{A.~F.} \bibnamefont{Zinhailo}}, \bibinfo{journal}{Phys. Rev. D} \textbf{\bibinfo{volume}{107}}, \bibinfo{pages}{104050} (\bibinfo{year}{2023}), \eprint{2303.01987}.

\bibitem[{\citenamefont{Flachi and Lemos}(2013)}]{Flachi2013}
\bibinfo{author}{\bibfnamefont{A.}~\bibnamefont{Flachi}} \bibnamefont{and} \bibinfo{author}{\bibfnamefont{J.~P.~S.} \bibnamefont{Lemos}}, \bibinfo{journal}{Phys. Rev. D} \textbf{\bibinfo{volume}{87}}, \bibinfo{pages}{024034} (\bibinfo{year}{2013}), \eprint{1211.6212}.

\bibitem[{\citenamefont{Dutta~Roy and Kar}(2022)}]{DuttaRoy2022}
\bibinfo{author}{\bibfnamefont{P.}~\bibnamefont{Dutta~Roy}} \bibnamefont{and} \bibinfo{author}{\bibfnamefont{S.}~\bibnamefont{Kar}}, \bibinfo{journal}{Phys. Rev. D} \textbf{\bibinfo{volume}{106}}, \bibinfo{pages}{044028} (\bibinfo{year}{2022}), \eprint{2206.04505}.

\bibitem[{\citenamefont{Batic and Dutykh}(2024{\natexlab{a}})}]{Batic2024CQG}
\bibinfo{author}{\bibfnamefont{D.}~\bibnamefont{Batic}} \bibnamefont{and} \bibinfo{author}{\bibfnamefont{D.}~\bibnamefont{Dutykh}}, \bibinfo{journal}{Class. Quantum Grav.} \textbf{\bibinfo{volume}{41}}, \bibinfo{pages}{215003} (\bibinfo{year}{2024}{\natexlab{a}}).

\bibitem[{\citenamefont{Batic and Dutykh}(2024{\natexlab{b}})}]{Batic2024EPJC}
\bibinfo{author}{\bibfnamefont{D.}~\bibnamefont{Batic}} \bibnamefont{and} \bibinfo{author}{\bibfnamefont{D.}~\bibnamefont{Dutykh}}, \bibinfo{journal}{Eur. Phys. J. C} \textbf{\bibinfo{volume}{84}}, \bibinfo{pages}{622} (\bibinfo{year}{2024}{\natexlab{b}}).

\bibitem[{\citenamefont{Batic et~al.}(2024)\citenamefont{Batic, Dutykh, and Giacchini}}]{Batic2024PRD}
\bibinfo{author}{\bibfnamefont{D.}~\bibnamefont{Batic}}, \bibinfo{author}{\bibfnamefont{D.}~\bibnamefont{Dutykh}}, \bibnamefont{and} \bibinfo{author}{\bibfnamefont{B.~L.} \bibnamefont{Giacchini}}, \bibinfo{journal}{Phys. Rev. D} \textbf{\bibinfo{volume}{110}}, \bibinfo{pages}{084032} (\bibinfo{year}{2024}).

\bibitem[{\citenamefont{Batic and Dutykh}(2025)}]{Batic2025EPJC}
\bibinfo{author}{\bibfnamefont{D.}~\bibnamefont{Batic}} \bibnamefont{and} \bibinfo{author}{\bibfnamefont{D.}~\bibnamefont{Dutykh}}, \bibinfo{journal}{Eur. Phys. J. C} \textbf{\bibinfo{volume}{85}}, \bibinfo{pages}{144} (\bibinfo{year}{2025}).

\bibitem[{\citenamefont{Batic et~al.}(2025)\citenamefont{Batic, Dutykh, and Beek}}]{Batic2025CQG}
\bibinfo{author}{\bibfnamefont{D.}~\bibnamefont{Batic}}, \bibinfo{author}{\bibfnamefont{D.}~\bibnamefont{Dutykh}}, \bibnamefont{and} \bibinfo{author}{\bibfnamefont{J.~J.} \bibnamefont{Beek}}, \bibinfo{journal}{Class. Quantum Grav.} \textbf{\bibinfo{volume}{42}}, \bibinfo{pages}{085003} (\bibinfo{year}{2025}).

\bibitem[{\citenamefont{Lambiase et~al.}(2023)\citenamefont{Lambiase, Pantig, Gogoi, and Övgün}}]{Lambiase2023EPJC}
\bibinfo{author}{\bibfnamefont{G.}~\bibnamefont{Lambiase}}, \bibinfo{author}{\bibfnamefont{R.~C.} \bibnamefont{Pantig}}, \bibinfo{author}{\bibfnamefont{D.~J.} \bibnamefont{Gogoi}}, \bibnamefont{and} \bibinfo{author}{\bibfnamefont{A.}~\bibnamefont{Övgün}}, \bibinfo{journal}{Eur. Phys. J. C} \textbf{\bibinfo{volume}{83}}, \bibinfo{pages}{679} (\bibinfo{year}{2023}).

\bibitem[{\citenamefont{Malik}(2024)}]{Malik2024EPL}
\bibinfo{author}{\bibfnamefont{Z.}~\bibnamefont{Malik}}, \bibinfo{journal}{EPL} \textbf{\bibinfo{volume}{147}}, \bibinfo{pages}{69001} (\bibinfo{year}{2024}).

\bibitem[{\citenamefont{Stashko}(2024)}]{Stashko2024PRD}
\bibinfo{author}{\bibfnamefont{O.}~\bibnamefont{Stashko}}, \bibinfo{journal}{Phys. Rev. D} \textbf{\bibinfo{volume}{110}}, \bibinfo{pages}{084016} (\bibinfo{year}{2024}).

\bibitem[{\citenamefont{Lambiase and Scardigli}(2022)}]{Lambiase2022PRD}
\bibinfo{author}{\bibfnamefont{G.}~\bibnamefont{Lambiase}} \bibnamefont{and} \bibinfo{author}{\bibfnamefont{F.}~\bibnamefont{Scardigli}}, \bibinfo{journal}{Phys. Rev. D} \textbf{\bibinfo{volume}{105}}, \bibinfo{pages}{124054} (\bibinfo{year}{2022}).

\bibitem[{\citenamefont{Bonanno and Reuter}(2004)}]{Bonanno2004}
\bibinfo{author}{\bibfnamefont{A.}~\bibnamefont{Bonanno}} \bibnamefont{and} \bibinfo{author}{\bibfnamefont{M.}~\bibnamefont{Reuter}}, \bibinfo{journal}{Int. J. Mod. Phys. D} \textbf{\bibinfo{volume}{13}}, \bibinfo{pages}{107} (\bibinfo{year}{2004}), \eprint{astro-ph/0210472}.

\bibitem[{\citenamefont{Scardigli and Lambiase}(2023)}]{Scardigli2023PRD}
\bibinfo{author}{\bibfnamefont{F.}~\bibnamefont{Scardigli}} \bibnamefont{and} \bibinfo{author}{\bibfnamefont{G.}~\bibnamefont{Lambiase}}, \bibinfo{journal}{Phys. Rev. D} \textbf{\bibinfo{volume}{107}}, \bibinfo{pages}{104001} (\bibinfo{year}{2023}).

\bibitem[{\citenamefont{Hamber and Liu}(1995)}]{Hamber1995PLB}
\bibinfo{author}{\bibfnamefont{H.~W.} \bibnamefont{Hamber}} \bibnamefont{and} \bibinfo{author}{\bibfnamefont{S.}~\bibnamefont{Liu}}, \bibinfo{journal}{Phys. Lett. B} \textbf{\bibinfo{volume}{357}}, \bibinfo{pages}{51} (\bibinfo{year}{1995}).

\bibitem[{\citenamefont{Bjerrum-Bohr et~al.}(2003{\natexlab{a}})\citenamefont{Bjerrum-Bohr, Donoghue, and Holstein}}]{Bjerrum2003PRD}
\bibinfo{author}{\bibfnamefont{N.~E.~J.} \bibnamefont{Bjerrum-Bohr}}, \bibinfo{author}{\bibfnamefont{J.~F.} \bibnamefont{Donoghue}}, \bibnamefont{and} \bibinfo{author}{\bibfnamefont{B.~R.} \bibnamefont{Holstein}}, \bibinfo{journal}{Phys. Rev. D} \textbf{\bibinfo{volume}{68}}, \bibinfo{pages}{084005} (\bibinfo{year}{2003}{\natexlab{a}}), \bibinfo{note}{erratum-ibid: Phys. Rev. D {\bf 71}, 069904 (2005)}.

\bibitem[{\citenamefont{Khriplovich and Kirilin}(2002)}]{Khriplovich2002JETP}
\bibinfo{author}{\bibfnamefont{I.~B.} \bibnamefont{Khriplovich}} \bibnamefont{and} \bibinfo{author}{\bibfnamefont{G.~G.} \bibnamefont{Kirilin}}, \bibinfo{journal}{J. Exp. Theor. Phys.} \textbf{\bibinfo{volume}{95}}, \bibinfo{pages}{981} (\bibinfo{year}{2002}).

\bibitem[{\citenamefont{Bjerrum-Bohr et~al.}(2003{\natexlab{b}})\citenamefont{Bjerrum-Bohr, Donoghue, and Holstein}}]{Bjerrum2003PRDa}
\bibinfo{author}{\bibfnamefont{N.~E.~J.} \bibnamefont{Bjerrum-Bohr}}, \bibinfo{author}{\bibfnamefont{J.~F.} \bibnamefont{Donoghue}}, \bibnamefont{and} \bibinfo{author}{\bibfnamefont{B.~R.} \bibnamefont{Holstein}}, \bibinfo{journal}{Phys. Rev. D} \textbf{\bibinfo{volume}{67}}, \bibinfo{pages}{084033} (\bibinfo{year}{2003}{\natexlab{b}}), \bibinfo{note}{erratum-ibid: Phys. Rev. D {\bf 71}, 069903 (2005)}.

\bibitem[{\citenamefont{Khriplovich and Kirilin}(2004)}]{Khriplovich2004JETP}
\bibinfo{author}{\bibfnamefont{I.~B.} \bibnamefont{Khriplovich}} \bibnamefont{and} \bibinfo{author}{\bibfnamefont{G.~G.} \bibnamefont{Kirilin}}, \bibinfo{journal}{J. Exp. Theor. Phys.} \textbf{\bibinfo{volume}{98}}, \bibinfo{pages}{1063} (\bibinfo{year}{2004}).

\bibitem[{\citenamefont{Akhundov and Shiekh}(2008)}]{Akhundov2008EJTP}
\bibinfo{author}{\bibfnamefont{A.}~\bibnamefont{Akhundov}} \bibnamefont{and} \bibinfo{author}{\bibfnamefont{A.}~\bibnamefont{Shiekh}}, \bibinfo{journal}{EJTP} \textbf{\bibinfo{volume}{5}}, \bibinfo{pages}{1} (\bibinfo{year}{2008}).

\bibitem[{\citenamefont{Kiefer}(2013)}]{Kiefer}
\bibinfo{author}{\bibfnamefont{C.}~\bibnamefont{Kiefer}}, \bibinfo{journal}{J. Phys. Conf. Ser.} \textbf{\bibinfo{volume}{442}}, \bibinfo{pages}{012025} (\bibinfo{year}{2013}).

\bibitem[{\citenamefont{Bjerrum-Bohr et~al.}(2015)\citenamefont{Bjerrum-Bohr, Donoghue, Holstein, Plant$\acute{e}$, and Vanhove}}]{Bjerrum2015PRL}
\bibinfo{author}{\bibfnamefont{N.~E.~J.} \bibnamefont{Bjerrum-Bohr}}, \bibinfo{author}{\bibfnamefont{J.~F.} \bibnamefont{Donoghue}}, \bibinfo{author}{\bibfnamefont{B.~R.} \bibnamefont{Holstein}}, \bibinfo{author}{\bibfnamefont{L.}~\bibnamefont{Plant$\acute{e}$}}, \bibnamefont{and} \bibinfo{author}{\bibfnamefont{P.}~\bibnamefont{Vanhove}}, \bibinfo{journal}{Phys. Rev. Lett.} \textbf{\bibinfo{volume}{114}}, \bibinfo{pages}{061301} (\bibinfo{year}{2015}).

\bibitem[{\citenamefont{Donoghue and Holstein}(2015)}]{Donoghue2015JPG}
\bibinfo{author}{\bibfnamefont{J.~F.} \bibnamefont{Donoghue}} \bibnamefont{and} \bibinfo{author}{\bibfnamefont{B.~R.} \bibnamefont{Holstein}}, \bibinfo{journal}{J. Phys. G: Nucl. Part. Phys.} \textbf{\bibinfo{volume}{42}}, \bibinfo{pages}{103102} (\bibinfo{year}{2015}).

\bibitem[{\citenamefont{Bargueño et~al.}(2016)\citenamefont{Bargueño, Medina, Nowakowski, and Batic}}]{Batic2016EPJC}
\bibinfo{author}{\bibfnamefont{P.}~\bibnamefont{Bargueño}}, \bibinfo{author}{\bibfnamefont{S.~B.} \bibnamefont{Medina}}, \bibinfo{author}{\bibfnamefont{M.}~\bibnamefont{Nowakowski}}, \bibnamefont{and} \bibinfo{author}{\bibfnamefont{D.}~\bibnamefont{Batic}}, \bibinfo{journal}{Eur. Phys. J. C} \textbf{\bibinfo{volume}{76}}, \bibinfo{pages}{543} (\bibinfo{year}{2016}).

\bibitem[{\citenamefont{Bargueño et~al.}(2017)\citenamefont{Bargueño, Medina, Nowakowski, and Batic}}]{Batic2017EPL}
\bibinfo{author}{\bibfnamefont{P.}~\bibnamefont{Bargueño}}, \bibinfo{author}{\bibfnamefont{S.~B.} \bibnamefont{Medina}}, \bibinfo{author}{\bibfnamefont{M.}~\bibnamefont{Nowakowski}}, \bibnamefont{and} \bibinfo{author}{\bibfnamefont{D.}~\bibnamefont{Batic}}, \bibinfo{journal}{EPL} \textbf{\bibinfo{volume}{117}}, \bibinfo{pages}{6006} (\bibinfo{year}{2017}).

\bibitem[{\citenamefont{Batic et~al.}(2019)\citenamefont{Batic, Kelkar, Nowakowski, and Redway}}]{Batic2019EPJC}
\bibinfo{author}{\bibfnamefont{D.}~\bibnamefont{Batic}}, \bibinfo{author}{\bibfnamefont{N.~G.} \bibnamefont{Kelkar}}, \bibinfo{author}{\bibfnamefont{M.}~\bibnamefont{Nowakowski}}, \bibnamefont{and} \bibinfo{author}{\bibfnamefont{K.}~\bibnamefont{Redway}}, \bibinfo{journal}{Eur. Phys. J. C} \textbf{\bibinfo{volume}{79}}, \bibinfo{pages}{581} (\bibinfo{year}{2019}).

\bibitem[{\citenamefont{Ince}(1956)}]{Ince1956}
\bibinfo{author}{\bibfnamefont{E.~L.} \bibnamefont{Ince}}, \emph{\bibinfo{title}{Ordinary Differential Equations}} (\bibinfo{publisher}{Dover: New York}, \bibinfo{year}{1956}).

\bibitem[{\citenamefont{Olver}(1994)}]{Olver1994MAA}
\bibinfo{author}{\bibfnamefont{F.~W.~J.} \bibnamefont{Olver}}, \bibinfo{journal}{Methods Appl. Anal.} \textbf{\bibinfo{volume}{1}}, \bibinfo{pages}{1} (\bibinfo{year}{1994}).

\bibitem[{\citenamefont{Bender and Orszag}(1999)}]{Bender1999}
\bibinfo{author}{\bibfnamefont{C.~M.} \bibnamefont{Bender}} \bibnamefont{and} \bibinfo{author}{\bibfnamefont{S.~A.} \bibnamefont{Orszag}}, \emph{\bibinfo{title}{Advanced Mathematical Methods for Scientists and Engineers I: Asymptotic Methods and Perturbation Theory}} (\bibinfo{publisher}{Springer: New York}, \bibinfo{year}{1999}).

\bibitem[{\citenamefont{Trefethen}(2000)}]{Trefethen2000}
\bibinfo{author}{\bibfnamefont{L.~N.} \bibnamefont{Trefethen}}, \emph{\bibinfo{title}{Spectral methods in MatLab}} (\bibinfo{publisher}{Society for Industrial and Applied Mathematics, Philadelphia, PA, USA}, \bibinfo{year}{2000}), \urlprefix\url{http://web.comlab.ox.ac.uk/oucl/work/nick.trefethen/spectral.html}.

\bibitem[{\citenamefont{Boyd}(2000)}]{Boyd2000}
\bibinfo{author}{\bibfnamefont{J.~P.} \bibnamefont{Boyd}}, \emph{\bibinfo{title}{Chebyshev and Fourier Spectral Methods}} (\bibinfo{publisher}{Dover Publications, New York}, \bibinfo{year}{2000}), \bibinfo{edition}{2nd} ed.

\bibitem[{\citenamefont{Fox and Parker}(1968)}]{Fox1968}
\bibinfo{author}{\bibfnamefont{L.}~\bibnamefont{Fox}} \bibnamefont{and} \bibinfo{author}{\bibfnamefont{I.~B.} \bibnamefont{Parker}}, \emph{\bibinfo{title}{Chebyshev Polynomials in Numerical Analysis}} (\bibinfo{publisher}{Oxford University Press}, \bibinfo{year}{1968}).

\bibitem[{\citenamefont{Tisseur and Meerbergen}(2001)}]{Tisseur2001}
\bibinfo{author}{\bibfnamefont{F.}~\bibnamefont{Tisseur}} \bibnamefont{and} \bibinfo{author}{\bibfnamefont{K.}~\bibnamefont{Meerbergen}}, \bibinfo{journal}{SIAM Review} \textbf{\bibinfo{volume}{43}}, \bibinfo{pages}{235} (\bibinfo{year}{2001}), ISSN \bibinfo{issn}{0036-1445}.

\bibitem[{\citenamefont{Holodoborodko}(2023)}]{mct2015}
\bibinfo{author}{\bibfnamefont{P.}~\bibnamefont{Holodoborodko}}, \emph{\bibinfo{title}{{Multiprecision Computing Toolbox for MATLAB 5.2.7.15522}}} (\bibinfo{publisher}{Advanpix LLC.}, \bibinfo{address}{Yokohama, Japan}, \bibinfo{year}{2023}).

\bibitem[{\citenamefont{Leaver}(1985)}]{Leaver1985PRSLA}
\bibinfo{author}{\bibfnamefont{E.~W.} \bibnamefont{Leaver}}, \bibinfo{journal}{Proc. R. Soc. Lond. A} \textbf{\bibinfo{volume}{402}}, \bibinfo{pages}{285} (\bibinfo{year}{1985}).

\bibitem[{\citenamefont{Nollert}(1993)}]{Nollert1993PRD}
\bibinfo{author}{\bibfnamefont{H.-P.} \bibnamefont{Nollert}}, \bibinfo{journal}{Phys. Rev. D} \textbf{\bibinfo{volume}{47}}, \bibinfo{pages}{5253} (\bibinfo{year}{1993}).

\bibitem[{\citenamefont{Hod}(1998)}]{Hod1998PRL}
\bibinfo{author}{\bibfnamefont{S.}~\bibnamefont{Hod}}, \bibinfo{journal}{Phys Rev. Lett.} \textbf{\bibinfo{volume}{81}}, \bibinfo{pages}{4293} (\bibinfo{year}{1998}).

\bibitem[{\citenamefont{Motl and Neitzke}(2003)}]{Motl2003ATMP}
\bibinfo{author}{\bibfnamefont{L.}~\bibnamefont{Motl}} \bibnamefont{and} \bibinfo{author}{\bibfnamefont{A.}~\bibnamefont{Neitzke}}, \bibinfo{journal}{Adv. Theor. Math. Phys.} \textbf{\bibinfo{volume}{7}}, \bibinfo{pages}{307} (\bibinfo{year}{2003}).

\bibitem[{\citenamefont{Natário and Schiappa}(2004)}]{Natario2004ATMP}
\bibinfo{author}{\bibfnamefont{J.}~\bibnamefont{Natário}} \bibnamefont{and} \bibinfo{author}{\bibfnamefont{R.}~\bibnamefont{Schiappa}}, \bibinfo{journal}{Adv. Theor. Math. Phys.} \textbf{\bibinfo{volume}{8}}, \bibinfo{pages}{1001} (\bibinfo{year}{2004}).

\bibitem[{\citenamefont{Maggiore}(2008)}]{Maggiore2008PRL}
\bibinfo{author}{\bibfnamefont{M.}~\bibnamefont{Maggiore}}, \bibinfo{journal}{Phys. Rev. Lett.} \textbf{\bibinfo{volume}{100}}, \bibinfo{pages}{141301} (\bibinfo{year}{2008}).

\bibitem[{\citenamefont{Skakala}(2012)}]{Skakala2012JHEP}
\bibinfo{author}{\bibfnamefont{J.}~\bibnamefont{Skakala}}, \bibinfo{journal}{JHEP} \textbf{\bibinfo{volume}{06}}, \bibinfo{pages}{094} (\bibinfo{year}{2012}).

\bibitem[{\citenamefont{Yi et~al.}(2024)\citenamefont{Yi, Kuntz, Barausse, Berti, Cheung, Kritos, and Maselli}}]{Yi2024}
\bibinfo{author}{\bibfnamefont{S.}~\bibnamefont{Yi}}, \bibinfo{author}{\bibfnamefont{A.}~\bibnamefont{Kuntz}}, \bibinfo{author}{\bibfnamefont{E.}~\bibnamefont{Barausse}}, \bibinfo{author}{\bibfnamefont{E.}~\bibnamefont{Berti}}, \bibinfo{author}{\bibfnamefont{M.~H.-Y.} \bibnamefont{Cheung}}, \bibinfo{author}{\bibfnamefont{K.}~\bibnamefont{Kritos}}, \bibnamefont{and} \bibinfo{author}{\bibfnamefont{A.}~\bibnamefont{Maselli}}, \bibinfo{journal}{Phys. Rev. D} \textbf{\bibinfo{volume}{109}}, \bibinfo{pages}{124029} (\bibinfo{year}{2024}), \eprint{2403.09767}.

\bibitem[{\citenamefont{Berti et~al.}(2025)}]{Berti2025}
\bibinfo{author}{\bibfnamefont{E.}~\bibnamefont{Berti}} \bibnamefont{et~al.} (\bibinfo{year}{2025}), \eprint{2505.23895}.

\bibitem[{\citenamefont{CERN}(2025)}]{FCC}
\bibinfo{author}{\bibnamefont{CERN}}, \emph{\bibinfo{title}{Future circular collider}} (\bibinfo{year}{2025}), \urlprefix\url{https://home.cern/science/accelerators/future-circular-collider}.

\bibitem[{\citenamefont{Scardigli et~al.}(2011)\citenamefont{Scardigli, Gruber, and Chen}}]{Scardigli2010}
\bibinfo{author}{\bibfnamefont{F.}~\bibnamefont{Scardigli}}, \bibinfo{author}{\bibfnamefont{C.}~\bibnamefont{Gruber}}, \bibnamefont{and} \bibinfo{author}{\bibfnamefont{P.}~\bibnamefont{Chen}}, \bibinfo{journal}{Phys. Rev. D} \textbf{\bibinfo{volume}{83}}, \bibinfo{pages}{063507} (\bibinfo{year}{2011}), \eprint{1009.0882}.

\bibitem[{\citenamefont{LISA}(2025)}]{LISA}
\bibinfo{author}{\bibnamefont{LISA}}, \emph{\bibinfo{title}{Laser interferometer space antenna}} (\bibinfo{year}{2025}), \urlprefix\url{https://lisa.nasa.gov/}.

\bibitem[{\citenamefont{ET}(2025)}]{ET}
\bibinfo{author}{\bibnamefont{ET}}, \emph{\bibinfo{title}{Einstein telescope}} (\bibinfo{year}{2025}), \urlprefix\url{https://www.einstein-telescope.it/en/home-en/}.

\end{thebibliography}
\end{document}